\documentclass[
  twocolumn,
  final,
  ]{elsarticle}

\usepackage{framed,multirow}

\usepackage[utf8]{inputenc}
\usepackage[T1]{fontenc}
\usepackage{microtype}

\usepackage{geometry}
\global\bibsep=0pt
\input{fleqn.clo}

\usepackage{times}

\usepackage{amsmath,amsfonts,amssymb}
\usepackage{bm}
\usepackage{nicefrac}
\usepackage{latexsym}
\usepackage{tabularx}
\usepackage{booktabs}

\usepackage{url}
\usepackage[dvipsnames]{xcolor}

\usepackage[
    colorlinks
]{hyperref}

\usepackage{tikz}
\usepackage{multirow}

\DeclareMathOperator*{\argmin}{arg\,min~}
\DeclareMathOperator*{\argmax}{arg\,max~}
\DeclareMathOperator{\Forall}{\forall}

\newcommand{\given}{\,\vert\,}
\newcommand{\Given}{\,\Vert\,}
\DeclareMathOperator{\kl}{KL}
\DeclareMathOperator{\E}{\mathbb{E}}

\renewcommand{\cite}{\citep}

\journal{Medical Image Analysis}
\date{February 1, 2022}

\begin{document}

\begin{frontmatter}

\title{Posterior temperature optimized Bayesian models for inverse problems in medical imaging}

\author[1]{Max-Heinrich Laves\corref{cor1}\fnref{fn1}}
\author[2]{Malte T\"olle\fnref{fn1}}
\cortext[cor1]{Corresponding author: 
  Tel.: +49 40 42878 3389;  
  E-mail: max-heinrich.laves@tuhh.de}
\fntext[fn1]{Contributed equally.}
\author[1]{Alexander Schlaefer}
\author[2]{Sandy Engelhardt}

\address[1]{Institute of Medical Technology and Intelligent Systems, Hamburg University of Technology, Am Schwarzenberg-Campus 3, 21073 Hamburg, Germany}
\address[2]{Group Artificial Intelligence in Cardiovascular Medicine, Heidelberg University Hospital, Im Neuenheimer Feld 410, 69120 Heidelberg, Germany}

\begin{abstract}
We present Posterior Temperature Optimized Bayesian Inverse Models (POTOBIM), an unsupervised Bayesian approach to inverse problems in medical imaging using mean-field variational inference with a fully tempered posterior.
Bayesian methods exhibit useful properties for approaching inverse tasks, such as tomographic reconstruction or image denoising.
A suitable prior distribution introduces regularization, which is needed to solve the ill-posed problem and reduces overfitting the data.
In practice, however, this often results in a suboptimal posterior temperature, and the full potential of the Bayesian approach is not being exploited.
In POTOBIM, we optimize both the parameters of the prior distribution and the posterior temperature with respect to reconstruction accuracy using Bayesian optimization with Gaussian process regression.
Our method is extensively evaluated on four different inverse tasks on a variety of modalities with images from public data sets and we demonstrate that an optimized posterior temperature outperforms both non-Bayesian and Bayesian approaches without temperature optimization.
The use of an optimized prior distribution and posterior temperature leads to improved accuracy and uncertainty estimation and we show that it is sufficient to find these hyperparameters per task domain.
Well-tempered posteriors yield calibrated uncertainty, which increases the reliability in the predictions.
Our source code is publicly available at \href{https://github.com/Cardio-AI/mfvi-dip-mia}{github.com/Cardio-AI/mfvi-dip-mia}.
\end{abstract}


\end{frontmatter}

\section{Introduction}

Automated methods for improving image quality have several applications in medical imaging, as acquiring high-quality images is time-consuming, costly, or entails a considerable radiation dose to the patient and medical personnel.
Use cases include post-processing methods such as denoising and artifact removal in low-dose computed tomography (CT) \cite{Yang2018,Ma2020,Wang2018}, despeckling in ultrasound or optical coherence tomography \cite{Michailovich2006,Bernardes2010}, super-resolution of magnetic resonance imaging (MRI) \cite{Tanno2017}, or inpainting for hair removal in dermoscopy images \cite{Abbas2011}.
Other approaches try to enhance image quality at the reconstruction level, e.g., sparse-view CT reconstruction \cite{Lee2018deep} or reconstruction from undersampled measurements \cite{Tezcan2018} and motion artifact removal in MRI \cite{Huang2019,Huang2021}.
Improving medical images of poor quality is a fundamental step for better diagnosis or subsequent image analysis.

\begin{figure*}[t]
    \centering
    \begin{tikzpicture}
            \node[] {\includegraphics[width=\textwidth]{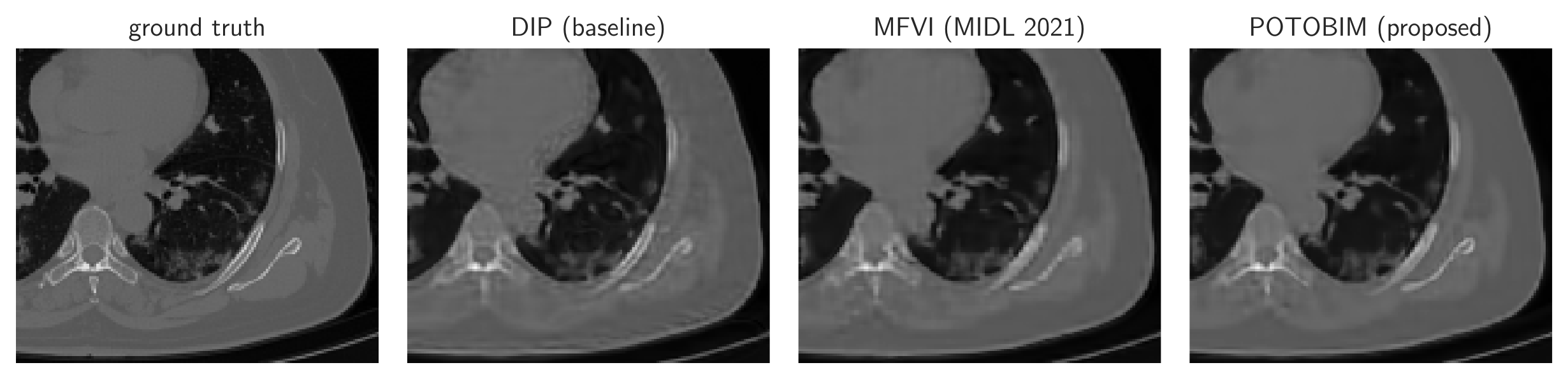}};
            \draw [-latex,BurntOrange,very thick] (-0.8,-2.0) -- (-1.0,-1.65);
            \draw [-latex,BurntOrange,very thick] (-3.2,0.9) -- (-2.9,0.6);
            \draw [-latex,BurntOrange,very thick] (-1.9,1.1) -- (-1.55,1.30);
            \draw [-latex,Green,very thick] (0.4,-0.5) -- (0.55,-0.9);
            \draw [-latex,Green,very thick] (3.7,-1.2) -- (3.48,-0.84);
            \draw [-latex,Green,very thick] (5.0,-0.5) -- (5.15,-0.9);
            \draw [-latex,Green,very thick] (8.3,-1.15) -- (8.08,-0.79);
        \end{tikzpicture}
    \caption{%
    Sparse-view CT reconstruction. Our proposed posterior temperature optimized Bayesian inverse model does not show sparse-view related patterns due to overfitting, as the non-Bayesian deep image prior baseline (orange arrows). Additionally, it does not smooth out fine details as our previous mean-field variational inference approach (green arrows) and provides consistent pixel-wise uncertainty maps.}
    \label{fig:opener_results}
\end{figure*}
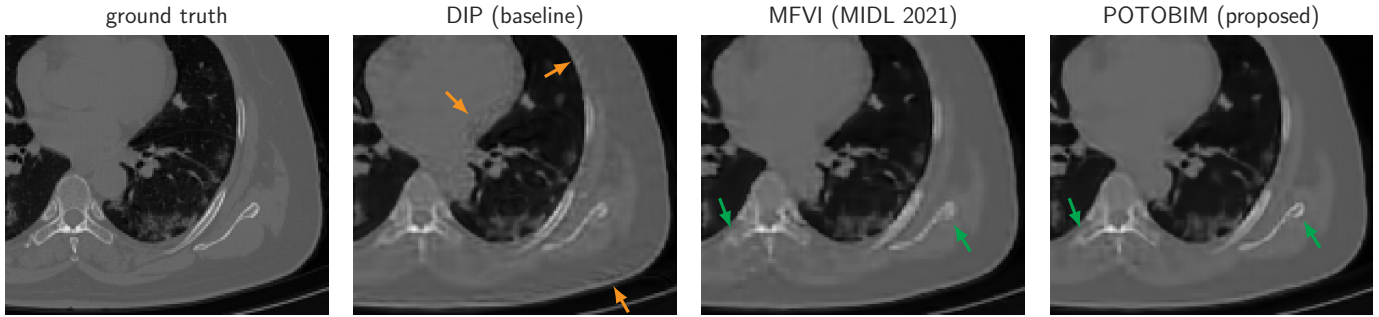

The aforementioned methods involve solving an inverse imaging problem by attempting to reconstruct a high-quality image $ \hat{\bm{x}} $ from a low-quality observation $ \bm{y} = \mathcal{F} [ \bm{x} ] $ of the true, but unknown image $ \bm{x} $ affected by some forward process $ \mathcal{F} $.
Possible forward processes include undersampling, image corruption by occlusion and additive noise, or the Radon transform. 
As the forward operator $ \mathcal{F} $ is typically not invertible directly, the reconstruction of $ \bm{x} $ comprises minimization of an ill-posed objective function
\begin{equation}
    \hat{\bm{x}} = \argmin
    \Big\{
    \mathcal{L}(\bm{y}, \mathcal{F} [ \hat{\bm{x}} ] ) + \lambda \mathcal{R}(\hat{\bm{x}})
    \Big\} ~ ,
    \label{eq:inverse_problem}
\end{equation}
governed by a similarity measure $ \mathcal{L} $ and some regularizing image prior $ \mathcal{R} $, weighted by a factor $ \lambda $ \cite{Sotiras2013}.
Common priors for image quality enhancement are total variation or penalization of first and higher order spatial derivatives \cite{rudin1992}.
The prior is of particular importance as it is responsible for the properties of the enhanced image; its manual selection is a delicate task.

More recently, deep-learning-based convolutional autoencoders have been trained to enhance images using sets of corrupted and uncorrupted data pairs \cite{Jain2009}.
Autoencoders extract important visual features from the corrupted input image and reconstruct the input from the extracted features using learned image statistics.
Through this, the neural networks implicitly learn regularization priors from data.

However, deep-learning-based methods show insufficient robustness to input data that lay outside their training domain.
\citet{Antun2020} have demonstrated that state-of-the-art deep learning methods for CT and MR image reconstruction, such as AUTOMAP \cite{Zhu2018}, show severe instabilities to tiny perturbations in the input data, which causes the reconstructions to contain considerable artifacts.
These artifacts are composed out of learned image statistics, which can lead to false anatomical structures being embedded in the reconstruction that are not present in the imaged object \cite{Bhadra2020}.
This phenomenon is referred to as \emph{hallucination} and is not limited to tomographic reconstruction but also happens in other deep-learning-based inverse image tasks \cite{Laves2020MCDIP}.
Hallucinations can result in misdiagnosis and must be avoided at all costs in medical imaging.
Motivated by the need for preventing the occurrence of hallucinations, we utilize Bayesian methods that incorporate predictive uncertainty in their reconstruction, with which, given the estimates are well calibrated, unnatural deviations in the reconstructions, i.e., hallucinations can be detected.

\paragraph{Contributions}

Our contribution is the Posterior Temperature Optimized Bayesian Inverse Model (POTOBIM), a novel approximate Bayesian approach to inverse problems in medical imaging by employing mean-field variational inference (MFVI) with posterior temperature optimization.
We use Bayesian optimization (BO) to tune the temperature and the parameters of the prior distribution on a per-task level and show its superiority to similar approaches on a variety of inverse medical image problems and modalities.
Besides improved reconstruction accuracy, POTOBIM provides well-calibrated pixel-wise uncertainty maps.
In our preliminary work presented at the ``Fourth Conference on Medical Imaging with Deep Learning'' (MIDL 2021) \cite{Toelle2021}, we used Bayesian optimization for an empirical improvement of the right amount of induced regularization lacking the theoretical background provided in this paper by the cold posterior effect (see Fig.~\ref{fig:opener_results}).
We extend the BO introducing full posterior temperature optimization as a new method and, thus, provide new experimental results on a more diverse set of inverse imaging postprocessing tasks, i.e., CT reconstruction for an increased number of test images.
We further perform assessment of uncertainty calibration in particular tackling the possible occurrence of hallucinations.
Our code is publicly available at \href{https://github.com/Cardio-AI/mfvi-dip-mia}{github.com/Cardio-AI/mfvi-dip-mia}.

\subsection{Related Work}

Deep learning based methods have seen increased use for medical image post-processing or tomographic reconstruction.
Most approaches rely on supervised learning from data sets of input and output image pairs.
Generative adversarial networks (GANs) are popular among enhancement methods in the image domain, such as noise reduction in low-dose CT \cite{Wolterink2017}.
More recent works have adapted to more advanced networks, such as conditional Wasserstein GANs \cite{Yi2018}.
Usually, the networks are trained using data pairs of low-dose and clinical-dose CTs from the same patient.
GANs have also been used for de-aliasing in MRI reconstruction \cite{Yang2018} and quality enhancement in low-dose PET imaging \cite{Wang2018}.
Bayesian approaches have been used by applying variational dropout to subpixel convolutional networks for MRI super-resolution \cite{Tanno2017} and Bayesian inversion with conditional Wasserstein GANs for upsampling extreme low-dose CT \cite{Adler2019}.
A prominent method using CNNs for CT image reconstruction from sinograms is AUTOMAP, which directly maps information from the sensor-domain to image-domain using supervised learning \cite{Zhu2018}.
The sinograms are fed into a CNN and the reconstruction is transformed back using the Radon transform to compute the loss between input and output.
\citet{Hammernik2018} proposed to learn accelerated MRI reconstruction by combining variational methods with deep learning.

The concept of deep image prior (DIP) for inverse tasks does not require supervised training, and thus it is not affected by the aforementioned instabilities and hallucinations.
As \citet{Lempitsky2018} have shown the structure of a CNN is sufficient to capture a great amount of image statistics and impose a strong prior to restore a high-quality image from a low-quality observation without having access to any data.
Therefore, in DIP, a convolutional autoencoder with skip-connections is interpreted as a parameterization of the image to be reconstructed.
Besides empirical evidence, the effectiveness of DIP can be explained by the spectral bias of deep networks \cite{Rahaman2019}.
An autoencoder network decouples the frequency components of an image, comparable to a Fourier transform \cite{Chakrabarty2019}.
During optimization, the frequency components are learned at different rates.
Lower frequencies are reconstructed first, which behaves like a low pass filter; image corruptions such as noise are usually encoded in the high-frequency components.
This makes early stopping in optimization a crucial step in order to not overfit the corrupting features (see Fig.\,\ref{fig:opener_results}).
An alternative to early stopping is carefully selecting the number of trainable parameters, which introduces an architectural form of regularization.
The deep decoder framework has demonstrated that manually fine-tuned under-parameterization of a decoder network can also address overfitting \cite{Heckel2020}.
DIP in a non-Bayesian setting was already used for unsupervised CT reconstruction \cite{Baguer2020}.

However, both early stopping and under-parameterization require expert human interaction.
We seek to find a more automated way to prevent DIP from overfitting in order to take advantage of its robustness towards hallucinations.
\citet{Cheng2019} presented a first Bayesian approach to DIP in the context of natural images, where a prior distribution is placed over the weights of the network and the posterior distribution is used to output the final image.
They derived a Monte Carlo (MC) sampler from DIP using stochastic gradient Langevin dynamics (SGLD) as Bayesian approximation, which uses injection of Gaussian noise into the gradients during each SGD step \cite{Welling2011}.
The authors claim to have solved the problem of overfitting and provide pixel-wise reconstruction uncertainty estimates.
SGLD DIP has already been applied to PET image reconstruction \cite{Carrillo2021}.
Prior to \cite{Toelle2021}, we have shown that DIP with SGLD shows almost unchanged overfitting behavior in the case of medical images \cite{Laves2020MCDIP}.
As a solution, we proposed a variational inference (VI) approach to DIP using Monte Carlo dropout (MCD) \cite{Gal2016}.

In this paper, we show that Bayesian approaches to DIP employing SGLD or MCD show overfitting on medical images at some point.
We attribute this to the misspecified posterior temperature and manual selection of the prior distribution of the weights.
At this point it is important to distinguish between DIP, which imposes a spectral bias towards lower frequencies, and the prior distribution over the weights of the network in Bayesian inference.
In SGLD and MC dropout, the prior is implicitly defined by weight decay or the dropout rate.
We hypothesize that the potential of DIP can be utilized in medical image enhancement using a well-defined posterior temperature and prior distribution in a Bayesian setting.

\subsection{Bayesian Deep Learning}

In Bayesian deep learning, a prior distribution $ p(\bm{w} \given \alpha) $ is placed over the weights $ \bm{w} $ of a neural network, governed by a hyperparameter $ \alpha $.
After observing the data $ \mathcal{D} = \{ \mathcal{X} ; \mathcal{Y} \} $ consisting of pairwise observations with $ \bm{x}_i \in \mathcal{X} $  and $ \bm{y}_i \in \mathcal{Y} $, we are interested in the posterior
\begin{equation*}
p(\bm{w} \given \mathcal{D}, \alpha) = p(\mathcal{D} \given \bm{w}, \alpha) p(\bm{w} \given \alpha) / p(\mathcal{D}) ~ ,
\end{equation*}
where we use $ p(\mathcal{D}) $ as abbreviation for $ p(\mathcal{Y} \given \mathcal{X}) $ for convenience.
However, this distribution is not tractable in general as the normalizing factor involves marginalization of the model likelihood over the prior
\begin{equation*}
p(\mathcal{D}) = \int p(\mathcal{D} \given \bm{w}, \alpha) p(\bm{w} \given \alpha) \, \mathrm{d}\bm{w} ~ .
\end{equation*}
Consequently, the posterior predictive distribution is intractable as well.
This gives rise to different approximate Bayesian inference techniques that rely on either sampling or VI.
SGLD is a framework that derives a Markov chain Monte Carlo (MCMC) sampler from SGD by injecting Gaussian noise into the gradients after each learning step \cite{Welling2011}.
Under suitable conditions (i.e., variance of injected noise and learning rate decay),
SGLD eventually converges to the posterior distribution.
When applied to inverse tasks, SGLD is very sensitive to the learning rate decay scheme.
Choosing a decrease that is too slow (i.e., $ \epsilon_{t} = 0.9999^{t} \epsilon_{0} $) can result in overfitting; a decrease that is too fast (i.e., $ \epsilon_{t} = 0.99^{t} \epsilon_{0} $) can result in convergence to a worse-than-possible reconstruction.

In VI, we try to find a simpler, variational approximation to the Bayesian posterior distribution.
VI uses optimization instead of sampling to find the member $ q_{\bm{\phi}}(\bm{w}) $ of a family of distributions (e.g., a multivariate Gaussian) that is close to the exact posterior, defined by the variational parameters $ \bm{\phi} $.
We optimize $ q_{\bm{\phi}} $ w.r.t.\ $ \bm{\phi} $, such that the Kullback-Leibler (KL) divergence is minimized with regard to the true posterior \cite{Blei2017}.
Two practical implementations are MC dropout \cite{Gal2015Bernoulli} and Bayes by backprop \cite{Blundell2015}.
The former uses dropout before every weight layer during training and at inference time, which allows sampling from the approximate posterior.
The latter assumes a fully factorized Gaussian distribution $ w_{ij} \sim \mathcal{N}(\mu_{ij}, \sigma^{2}_{ij}) $, also known as mean-field distribution, which treats the mean and variance of each weight of a multi-layer network as learnable parameter.
In contrast to MC dropout, MFVI allows us to directly compute the KL divergence between the variational posterior and the prior, which enables us to select other (non-Gaussian) prior distributions, where no closed form exists.

\subsection{Cold Posteriors}

In order to bring the variational distribution $ q_{\bm{\phi}}(\bm{w}) $ close to the true posterior in variational inference, a lower bound on the log-evidence (ELBO) is derived and maximized.
\citet{Graves2011} already suggested to reweight the complexity term in the ELBO using a factor $ \lambda $ to balance both terms in case of discrepancy between number of weights and training samples:
\begin{equation}
    \mathrm{ELBO}(q_{\bm{\phi}}(\bm{w})) = \mathbb{E}_{\bm{w} \sim q} [ \log p(\mathcal{D} \given \bm{w}) ] - \lambda \kl [q_{\bm{\phi}}(\bm{w}) \Given p(\bm{w})] ~ .
    \label{eq:elbo_lambda}
\end{equation}
It is common for Bayesian deep learning researchers to employ values of $ \lambda < 1 $ to achieve better predictive performance \cite{Ashukha2020,Blundell2015}.
While their main motivation was to qualitatively balance out discrepancies between number of model parameter and dataset size, the reweighting has recently been studied in more detail and described as the ``cold posterior'' effect \cite{Wilson2020}.
\citet{Wenzel2020} derived the tempered Bayesian posterior $ p(\bm{w} \given \mathcal{D}) \propto \exp(-U(\bm{w})/T) $ with posterior energy function $ U(\bm{w}) = - \log p(\mathcal{D} \given \bm{w} ) - \log p(\bm{w}) $ and have shown empirically that cold posteriors with $ T < 1 $ perform considerably better.
The authors also recover Eq.\,(\ref{eq:elbo_lambda}) and show that introducing $ \lambda $ into the ELBO is equivalent to a partially tempered posterior, where only the likelihood term is scaled.
In this paper, we will not argue whether cold posteriors invalidate Bayesian principles, as there is disagreement among researchers \cite{Wilson2020}, but use it in a directed way to increase predictive performance and uncertainty calibration.

\subsection{Deep Image Prior}

DIP uses a convolutional image-generating network $ \hat{\bm{x}} = \bm{f}_{\bm{w}}(\bm{z}) $ with randomly-initialized weights $ \bm{w} $ as neural parameterization of an image.
The input $ \bm{z} $ is sampled from a uniform distribution $ \bm{z} \in \mathbb{R}^{C \times H \times W} \sim \mathcal{U}(0, 0.1) $ and has the same spatial dimensions as $ \hat{\bm{x}} $ with channels $ C $, width $ W $ and height $ H $.
Given a low-quality target observation $ \bm{y} $ and corresponding forward process $ \mathcal{F} $, the reconstructed image $ \hat{\bm{x}} $ is obtained by minimizing the pixel-wise mean squared error $ \Vert \bm{y} - \mathcal{F} [ \hat{\bm{x}} ] \Vert^{2} $ w.r.t.\ the weights $ \bm{w} $.
Due to the spectral bias of convolutional networks towards lower frequencies, early stopping behaves like a low-pass filter \cite{Chakrabarty2019}, making it suitable for many inverse image tasks.

\subsection{Cold Posteriors in Inverse Modeling}

\citet{Wilson2020} advocate to scale the posterior for any model and data, as it would be highly surprising if $ T=1 $ would be the optimal value for this hyperparameter.
Deep models for inverse tasks do not scale their capacity to the available data and are most likely misspecified at $ T=1 $.
We hypothesize that a cold posterior would be beneficial in our case (rather than $ T \geq 1 $), as the incorporated deep image prior uses highly overparameterized networks for the task of reconstructing a single image.
A cold posterior implicates overcounting the data, which, if optimized as proposed below, reduces the regularization introduced by the Bayesian approach to the correct amount.
The model will neither overfit the noisy or sparse data, nor converge to a subpar solution.

Concurrent works have made similar findings in MRI reconstruction with MFVI \cite{Narnhofer2021} and variational autoencoders \cite{Edupuganti2021}, where scaling the KL improved the results.
However, these works only used a partially scaled posterior and did not optimize the scaling factor in a directed way.
We will use a temperature optimized posterior in conjunction with MFVI as methodological improvement.

\section{Methods}

\subsection{Mean-Field Variational Inference for Deep Image Prior}
\label{sec:methods_mfvi}

\begin{figure*}[ht]
    \centering
    \includegraphics{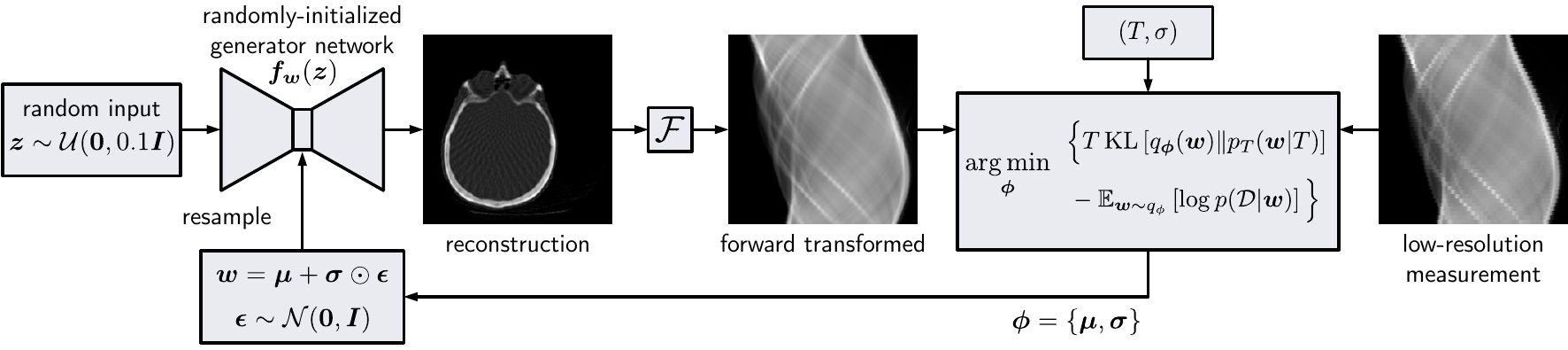}
    \caption{Conceptual overview of POTOBIM shown exemplarily for CT reconstruction. A randomly-initialized MFVI autoencoder network fed with uniform noise outputs a CT. The image reconstruction is performed iteratively by applying the forward transform $ \mathcal{F} $ and minimizing the fully tempered negative ELBO w.r.t.\ the variational parameters $ \bm{\phi} = \{ \bm{\mu}, \bm{\sigma} \} $ using gradient descent. The posterior temperature $ T $ and prior standard deviation $ \sigma $ are found using Bayesian optimization.}
    \label{fig:concept}
\end{figure*}
\begin{figure*}
    \centering
    \includegraphics[width=7cm]{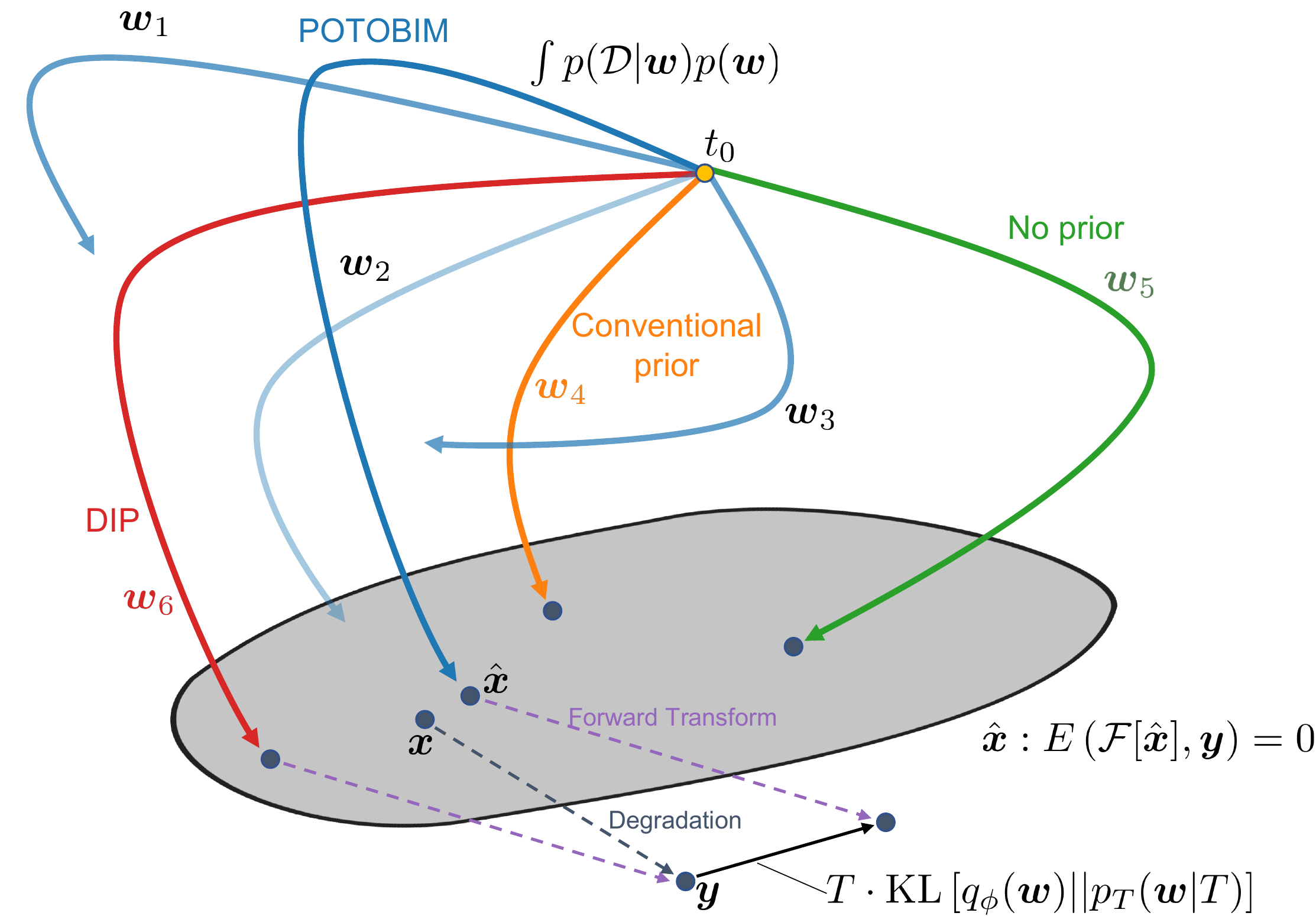} 
    \includegraphics[width=9cm]{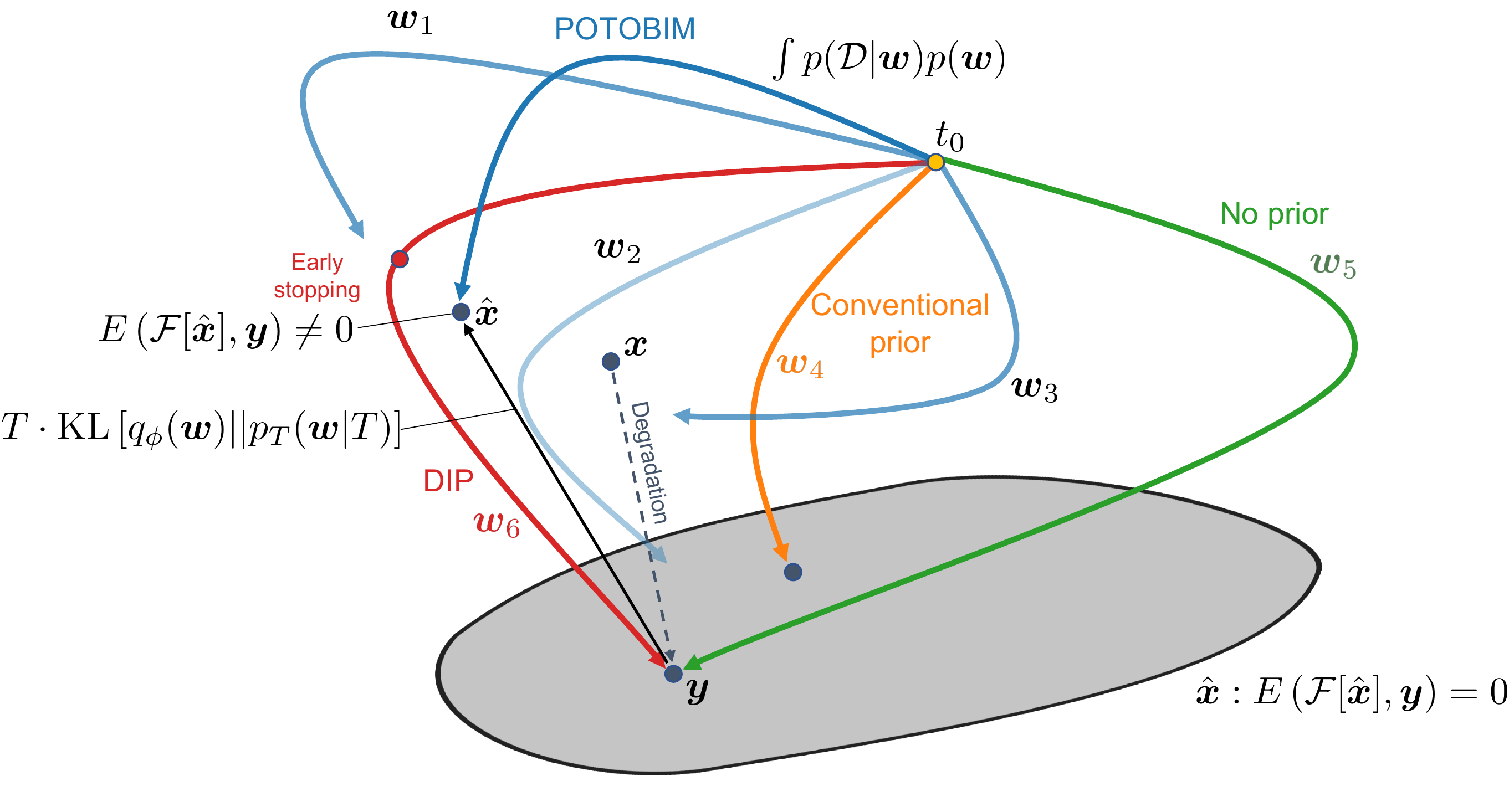}
    \caption{Restoration with POTOBIM in image space. Similar to \citet{Lempitsky2018} we present the optimization curves of standard restoration techniques with (yellow) and without prior (green) and non-Bayesian DIP (red). Additionally we show the learning paths of POTOBIM using MFVI (blue). We distinguish two cases for reconstructing clean images from corrupted measurements. On the left we show the example in which we have a manifold of points $\hat{\bm{x}}$ that exhibit zero energy and can all be mapped to the same same degraded measurement $\bm{y}$ with some forward operator $E\left(\mathcal{F}[\hat{\bm{x}}],\bm{y}\right) = 0$ (e.g.\ Radon transform in CT reconstruction). Non-Bayesian DIP achieves its superior results to conventional reconstruction techniques with and without prior by re-parameterisation of the optimization curve. While the resulting reconstruction can exhibit zero cost compared to the degraded measurement, it can still be far off the original image. We circumvent this problem by employing a distance by means of KL divergence, essentially leading to higher costs compared to the degraded measurement, but getting a reconstruction closer to the true image by taking into account all possible optimization curves indicated by the blue curves. Note that we do also take the optimization curves of non-Bayesian DIP and reconstruction with and without prior into account but weigh them with probability $p(\bm{w})$. In the second case on the right the true image does not exhibit zero cost $E\left(\mathcal{F}[\hat{\bm{x}}] , \bm{y}\right) > 0$ as e.g.\ in denoising. The forward operator is the identity mapping in this case. Instead of applying early stopping to get the optimal result, as must be done for DIP, we use the KL divergence to obtain a distance close to the original degradation process in an unsupervised fashion.}
    \label{fig:mfvi-dip-image-space}
\end{figure*}

During optimization, DIP aims at finding the optimal weight point estimate $ \hat{\bm{w}} $ by maximum likelihood estimation (MLE) with gradient descent.
Before turning to a Bayesian approach, we model heteroscedastic reconstruction uncertainty by assuming that the low-quality observation $ \bm{y} $ is sampled from a spatial random process and that each pixel $ i $ follows a Gaussian distribution $ \mathcal{N}(y_{i}; \hat{y}_{i}, \hat{s}^{2}_{i}) $ with mean $ \hat{y}_{i} $ and variance $ \hat{s}^{2}_{i} $.
We extend the last layer such that the network outputs these values for each pixel of the reconstructed image
$
    \bm{f}_{\bm{w}}(\bm{z}) = \{ \hat{\bm{x}}, \hat{\bm{\sigma}}^{2} \}
$
and apply the forward operator to the parameters, resulting in $ \hat{\bm{y}} = \mathcal{F} [ \hat{\bm{x}} ] $ and $ \hat{\bm{s}}^{2} = \mathcal{F} [ \hat{\bm{\sigma}}^{2} ] $.
This works in cases where the forward transform of a Gaussian remains to be a Gaussian (cf.\ § \ref{sec:experiments}).
Maximum posterior estimation is performed by minimizing the negative log-likelihood, which leads to the following optimization criterion
\begin{equation}
    \mathcal{L}(\bm{w}) = \frac{1}{N}\sum_{i=1}^{N} \hat{s}_{i}^{-2} \big\Vert y_{i} - \hat{y}_{i} \big\Vert^{2} + \log \hat{s}_{i}^{2} ~ ,
    \label{eq:nll}
\end{equation}
where $ N $ is the number of pixels per image.
In this case, $ \hat{\bm{s}}^{2} $ captures the pixel-wise aleatoric uncertainty and is jointly estimated with $ \hat{\bm{y}} $ by finding $ \bm{w} $ that minimizes Eq.~(\ref{eq:nll}) with gradient descent.
For numerical stability, Eq.~(\ref{eq:nll}) is implemented such that the network directly outputs $ -\log \hat{\bm{\sigma}}^{2} $, if the forward operator does not involve adding pixel values (see §~\ref{sec:experiments}).

Next, we employ a MFVI approach to DIP by assuming that the variational posterior can be factorized as $ q_{\bm{\phi}}(\bm{w}) = \prod_{i=1}^{L} \mathcal{N}(w_{i} \given \mu_{i}, \sigma_{i}^{2}) $, with number of layers $ L $.
In each forward pass, the weights are sampled using reparameterization $ \bm{w} = \bm{\mu} + \bm{\sigma} \odot \bm{\epsilon} $ with $ \bm{\epsilon} \sim \mathcal{N}(\bm{0}, \bm{I}) $, where $ \odot $ denotes element-wise multiplication.
The variational parameters $ \bm{\phi} = \{ \bm{\mu}, \bm{\sigma} \} $ are optimized by minimizing the negative log evidence lower bound (ELBO)
\begin{equation}
    \bm{\phi}^{\ast} = \argmin_{\bm{\phi}} \kl [q_{\bm{\phi}}(\bm{w}) \Given p(\bm{w})] - \mathbb{E}_{\bm{w} \sim q_{\bm{\phi}}} [ \log p(\mathcal{D} \given \bm{w}) ]
    \label{eq:elbo}
\end{equation}
using backpropagation without weight decay.
This effectively doubles the number of trainable parameters and is known as Bayes by backprop \cite{Blundell2015}.
The first term in Eq.\,(\ref{eq:elbo}) is usually approximated with MC integration by
\begin{equation}
    \kl [q \Vert p] \approx \frac{1}{T} \sum_{i=1}^{T} \log q_{\bm{\phi}}(\bm{w}_{i}) - \log p(\bm{w}_{i}) ~ ,
    \label{eq:kl_mc}
\end{equation}
with $ T $ Monte Carlo samples $ \bm{w}_{i} $ drawn from the variational posterior $ q_{\bm{\phi}}(\bm{w}) $.
In case of a Gaussian prior, it can be implemented in closed form accelerating training by omitting the need for drawing MC samples (see derivation in supplemental material).
The second term in Eq.\,(\ref{eq:elbo}), the log-likelihood, is implemented using Eq.\,(\ref{eq:nll}) in the same MC fashion over draws from the posterior
\begin{equation}
    -\mathbb{E}_{\bm{w} \sim q_{\bm{\phi}}} [ \log p(\mathcal{D} \given \bm{w}) ] \approx \frac{1}{T} \sum_{i=1}^{T} \bm{s}^{-2}_{\bm{w}_{i}} \Vert \bm{y} - \hat{\bm{y}}_{\bm{w}_{i}} \Vert^{2} + \log \bm{s}^{2}_{\bm{w}_{i}} ~ .
    \label{eq:nll_mc}
\end{equation}

The mean-field approximation to the true posterior has the same mean or mode (depending on the direction of the KL divergence), but different shape \cite{Blei2017}.
In practice, the approximate posterior is usually narrower and the scale of the posterior Gaussians is underestimated.
This results in an underestimation of uncertainty in the predictive distribution, which could be fixed with post-hoc calibration.

After convergence, we obtain the high-quality reconstruction $ \mathbb{E} [ \hat{\bm{x}} ] $ 
\begin{equation}
    \mathbb{E}_{\bm{w} \sim q_{\bm{\phi}}} [ \hat{\bm{x}} ] \approx \frac{1}{T} \sum_{i=1}^{T} \hat{\bm{x}}_{\bm{w}_{i}}
    \label{eq:mc_mean}
\end{equation}
and the accompanying pixel-wise uncertainty $ \mathrm{Var} [ \hat{\bm{x}} ] $
\begin{equation}
    \quad \mathrm{Var}_{\bm{w} \sim q_{\bm{\phi}}} [ \hat{\bm{x}} ] \approx \frac{1}{T} \sum_{i=1}^{T}  \hat{\bm{x}}_{\bm{w}_{i}}^{2} - \left( \frac{1}{T} \sum_{i=1}^{T} \hat{\bm{x}}_{\bm{w}_{i}} \right)^{2} + \frac{1}{T} \sum_{i=1}^{T} \hat{\bm{\sigma}}^{2}_{\bm{w}_{i}}
    \label{eq:mc_var}
\end{equation}
by integrating MC samples from the predictive posterior \cite{Kendall2017}.

\subsection{Temperature-scaled Posterior}

In the following, the ELBO for a fully temperature-scaled Bayesian posterior in variational inference is derived.
Let $ p_{T}(\bm{w} \given \mathcal{D}) $ be the fully tempered posterior \cite{Wenzel2020}:
\begin{align}
    & \kl \left[ q_{\phi}(\bm{w}) \Given p_{T}(\bm{w} \given \mathcal{D}) \right] \\
    &= \E_{\bm{w}} \left[ \log q_{\phi}(\bm{w}) - \log p_{T} (\bm{w} \given \mathcal{D}) \right] \\
    &= \E_{\bm{w}} \left[ \log q_{\phi}(\bm{w}) - \log \frac{(p(\mathcal{D} \given \bm{w}) p(\bm{w}))^{1/T}}{\int (p(\mathcal{D} \given \bm{w}') p(\bm{w}'))^{1/T} \, \mathrm{d} \bm{w}'} \right] \\
    \begin{split}
    &= \E_{\bm{w}} \left[ \log q_{\phi}(\bm{w}) - \log (p(\mathcal{D} \given \bm{w}) p(\bm{w}))^{1/T} \right] \\
    &\quad + \underbrace{\log \int (p(\mathcal{D} \given \bm{w}) p(\bm{w}))^{1/T} \, \mathrm{d} \bm{w}}_{\mathrm{const.\,w.r.t.\,} \bm{w},~ =: \log E_{T}}
    \end{split} \\
    \begin{split}
    &= \underbrace{\E_{\bm{w}} \left[ \log q_{\phi}(\bm{w}) - \tfrac{1}{T} \log p(\bm{w}) \right] - \E_{\bm{w}} \left[ \tfrac{1}{T} \log p (\mathcal{D} \given \bm{w}) \right]}_{=: \mathrm{ELBO}_{T}(q_{\bm{\phi}}(\bm{w}))} \\
    &\quad + \log E_{T}
    \end{split} \\
    &\Rightarrow \log E_{T} = \kl \left[ q_{\bm{\phi}}(\bm{w}) \Given p_{T}(\bm{w} \given \mathcal{D}) \right] + \mathrm{ELBO}_{T} (q_{\bm{\phi}}(\bm{w}))
\end{align}
As the tempered evidence $ E_{T} $ is constant, maximizing $ \mathrm{ELBO}_{T} $ minimizes the KL, thus bringing the variational distribution $ q_{\bm{\phi}}(\bm{w}) $ closer to the fully tempered posterior $ p_{T}(\bm{w} \given \mathcal{D}) $:
\begin{align}
    &\mathrm{ELBO}_{T}(q_{\bm{\phi}}(\bm{w})) \\
    &\quad = - \E_{\bm{w}} \left[ \log q_{\phi}(\bm{w}) - \tfrac{1}{T} \log p(\bm{w}) \right] + \E_{\bm{w}} \left[ \tfrac{1}{T} \log p (\mathcal{D} \given \bm{w}) \right] \\
    &\quad = - \kl \left[ q_{\bm{\phi}} (\bm{w}) \Given p(\bm{w})^{\nicefrac{1}{T}} \right] + \E_{\bm{w}} \left[ \tfrac{1}{T} \log p (\mathcal{D} \given \bm{w}) \right] ~ .
\end{align}
In practice, the scaled log-likelihood is straightforward to implement.
The scaled KL contains the scaled prior $ p_{T}(\bm{w}) \propto p(\bm{w})^{\nicefrac{1}{T}} $, which will have the same mean, but different variance as the unscaled prior.
In case of a Normal prior, this is equivalent to a scaled prior variance \cite{Aitchison2021}:
\begin{gather}
	p(\bm{w}) \propto \exp(-\Vert \bm{w} \Vert^{2} / 2 \sigma^{2}) , \\
	p(\bm{w})^{1/T} \propto \exp(-\Vert \bm{w} \Vert^{2} / 2 \sigma_{T}^{2}) , \\
	\sigma_T = \sqrt{T} \sigma ~ .
\end{gather}
We therefore set $ p_{T} (\bm{w}) = \mathcal{N}(\bm{0}, \frac{\sigma^{2}}{T} \bm{I}^{2}) $, which results in the following minimization criterion (with scaling by $ -T $)
\begin{equation}
    \argmin_{\bm{\phi}} T \cdot \kl \left[ q_{\bm{\phi}} (\bm{w}) \Given p_{T}(\bm{w} \given T) \right] - \E_{\bm{w}} \left[ \log p (\mathcal{D} \given \bm{w}) \right] ~ ,
    \label{eq:scaled_criterion}
\end{equation}
which, in contrast to Eq.\,(\ref{eq:elbo_lambda}), optimizes the fully temperature-scaled $ \mathrm{ELBO}_{T} $.
A conceptual overview of POTOBIM utilizing the temperature-scaled posterior is visualized in Fig.~\ref{fig:concept}; the restoration process in image space is further explained in Fig.~\ref{fig:mfvi-dip-image-space}.

\subsection{Calibration of Uncertainty}

To assess the quality of uncertainty estimates of Bayesian neural networks, we use the uncertainty calibration error (UCE) for regression \citep{Laves2020,Laves2021}.
In the case of regression, we expect the predicted uncertainty to scale linearly with the predicted error
\begin{equation}
    \mathbb{E}_{\hat{\Sigma}^{2}} \left[ \big\vert \big( \mathbb{E}[(\hat{\bm{y}}-\bm{y})^{2}] \, \big\vert \, \hat{\Sigma}^{2} = \beta^{2} \big) - \beta^{2} \big\vert \right] ~ \Forall \left\{ \beta^{2} \in \mathbb{R} \, \vert \, \beta^{2} \geq 0 \right\} ~ ,
    \label{eq:uce}
\end{equation}
with $ \hat{\Sigma} = \mathrm{Var}[\hat{\bm{x}}] $.
More intuitively, if all pixels in an image were estimated with uncertainty 0.2, the second moment of the predictive error should also equal 0.2. 
The UCE involves binning the uncertainty values and computing a weighted average of absolute differences between error und uncertainty per bin. 
For an image with $n$ pixels and a set of input indices $B_m$ the UCE can be quantified as:
\begin{equation}
    \mathrm{UCE} := {\sum_{k=1}^{K}} \frac{\vert B_{k} \vert}{m} \big\vert {\mathrm{var}}(B_{k}) - \mathrm{uncert}(B_{k}) \big\vert ~ ,
\end{equation}
where $\mathrm{var}(B_k)$ represents the mean of the variance per bin and $\mathrm{uncert}(B_k)$ the mean of predicted uncertainties respectively.

\subsection{Posterior Temperature Optimization}
\label{sec:bo}

Instead of manually selecting the optimal posterior temperature using heuristics or inefficient grid search, we employ derivative-free Bayesian optimization (BO) in two dimensions to jointly find the posterior temperature $ T $ and prior scale $ \sigma $.
BO allows us to optimize black-box functions that are expensive to evaluate, such as the training of a deep network \cite{Snoek2015}.
It uses a computationally inexpensive surrogate model to retrieve a distribution over functions.

In this work, we apply optimization of the posterior temperature to multiple inverse post-processing problems in the medical domain and maximize the peak signal-to-noise ratio (PSNR) between the reconstructed image $ \hat{\bm{x}} $ and the ground truth image $ \bm{x} $ as a function of the hyperparameters $ T $ and $ \sigma $
\begin{equation}
    \max_{T \in \mathcal{T}, \sigma \in \mathcal{S}} ~ f(T, \sigma) = \max_{T \in \mathcal{T}, \sigma \in \mathcal{S}} ~  \mathrm{PSNR}(\hat{\bm{x}}(T, \sigma), \bm{x})
    \label{eq:bo_objective}
\end{equation}
using a Gaussian process (GP) as surrogate $ f \sim \mathcal{GP} $.
In each step of the BO, we evaluate our objective function $ f $ at the current candidates $ T^{\ast} $ and $ \sigma^{\ast} $ to increase the set of observations $ \mathcal{D}_{\mathrm{BO}} $ and update the posterior of the surrogate model.
Next, we maximize an acquisition function $ a(T, \sigma ; \mu_{\mathcal{GP}}, \sigma^{2}_{\mathcal{GP}}) $ using the current GP posterior mean $ \mu_{\mathcal{GP}} $ and variance $ \sigma^{2}_{\mathcal{GP}} $.
Its maximizing arguments $ T^{\ast}, \sigma^{\ast} \leftarrow \argmax a(T, \sigma ; \mu_{\mathcal{GP}}, \sigma^{2}_{\mathcal{GP}}) $ are used as candidates for the next iteration \cite{Frazier2018}.
We choose the commonly accepted expected improvement (EI) as acquisition function
\begin{equation}
    \begin{split}
    &a_{\mathrm{EI}}(T, \sigma ; \mu_{\mathcal{GP}}, \sigma^{2}_{\mathcal{GP}}) \\
    &\quad = \mathbb{E} \left[ \max (y - f^{\ast}), 0) \given y \sim \mathcal{N} ( \mu_{\mathcal{GP}}(T, \sigma), \sigma^{2}_{\mathcal{GP}}(T, \sigma) ) \right] ~ ,
    \end{split}
    \label{eq:bo_ei}
\end{equation}
where $ f^{\ast} = f(T_{\mathrm{best}}, \sigma_{\mathrm{best}}) $ is the minimal value of the objective function observed so far.
Eq.\,(\ref{eq:bo_ei}) can be solved analytically as shown in \cite{Jones1998}.
We utilize automatic differentiation from modern deep learning frameworks to optimize the acquisition function in order to get the next candidates $ T^{\ast} $ and $ \sigma^{\ast} $ \cite{Gardner2018}.

\section{Experiments}
\label{sec:experiments}

\begin{figure*}[h]
    \setlength\tabcolsep{1pt}
	\renewcommand{\arraystretch}{0}
	\newlength{\subfigwidtha}
	\setlength{\subfigwidtha}{4.4cm}
    \centering
    \small
    \begin{tabular}{ccccc}
          & \textsf{~denoising} & \textsf{~super-resolution} & \textsf{~inpainting} & \textsf{~CT reconstruction} \\
        \raisebox{2.2cm}{\rotatebox[origin=c]{90}{\textsf{MCD}}}
        & \includegraphics[width=\subfigwidtha]{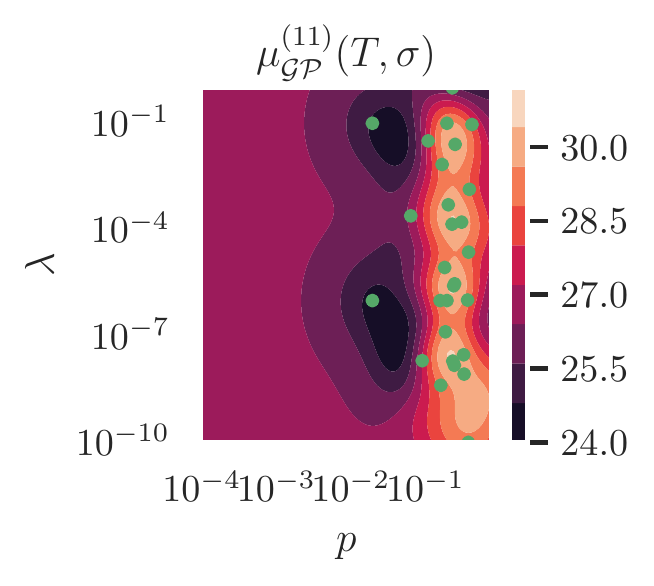}
        & \includegraphics[width=\subfigwidtha]{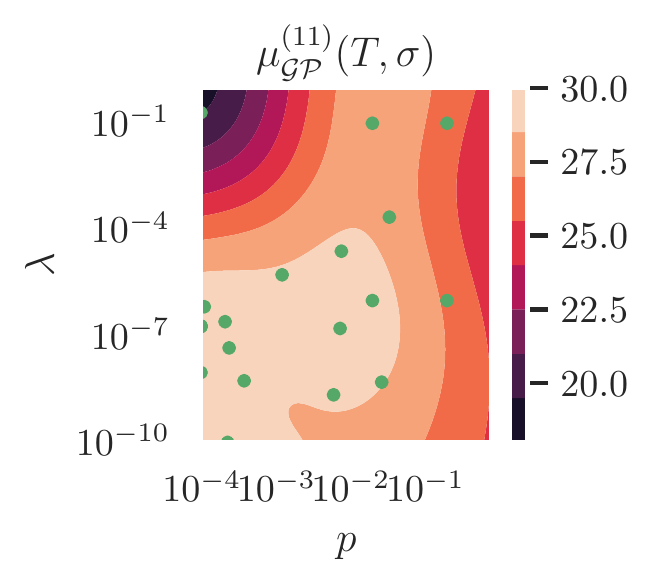}
        & \includegraphics[width=\subfigwidtha]{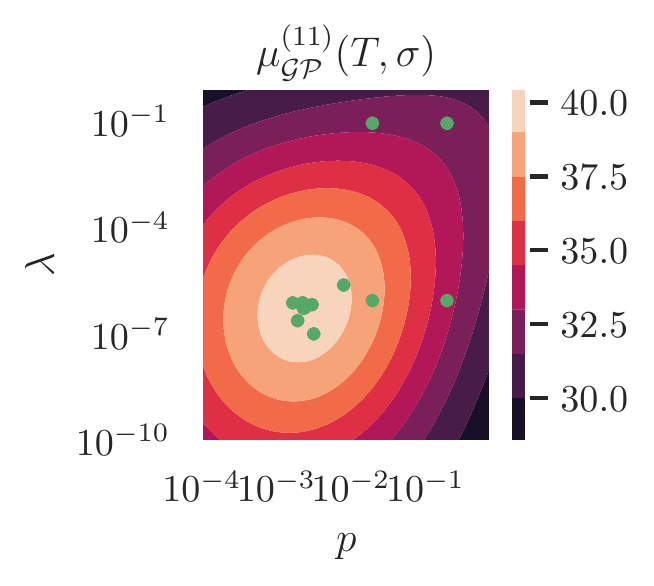}
        & \includegraphics[width=\subfigwidtha]{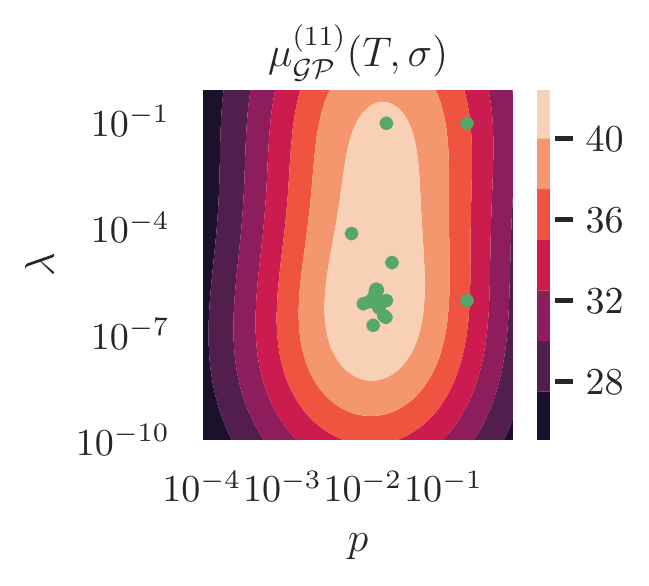} \\
        \raisebox{2.0cm}{\rotatebox[origin=c]{90}{\textsf{SGLD}}}
        & \includegraphics[width=\subfigwidtha]{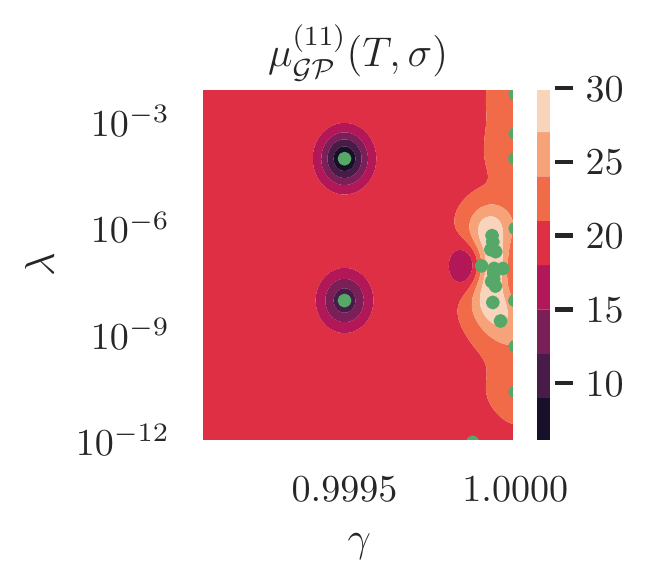}
        & \includegraphics[width=\subfigwidtha]{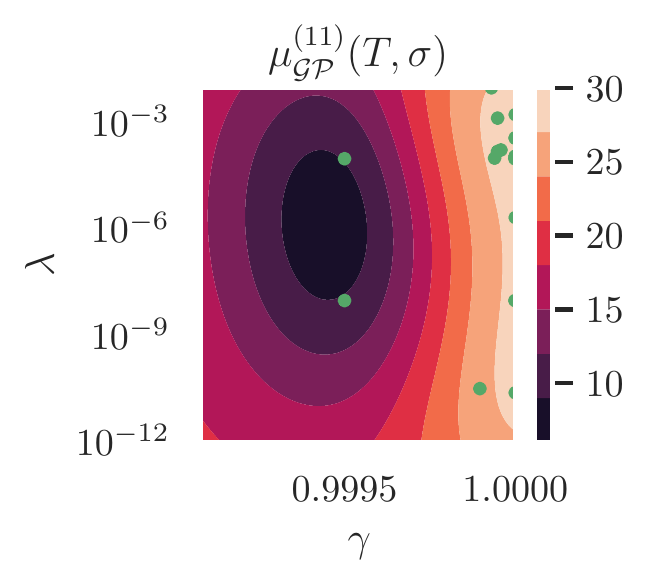}
        & \includegraphics[width=\subfigwidtha]{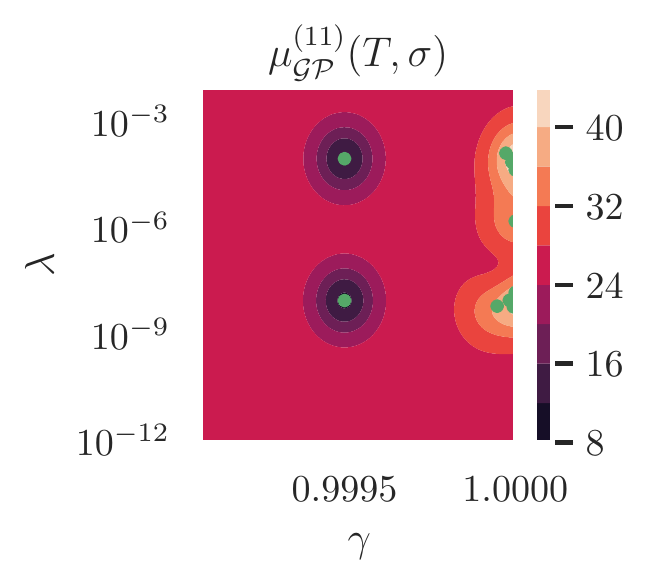}
        & \includegraphics[width=\subfigwidtha]{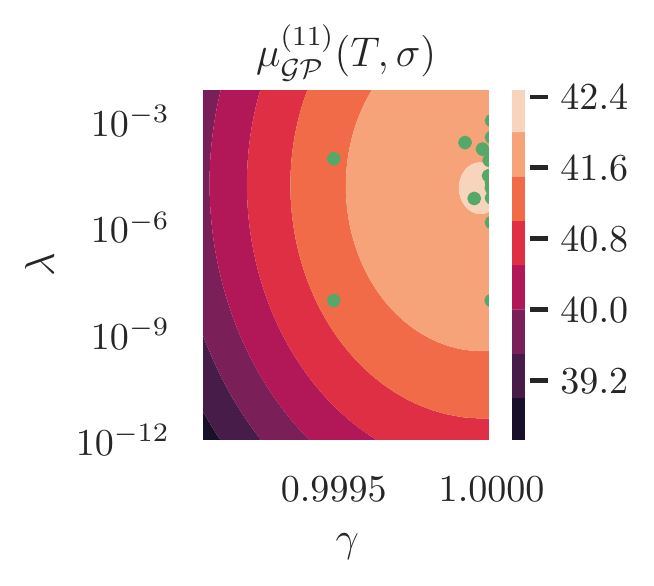} \\
        \raisebox{2.0cm}{\rotatebox[origin=c]{90}{\textsf{POTOBIM}}}
        & \includegraphics[width=\subfigwidtha]{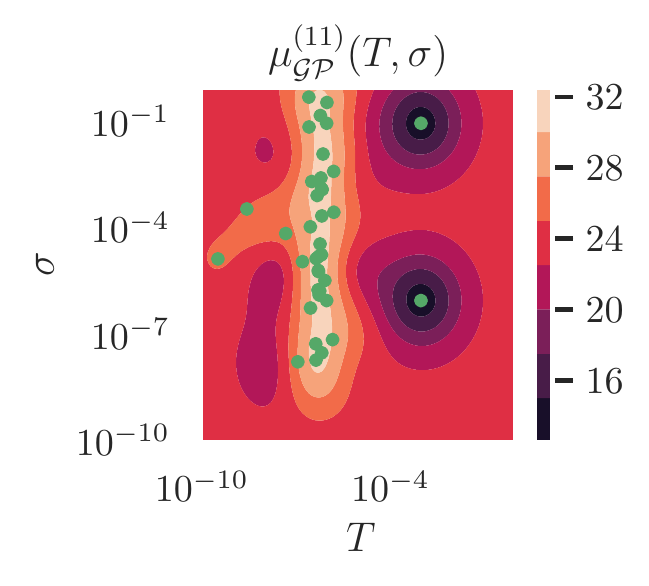}
        & \includegraphics[width=\subfigwidtha]{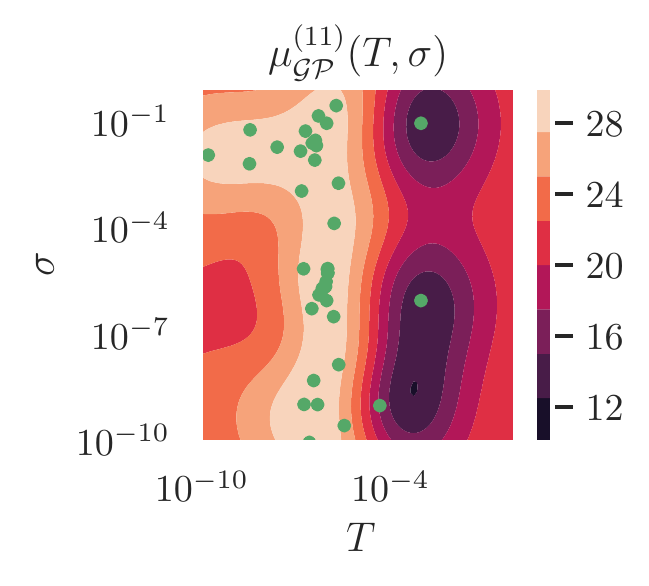}
        & \includegraphics[width=\subfigwidtha]{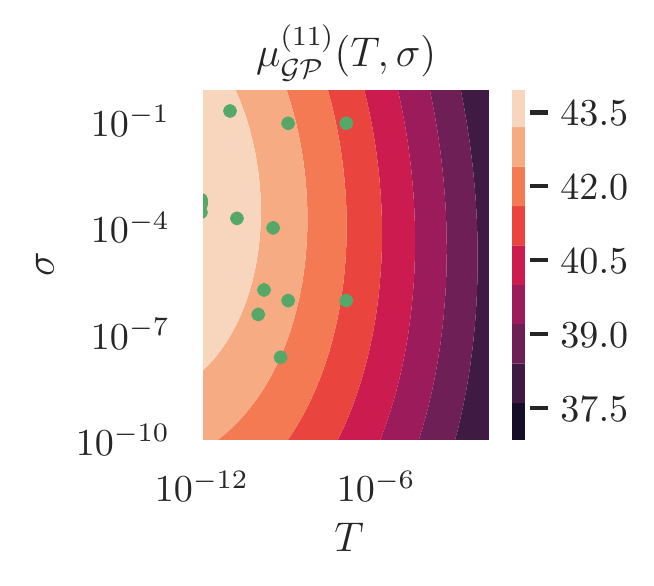}
        & \includegraphics[width=\subfigwidtha]{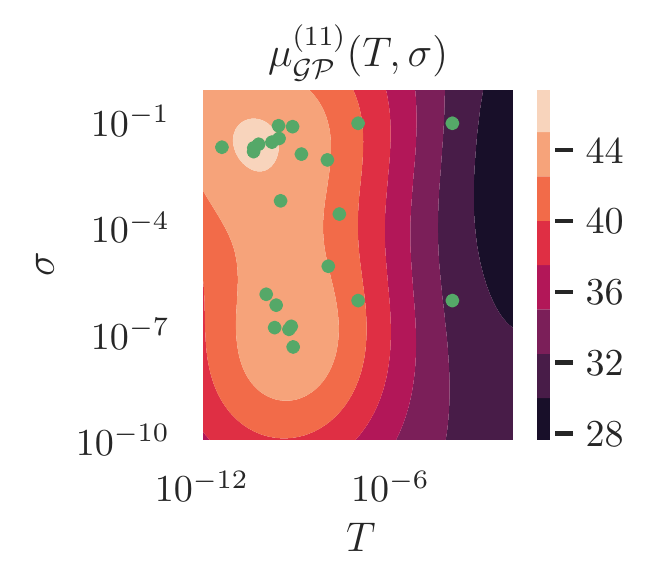} 
    \end{tabular}
    \caption{Gaussian process posterior mean $ \mu_{\mathcal{GP}}^{(11)} $ after BO termination for all Bayesian methods and inverse tasks. The colorbars denote reconstruction accuracy measured with PSNR. The BO is initialized with four uniformly distributed points within the selected bounds of the hyperparameters. Optimizing the respective hyperparameters has a considerable impact on the resulting reconstruction accuracy. The tasks of denoising and super-resolution exhibit larger plateaus, where optimal PSNR values are achieved, whereas for inpainting and CT reconstruction, more distinct spots are found. Green dots denote visited points used for training the GP during BO. See Tab.~\ref{tab:bo_parameters} for the parameter bounds and selected optimal values used to generate the test results. More results from intermediate BO steps can be found in the supplemental material.}
    \label{fig:bo_gp_mean}
\end{figure*}

\begin{table*}[h]
    \centering
    \caption{Hyperparameter bounds and results from BO per inverse task and Bayesian method. The full GP posterior mean is shown in Fig.~\ref{fig:bo_gp_mean}.}
    \begin{tabular}{lcccccc}
    \toprule
    Method & Parameter & $\log_{10}$-Bounds & Denoising & Super-Resolution & Inpainting & CT Reconstruction \\
    \midrule
    \multirow{ 2}{*}{MCD}  & $\lambda$ & $ [-10, 0] $ & $1.5\mathrm{e}{-8}$ & $1.3\mathrm{e}{-7}$ & $6.4\mathrm{e}{-7}$ & $1.5\mathrm{e}{-6}$ \\
                           & $p$ & $ [-4, -0.1] $       & $0.251$  & $0.028$ & $0.0025$  & $0.014$ \\
    \multirow{ 2}{*}{SGLD} & $ \lambda $ & $ [-12, -2] $ & $8.8\mathrm{e}{-9}$ & $1.6\mathrm{e}{-4}$   & $9.1\mathrm{e}{-5}$ & $8.7\mathrm{e}{-4}$ \\
                             & $\gamma$ & $ [-4\mathrm{e}{-4}, 0] $ & $0.99993$ & $0.99995$ & $1.0$ & $0.99989$ \\
    \multirow{ 2}{*}{POTOBIM (ours)} & $T$ & $ [-12, -2] $      & $5.6\mathrm{e}{-7}$ & $4.4\mathrm{e}{-7}$ & $7.1\mathrm{e}{-9}$ & $2.2\mathrm{e}{-10}$ \\
                              & $\sigma$ & $ [-10, 0] $ & $1.5\mathrm{e}{-5}$ & $4.9\mathrm{e}{-8}$ & $1.3\mathrm{e}{-2}$ & $1.7\mathrm{e}{-7}$ \\
    \bottomrule
    \end{tabular}
    
    \label{tab:bo_parameters}
\end{table*}

We evaluate the performance of our POTOBIM approach on the following four inverse imaging tasks and compare it to non-Bayesian DIP \cite{Lempitsky2018}, Bayesian DIP with SGLD \cite{Cheng2019} and with MC dropout \cite{Laves2020MCDIP}, and to domain algorithms not based on deep learning methods.
We apply BO to optimize the posterior temperature $ T $ and the variance $ \sigma $ of a zero-mean Gaussian prior per inverse task.
Non-negativity constrains on $ T $ and $ \sigma $ are implemented by performing the BO in $ \log $ space.
To ensure a fair comparison, the other Bayesian DIP methods receive the same amount of hyperparameter optimization by additionally performing BO to find optimal values for SGLD and MC dropout.
We choose to optimize the learning rate decay parameter $ \gamma $, with learning rate $ \epsilon_{t} = \gamma^{t} \epsilon_{0} $ at iteration step $ t $, for SGLD, the dropout rate $ p $ for MC dropout, and the amount of decoupled weight decay $ \lambda $, with parameter update step $ \bm{w}_{t+1} = (1-\lambda) \bm{w}_{t} - \epsilon \nabla \mathcal{L}_{t} (\bm{w}_{t}) $, for both methods \cite{Loshchilov2019}.
In the following experiments, we use the same convolutional encoder-decoder network architecture with skip-connections as proposed by \citet{Lempitsky2018}.

\paragraph{CT Reconstruction}

In computed tomography, multiple projections are created by passing X-rays through an object from a large number of different viewing angles.
Many algorithms exist to reconstruct the scanned object from the projections, e.g., the filtered back-projection (FBP) or algorithms that aim at solving Eq.\ (\ref{eq:inverse_problem}) iteratively, where the Radon transform is used as forward operator.

One practical way to reduce the radiation dose and scanning time is to acquire less projections per slice, which is referred to as sparse-view CT.
In general, (dense-view) CTs are computed from 1,000--2,000 projections per rotation and sparse-view CT refers to 10--100 projections per rotation \cite{Kudo2013}.
In this experiment, we simulate sparse-view CT by computing only 45 projections from lung CTs of COVID-19 patients using the parallel-ray forward Radon transform.
This equals to one X-ray every 4° per 180° rotation per slice.
As for super-resolution, this involves adding adjacent pixels, which are dependent random variables, rendering estimation of aleatoric uncertainty impossible.
Therefore, all networks in this experiment are trained by minimizing the pixel-wise mean-squared error only.
We use publicly available data from \href{https://coronacases.org}{https://coronacases.org}.
Unfortunately, we do not know the exact number of projections used in the employed public data and estimate the sparsification to be between 4.5\,\% and 2.25\,\% of the original data.
A single CT slice from the center of the 3D CT volume is used and rescaled to $ 256 \times 256 $ pixels.
The Houndsfield units are converted to floating point values and normalized.
The setup of this experiment is visualized in Fig.~\ref{fig:concept}.

\paragraph{Super-Resolution}

In CT and MRI, the sampling frequency is limited due to inherent physical limitations of the imaging utility, i.e., the pitch or spacing of the detector \cite{Greenspan2009}.
The resolution can be enhanced by reducing the size of detectors, but this comes at the expense of increased noise.
Since imaging devices are usually tuned towards low noise and short acquisition time, part of the resolution is sacrificed.
This motivates resolution-enhancing inverse post-processing methods using a single image.
We use slices of a high-resolution T1-weighted in vivo whole brain MRI with isotropic resolution of 250\,\textmu m \cite{T1MRI2018}
from public data sets.
The $ 512 \times 448 $ pixel full-resolution images act as ground truth $ \bm{x} $ and are downsampled by a factor of 4 to obtain low-resolution images $ \bm{y} $.
The image-generator network is optimized by applying a downsampling operator as forward process $ \mathcal{F} \colon \mathbb{R}^{4H \times 4W} \rightarrow \mathbb{R}^{H \times W} $ to its output $ \hat{\bm{x}} $.
Since there are many high-resolution images that reduce to the same low-resolution image, super-resolution is an ill-posed problem and choosing the downsampling operator is far from surjective.
Nearest neighbor downsampling preserves $ \hat{\bm{y}} $ to be Gaussian after performing the forward transform, as it is equivalent to an identity map for the remaining pixels.
Other methods, such as bilinear interpolation involve adding two or more highly dependent neighboring pixels.
The sum of two Gaussian random variables is not necessarily Gaussian, if the random variables are not independent, which prohibits direct estimation of aleatoric uncertainty as of Eq.~(\ref{eq:nll}).
We therefore use nearest neighbor downsampling as forward operator.

\paragraph{Denoising}

Optical coherence tomography (OCT) and ultrasound are prone to speckle noise due to interference phenomena, which can obscure small anatomical details and reduce image contrast.
Although speckle patterns contain information about the tissue microstructure, denoising of such images is desirable because this information is imperceptible to a human observer.
Speckle noise can be modeled as additive white Gaussian noise on log-transformed image intensities \cite{Michailovich2006}. 
Noise in low-dose X-ray originates from irregular photon density and can be modeled with Poisson noise \cite{Lee2018,Zabic2013}.
We approximate the Poisson noise with Gaussian noise since $ \mathsf{Poisson(\lambda)} $ approaches a Normal distribution as $ \lambda \rightarrow \infty $.
We first create a low-noise image $ \bm{x} $ by smoothing and downsampling the original image to $ 256 \times 256 $ pixel.
This averages over highly correlated neighboring pixels affected by uncorrelated noise and decreases the observation noise.
The downsampled image acts as ground truth and is corrupted by $ \bm{y} = \bm{x} + \mathcal{N}(\bm{0}, 0.1^{2}\bm{I}) $ using normal (X-ray) or log-transformed intensities (OCT).
The implemented forward operator for denoising is the identity mapping $ \mathcal{F} \colon \hat{\bm{x}} \mapsto \hat{\bm{y}} $ and thus, $ \hat{\bm{y}} $ remains to be Gaussian.
We use retinal OCT scans and pediatric pneumonia chest X-rays with native resolutions of $ 496 \times 496 $ and $ 1029 \times 1260 $ pixel from a public data set \cite{Kermany2018}.

\paragraph{Inpainting}

Applications of inpainting in medical imaging are hair removal in dermoscopy \cite{Abbas2011}, specular highlight removal in endoscopy \cite{Arnold2010}, or metal artifact removal in CT sinograms \cite{Peng2020} and MRI \cite{Armanious2020}.
In this paper, we focus on the former task and sample images from the HAM10000 data set \cite{Tschandl2018} showing different skin lesions with hair occlusions.
We manually mask the hair and optimize the fully-tempered ELBO with zero-weighting the masked pixels in the likelihood term.
The networks thus interpolate the masked areas.

\section{Results}

\begin{figure*}[h]
    \centering
    \includegraphics[width=5.8cm]{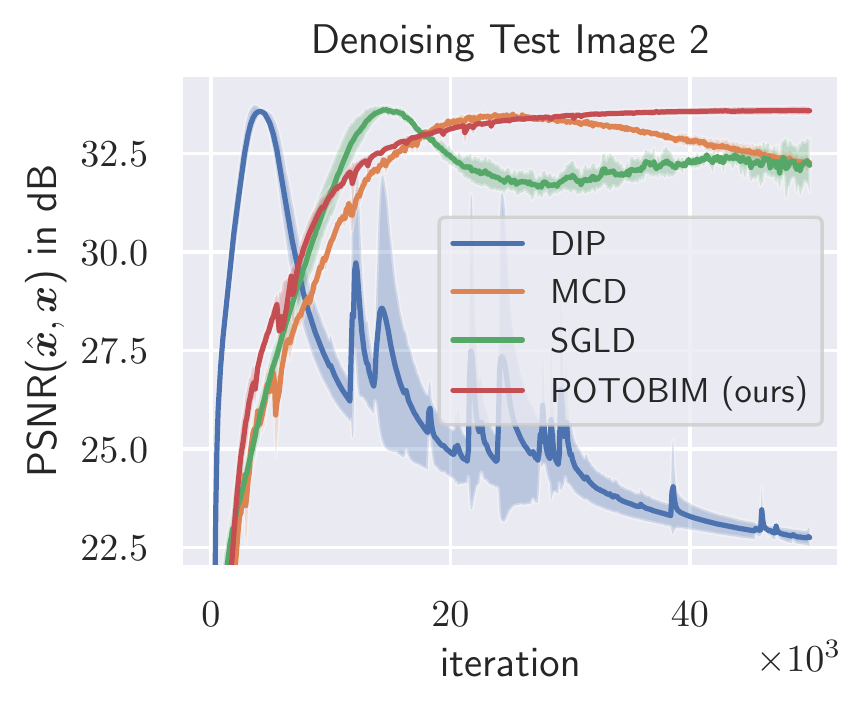} \hfill
    \includegraphics[width=5.8cm]{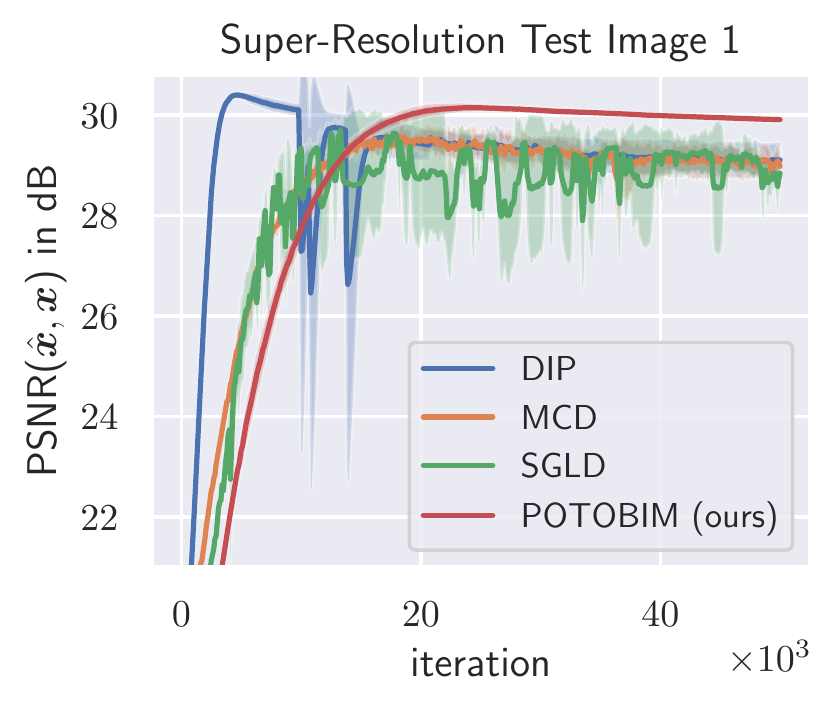} \hfill
    \includegraphics[width=5.8cm]{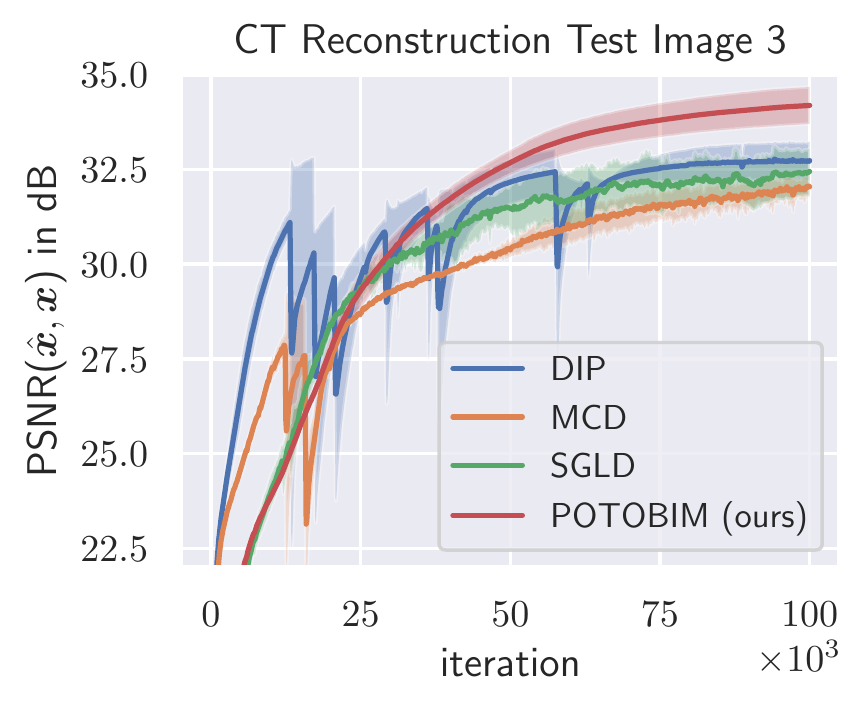}
    \caption{Our MFVI approach with an optimized prior does not overfit. The plots show $ \mu \pm \sigma $ from three runs with different random initialization. In denoising and super-resolution all methods except for POTOBIM exhibit overfitting and converge to sub-optimal results. POTOBIM on the other hand safely converges to its optimum. In CT reconstruction no method shows overfitting, but POTOBIM obtains the highest reconstruction accuracy.}
    \label{fig:overfitting}
\end{figure*}

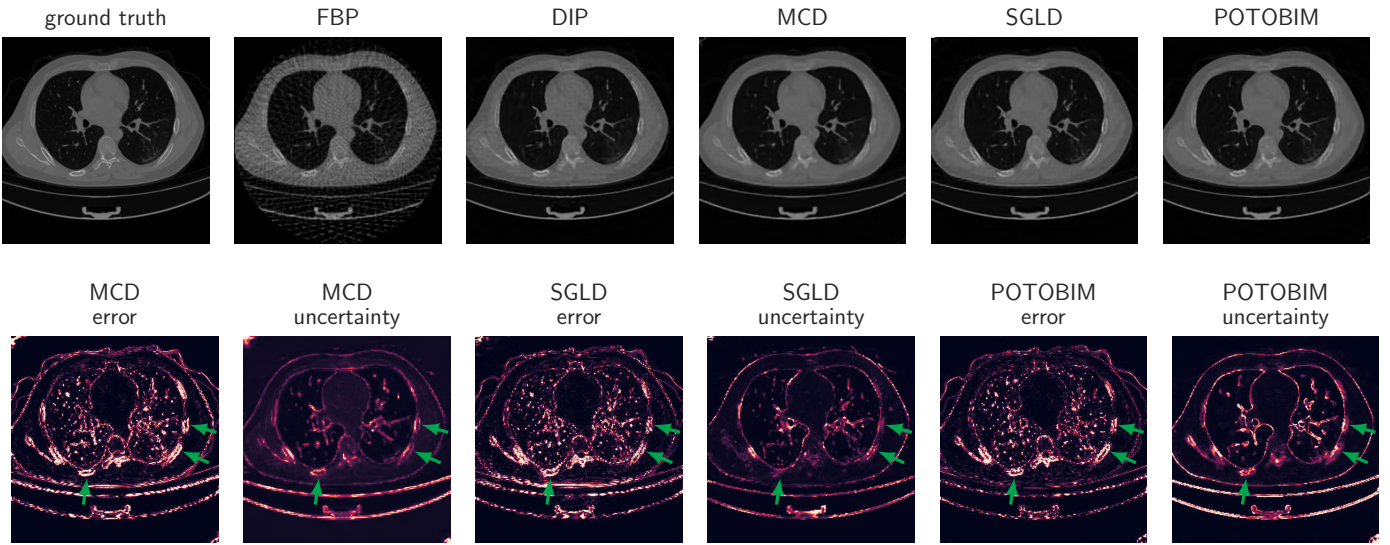
\begin{figure*}[h]
    \centering
    \includegraphics[width=1\textwidth]{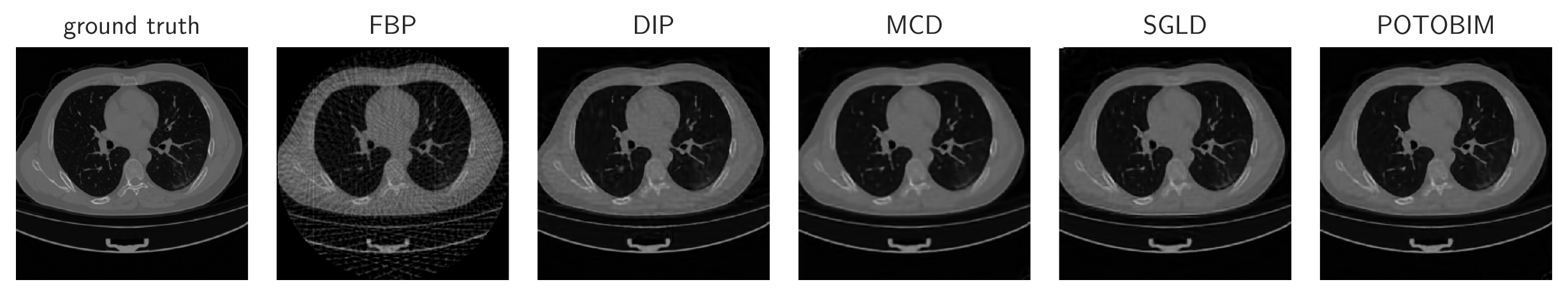}
    \begin{tikzpicture}
        \node[] {\includegraphics[width=1\textwidth]{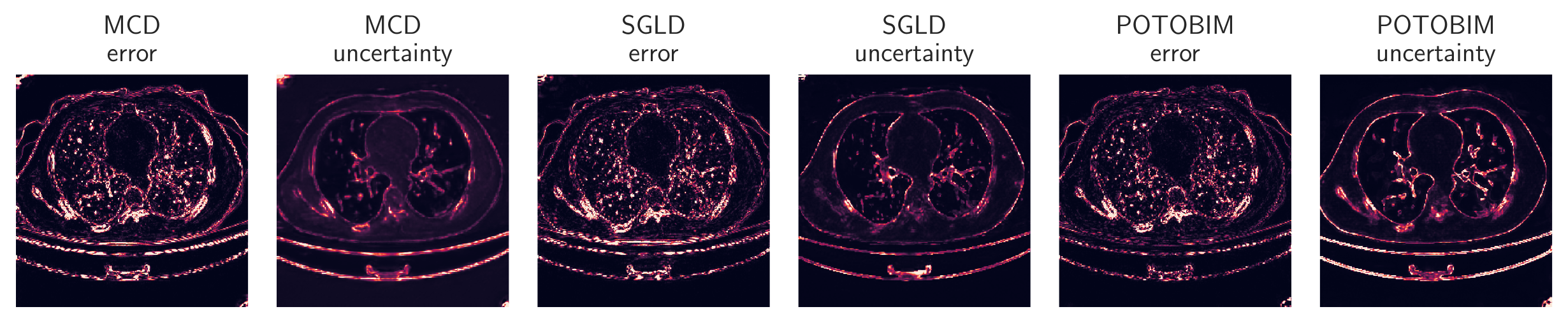}};
        \draw [-latex,Green,very thick] (8.95,-0.25) -- (8.6,-0.15);
        \draw [-latex,Green,very thick] (8.85,-0.65) -- (8.5,-0.5);
        \draw [-latex,Green,very thick] (7.2,-1.2) -- (7.25,-0.85);
        \draw [-latex,Green,very thick] (5.9,-0.25) -- (5.55,-0.15);
        \draw [-latex,Green,very thick] (5.8,-0.65) -- (5.45,-0.5);
        \draw [-latex,Green,very thick] (4.15,-1.2) -- (4.2,-0.85);
        \draw [-latex,Green,very thick] (2.85,-0.25) -- (2.5,-0.15);
        \draw [-latex,Green,very thick] (2.75,-0.65) -- (2.4,-0.5);
        \draw [-latex,Green,very thick] (1.07,-1.2) -- (1.12,-0.85);
        \draw [-latex,Green,very thick] (-0.2,-0.25) -- (-0.55,-0.15);
        \draw [-latex,Green,very thick] (-0.3,-0.65) -- (-0.65,-0.5);
        \draw [-latex,Green,very thick] (-1.95,-1.2) -- (-1.9,-0.85);
        \draw [-latex,Green,very thick] (-3.25,-0.25) -- (-3.6,-0.15);
        \draw [-latex,Green,very thick] (-3.35,-0.65) -- (-3.7,-0.5);
        \draw [-latex,Green,very thick] (-5.0,-1.2) -- (-4.95,-0.85);
        \draw [-latex,Green,very thick] (-6.3,-0.25) -- (-6.65,-0.15);
        \draw [-latex,Green,very thick] (-6.4,-0.65) -- (-6.75,-0.5);
        \draw [-latex,Green,very thick] (-8.05,-1.2) -- (-8.0,-0.85);
        \end{tikzpicture}
    \caption{Qualitative results for sparse-view CT reconstruction for COVID-19 test image 4 from \href{https://coronacases.org}{https://coronacases.org} using only 45 projections. (Top) Reconstruction results after 100k iterations and FBP as comparison. (Bottom) Squared error to ground truth and reconstruction uncertainty from Bayesian methods. POTOBIM yields highest reconstruction accuracy (e.g., see ribs, green arrows) without sparse-view related artifacts (cf.\ FBP and DIP) and provides best uncertainty estimates (e.g., high uncertainty around ribs correlates with the error). Best viewed with digital zoom.}
    \label{fig:uncertainty_ct}
\end{figure*}

\begin{table*}[h]
    \centering
    \caption{Quantitative results for sparse-view CT reconstruction using COVID-19 lung scans from \href{https://coronacases.org}{https://coronacases.org} as test data. Higher is better for reconstruction accuracy (PSNR and SSIM); lower is better for calibration error (UCE). We report means from three runs with different random initialization. Bold font denotes best values per test image.}
    \label{tab:quantitative_ct}
    \begin{tabular}{lcccccccccccc}
        \toprule
                & \multicolumn{3}{c}{Test Image 1} & \multicolumn{3}{c}{Test Image 2} & \multicolumn{3}{c}{Test Image 3} & \multicolumn{3}{c}{Test Image 4} \\
        \cmidrule(lr){2-4} \cmidrule(lr){5-7} \cmidrule(lr){8-10} \cmidrule(lr){11-13}
        Method  & PSNR  & SSIM  & UCE & PSNR & SSIM & UCE & PSNR & SSIM & UCE & PSNR & SSIM & UCE \\
        \midrule
        FBP            & 25.68 & 0.733          & ---   & 22.78 & 0.702 & ---   & 25.57 & 0.736  & ---  & 25.29 & 0.734 & --- \\
        DIP            & 34.07 & 0.936          & ---   & 30.71 & 0.865 & ---   & 32.72 & 0.898  & ---  & 34.22 & 0.907 & --- \\
        MCD            & 32.31 & 0.927 & 0.048 & 30.62 & 0.875 & 0.077 & 31.77 & 0.906 & 0.056 & 33.11 & 0.910 & 0.037 \\
        SGLD           & 32.89 & 0.931          & 0.045 & 30.46 & 0.868 & 0.085 & 32.43 & 0.906 & 0.053 & 33.34 & 0.915 & 0.037 \\
        POTOBIM (ours) & \textbf{34.51} & \textbf{0.943} & \textbf{0.034} & \textbf{31.17} & \textbf{0.882} & \textbf{0.074} & \textbf{34.18} & \textbf{0.924} & \textbf{0.043} & \textbf{35.05} & \textbf{0.924} & \textbf{0.029} \\
        \bottomrule
    \end{tabular}
\end{table*}

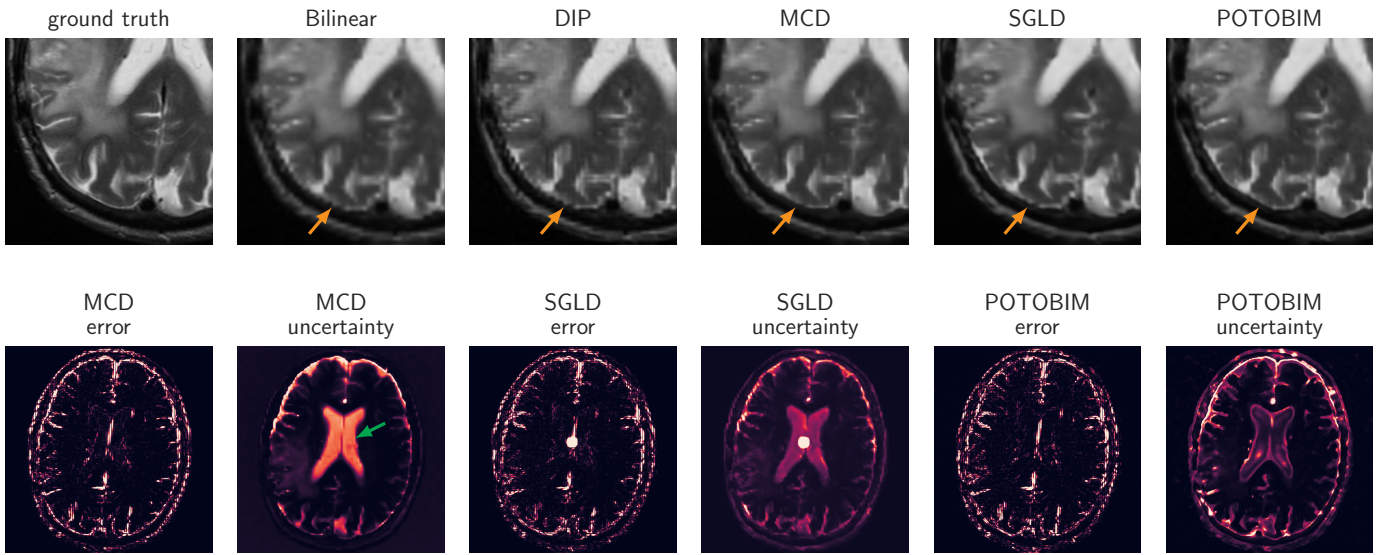
\begin{figure*}[h]
    \centering
    \begin{tikzpicture}
        \node[] { \includegraphics[width=1\textwidth]{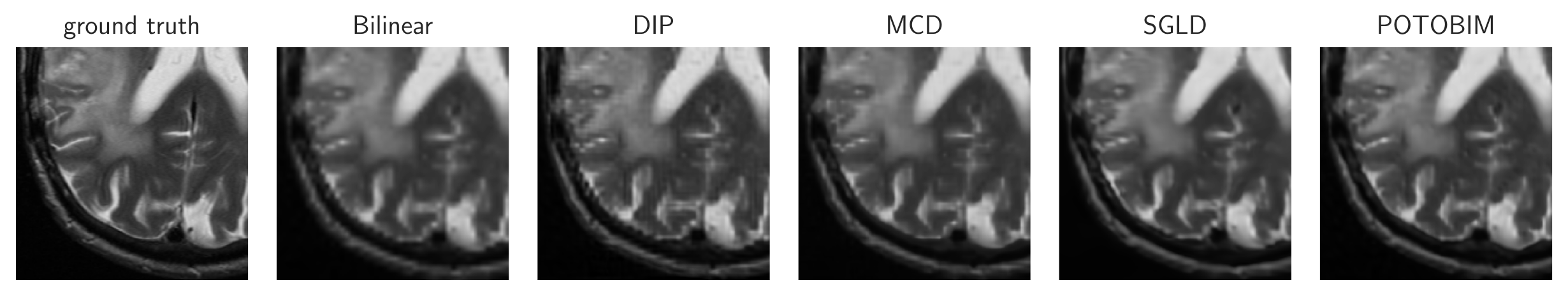}};
        \draw [-latex,BurntOrange,very thick] (7.2,-1.4) -- (7.5,-1.05);
        \draw [-latex,BurntOrange,very thick] (4.15,-1.4) -- (4.45,-1.05);
        \draw [-latex,BurntOrange,very thick] (1.1,-1.4) -- (1.4,-1.05);
        \draw [-latex,BurntOrange,very thick] (-1.95,-1.4) -- (-1.65,-1.05);
        \draw [-latex,BurntOrange,very thick] (-5.0,-1.4) -- (-4.7,-1.05);
    \end{tikzpicture}
    \begin{tikzpicture}
        \node[] {\includegraphics[width=1\textwidth]{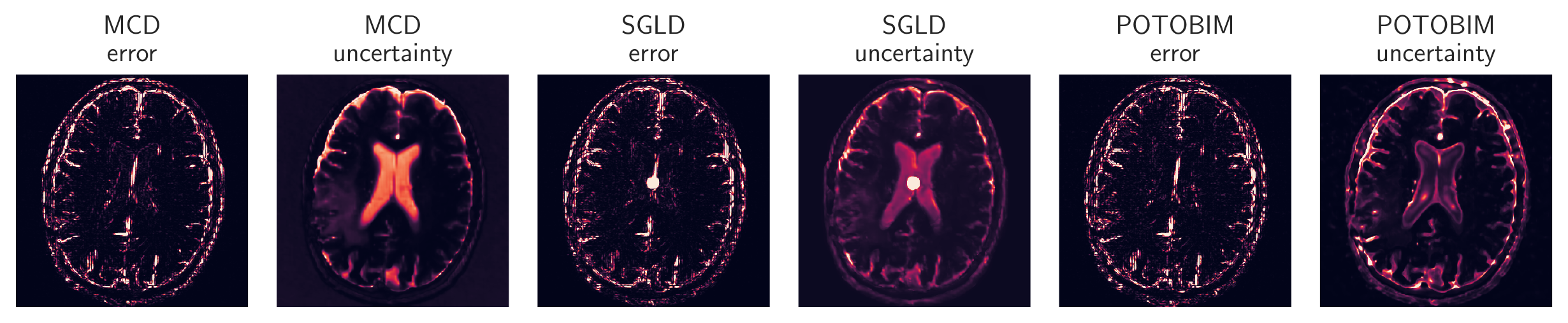}};
        \draw [-latex,Green,very thick] (-4.0,-0.0) -- (-4.4,-0.2);
    \end{tikzpicture}
    \caption{Qualitative results for MRI super-resolution test image 1. (Top) Reconstruction results after 50k iterations and bilinear upsampling as comparison. (Bottom) Squared error to ground truth and reconstruction uncertainty from Bayesian methods. POTOBOM shows the least aliasing artifacts (orange arrows), which can be caused by the employed nearest neighbor downsampling. MCD erroneously exhibits high uncertainty in the area of the lateral ventricles (green arrows), although the reconstruction error there is small. Best viewed with digital zoom.}
    \label{fig:uncertainty_sr}
\end{figure*}

\begin{table*}[h]
    \centering
    \caption{Quantitative results for MRI super-resolution. We use different slices from a high-resolution T1-weighted brain MRI as test data \citep{T1MRI2018}. Higher is better for reconstruction accuracy (PSNR and SSIM); lower is better for calibration error (UCE). We report mean values from three runs with different random initialization. Bold font denotes best values per test image.}
    \label{tab:quantitative_sr}
    \begin{tabular}{lcccccccccccc}
        \toprule
                & \multicolumn{3}{c}{Test Image 1} & \multicolumn{3}{c}{Test Image 2} & \multicolumn{3}{c}{Test Image 3} & \multicolumn{3}{c}{Test Image 4} \\
        \cmidrule(lr){2-4} \cmidrule(lr){5-7} \cmidrule(lr){8-10} \cmidrule(lr){11-13}
        Method  & PSNR  & SSIM  & UCE & PSNR & SSIM & UCE & PSNR & SSIM & UCE & PSNR & SSIM & UCE \\
        \midrule
        Bilinear       & 23.19 & 0.789 & --- & 23.19 & 0.788 & --- & 26.20 & 0.833 & --- & 25.49 & 0.818 & --- \\
        DIP            & 29.10 & 0.879 & ---  & 29.56 & 0.841 & --- & 29.74 & 0.855 & --- & \textbf{29.65} & \textbf{0.867} & --- \\
        MCD            & 29.03 & 0.882 & \textbf{0.068} & 29.50 & \textbf{0.845} & 0.107 & 29.55 & 0.855 & 0.106 & 29.23 & 0.861 & 0.114 \\
        SGLD           & 28.80 & 0.881 & 0.172 & 29.22 & 0.833 & 0.107 & 29.73 & 0.842 & 0.100 & 28.70 & 0.843 & 0.116 \\
        POTOBIM (ours) & \textbf{29.91} & \textbf{0.893} & 0.091 & \textbf{29.70} & 0.844 & \textbf{0.096} & \textbf{29.86} & \textbf{0.859} & \textbf{0.092} & 29.49 & 0.866 & \textbf{0.104} \\
        \bottomrule
    \end{tabular}
\end{table*}

\begin{figure*}[h]
    \centering
    \begin{tikzpicture}
            \node[] {\includegraphics[width=1\textwidth]{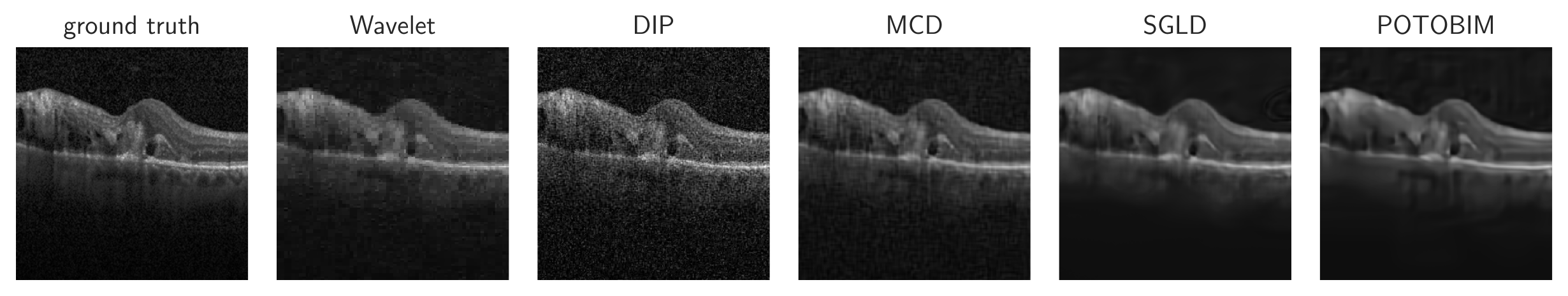}};
            \draw [-latex,Green,very thick] (8.7,0.75) -- (8.6,0.25);
            \draw [-latex,Green,very thick] (5.65,0.75) -- (5.55,0.25);
            \draw [-latex,Green,very thick] (2.6,0.75) -- (2.5,0.25);
            \draw [-latex,Green,very thick] (-0.45,0.75) -- (-0.55,0.25);
            \draw [-latex,Green,very thick] (-3.5,0.75) -- (-3.6,0.25);
    \end{tikzpicture}
    \includegraphics[width=1\textwidth]{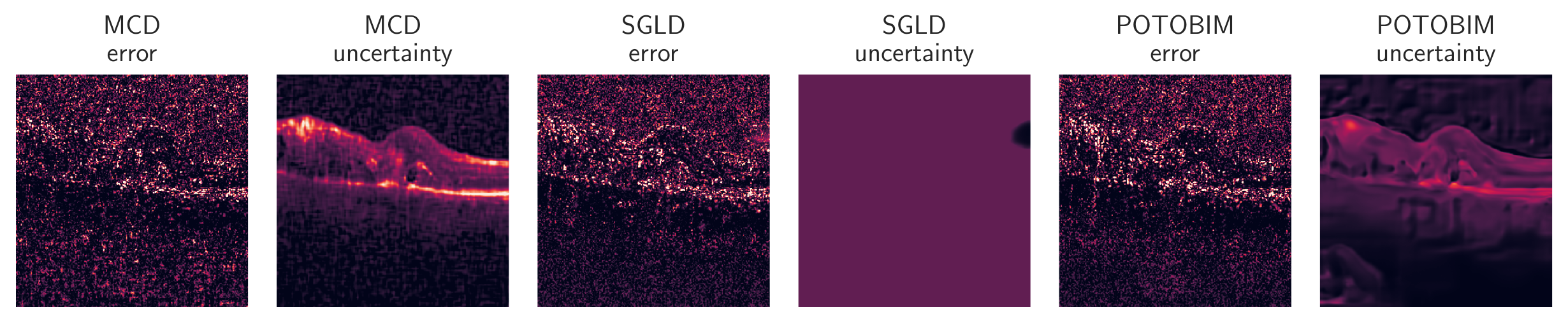}
    \caption{Qualitative results for denoising test image 4 (OCT). (Top) Reconstruction results after 100k iterations and wavelet denoising as comparison. (Bottom) Squared error to ground truth and reconstruction uncertainty from Bayesian methods. The distinct retinal layers on the right are best resolved by POTOBIM (green arrows). We observed a collapse of the aleatoric uncertainty from SGLD in all denoising test experiments.  Best viewed with digital zoom.}
    \label{fig:uncertainty_denoising}
\end{figure*}
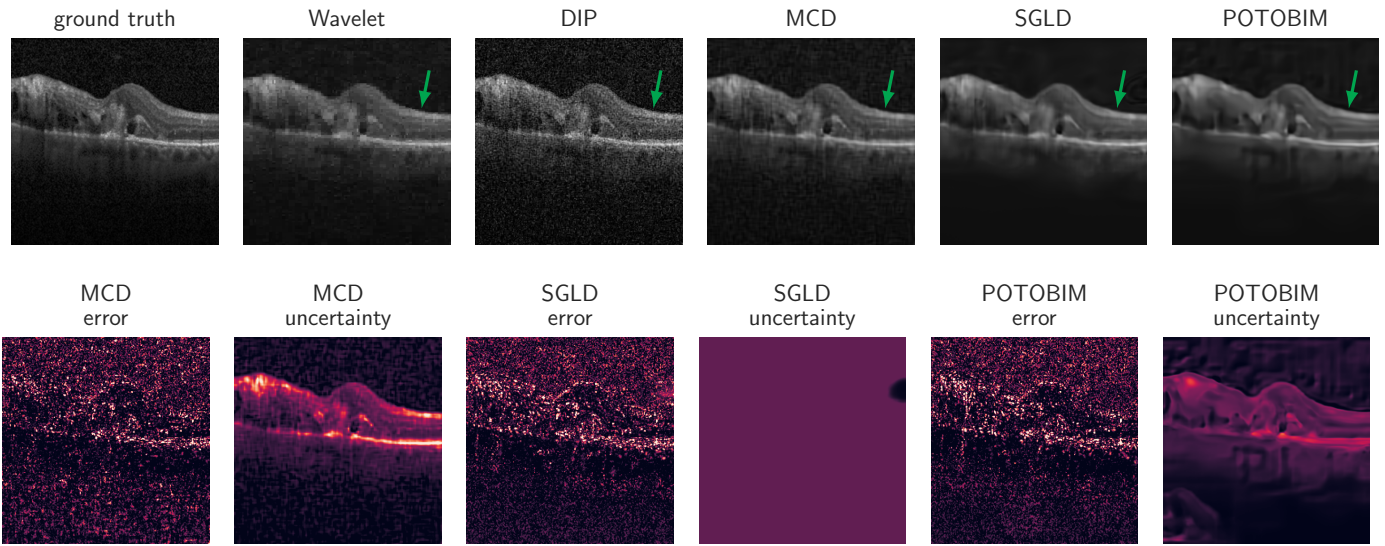

\begin{table*}[h]
    \centering
    \caption{Quantitative results for denoising. Test images 1--3 are pediatric pneumonia chest X-rays and test image 4 is a retinal OCT scan showing choroidal neovascularization. All images are sampled from a public data set presented by \citet{Kermany2018}. High UCE values for SGLD are caused by a collapse of the aleatoric uncertainty (cf.\ Fig.~\ref{fig:uncertainty_denoising}). Higher is better for reconstruction accuracy (PSNR and SSIM); lower is better for calibration error (UCE). We report mean values from three runs with different random initialization. Bold font denotes best values per test image.}
    \label{tab:quantitative_results_denoising}
    \begin{tabular}{lcccccccccccc}
        \toprule
                & \multicolumn{3}{c}{Test Image 1} & \multicolumn{3}{c}{Test Image 2} & \multicolumn{3}{c}{Test Image 3} & \multicolumn{3}{c}{Test Image 4} \\
        \cmidrule(lr){2-4} \cmidrule(lr){5-7} \cmidrule(lr){8-10} \cmidrule(lr){11-13}
        Method  & PSNR  & SSIM  & UCE & PSNR & SSIM & UCE & PSNR & SSIM & UCE & PSNR & SSIM & UCE \\
        \midrule
        Wavelet        & 28.73 & 0.861 & ---  & 31.22 & 0.897 & ---  & 28.73 & 0.861 & ---  & 27.56 & 0.768 & --- \\
        DIP            & 23.11 & 0.431 & ---  & 22.77 & 0.313 & ---  & 23.10 & 0.431 & ---  & 23.43 & 0.403 & --- \\
        MCD            & 30.87 & 0.840 & 1.12 & 32.47 & 0.831 & 1.05 & 30.82 & 0.839 & 1.11 & 27.97 & 0.582 & 0.70 \\
        SGLD           & 30.03 & 0.913 & 80.5 & 32.16 & 0.848 & 82.8 & 29.95 & 0.817 & 86.6 & 27.99 & 0.577 & 79.9 \\
        POTOBIM (ours) & \textbf{30.91} & \textbf{0.841} & \textbf{0.99} & \textbf{33.59} & \textbf{0.872} & \textbf{1.00} & \textbf{30.97} & \textbf{0.844} & \textbf{0.99} & \textbf{28.23} & \textbf{0.584} & \textbf{0.67} \\
        \bottomrule
    \end{tabular}
\end{table*}

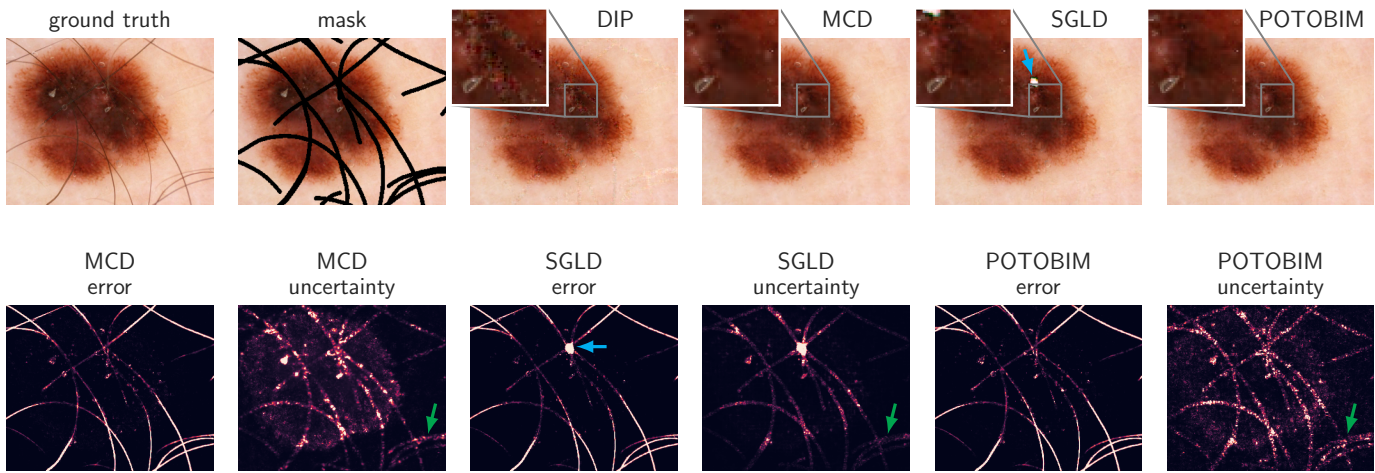
\begin{figure*}[h]
    \centering
    \begin{tikzpicture}
            \node[] {\includegraphics[width=1\textwidth]{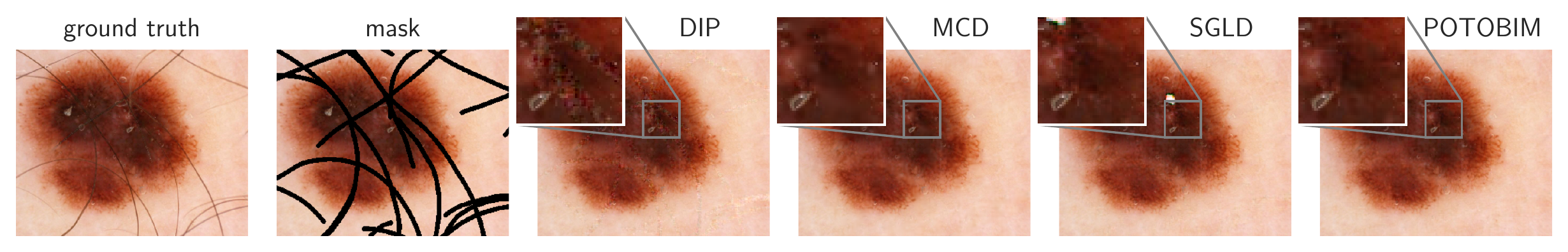}};
            \draw [-latex,Cyan,very thick] (4.4,0.75) -- (4.5,0.4);
    \end{tikzpicture}
    \begin{tikzpicture}
            \node[] {\includegraphics[width=1\textwidth]{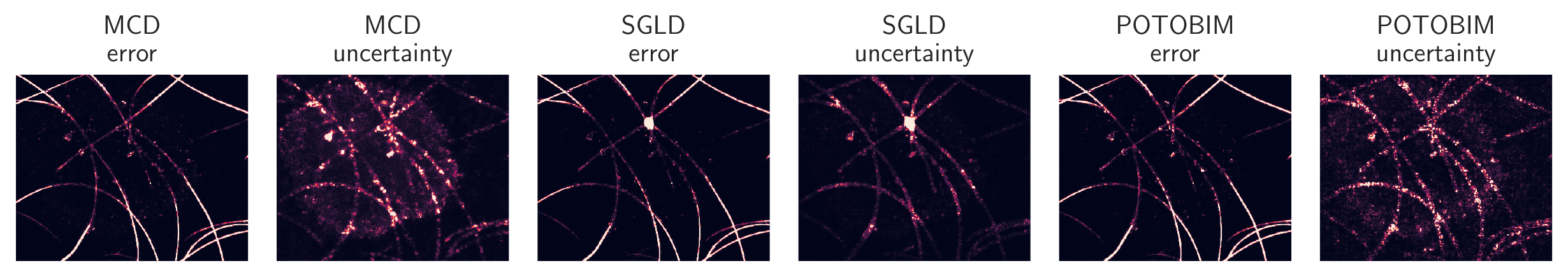}};
            \draw [-latex,Green,very thick] (8.75,-0.55) -- (8.65,-0.95);
            \draw [-latex,Green,very thick] (2.7,-0.55) -- (2.6,-0.95);
            \draw [-latex,Green,very thick] (-3.35,-0.55) -- (-3.45,-0.95);
            \draw [-latex,Cyan,very thick] (-1.1,0.2) -- (-1.5,0.2);
    \end{tikzpicture}
    \caption{Qualitative results for hair removal in dermoscopy images for skin lesion classification from the HAM10000 data set as test data of test image 1 \citep{Tschandl2018}. (Top) Reconstruction results after 50k iterations. (Bottom) Squared error to ground truth and reconstruction uncertainty from Bayesian methods. SGLD is prone to creating artifacts in the reconstruction (blue arrows) and MCD overly smooths out the inpainted areas (see zoomed window).
    POTOBIM yields highest reconstruction accuracy without artifacts and provides best uncertainty estimates (e.g., high uncertainty in regions corresponding to hair, green arrow). Best viewed with digital zoom.}
    \label{fig:uncertainty_inpainting}
\end{figure*}

\begin{figure*}[h]
    \centering
    \begin{tikzpicture}
            \node[] {\includegraphics[width=1\textwidth]{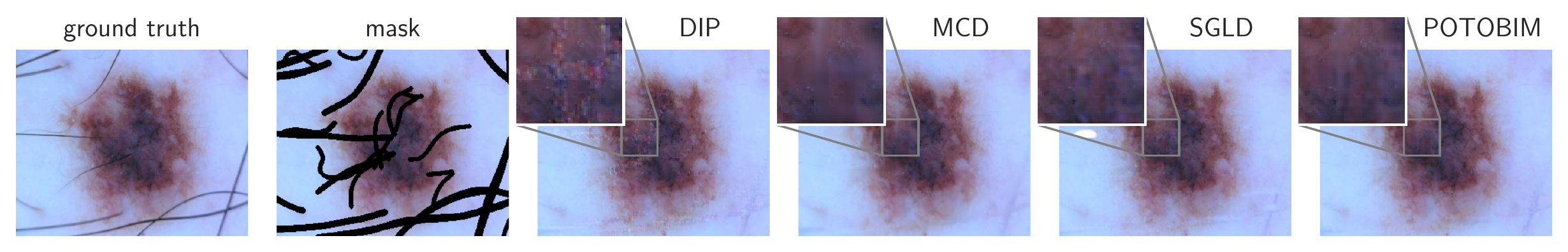}};
            \draw [-latex,NavyBlue,very thick] (3.4,-0.5) -- (3.5,-0.15);
    \end{tikzpicture}
    \begin{tikzpicture}
            \node[] {\includegraphics[width=1\textwidth]{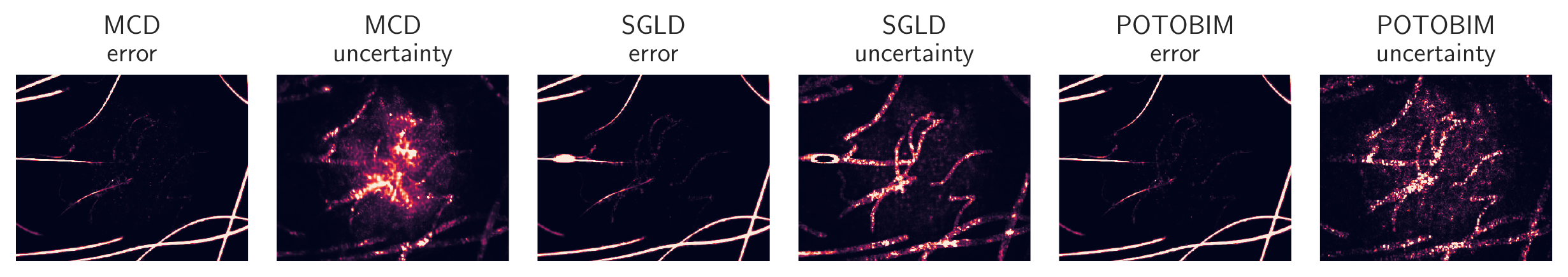}};
            \draw [-latex,NavyBlue,very thick] (-2.68,-0.65) -- (-2.58,-0.3);
    \end{tikzpicture}
    \caption{Qualitative results for hair removal in dermoscopy images for skin lesion classification from the HAM10000 data set as test data of test image 2. (Top) Reconstruction results after 50k iterations. (Bottom) Squared error to ground truth and reconstruction uncertainty from Bayesian methods.
    As on test image 1 (Fig.~\ref{fig:uncertainty_inpainting}), SGLD produces reconstruction artifacts (blue arrows).
    DIP shows color artifacts and MCD overly smooths the inpainted areas (see zoomed window).
    Best viewed with digital zoom.
    }
    \label{fig:uncertainty_inpainting2}
\end{figure*}

The results are presented as follows: 
First, we present the results of the BO used to optimize the posterior temperature $ T $ and the variance of the prior $\sigma$ of POTOBIM on a per task level.
We further show the results of BO for the other Bayesian methods, SGLD \citep{Cheng2019} and MC dropout \citep{Laves2020MCDIP}, where we optimized the amount of decoupled weight decay $ \lambda $ for both and the exponential learning rate decay parameter $ \gamma $ for SGLD and the dropout rate $ p $ for MC dropout respectively.
Second, we provide comparison of all four approaches on a variety of test images sampled from public data sets for four inverse imaging tasks.
We show that despite the optimized hyperparameters, DIP with SGLD and MC dropout overfit the corrupting patterns in inverse imaging tasks after sufficient enough iterations and converge to inferior reconstructions in terms of PSNR.
Our method on the other hand employing MFVI outperforms the aforementioned methods by means of reconstruction accuracy (PSNR) on all task and further provides well-calibrated predictive uncertainty maps.
For comparison we also provide the results of using the non-Bayesian DIP from \citet{Lempitsky2018} and another unsupervised reconstruction technique per task.

\paragraph{Bayesian Optimization}

Fig.~\ref{fig:bo_gp_mean} shows the Gaussian process posterior mean $ \mu_{\mathcal{GP}}^{(11)} $ after BO termination for all Bayesian methods and inverse tasks.
The quantitative values for each optimized parameter are presented in Tab.~\ref{tab:bo_parameters}.
For each task and Bayesian method, we find a distinct optimum in the search space.
The dropout rate $ p $ has a considerable effect on the performance of MC dropout, while the weight decay parameter $ \lambda $ contributes less (see, e.g., BO for denoising).
For SGLD, we often observe the learning rate decay parameter $ \gamma $ converging to 1.0, confirming the results of \citet{Brosse2018}.
The values for $ \lambda $ converge to different values compared to MC dropout, indicating different optimal prior distributions per task for the two methods.
In MFVI, we see higher contribution from the temperature $ T $ exemplified by the vertical band present in all BOs, which indicates a less prominent effect of the prior's variance $ \sigma $.
The temperature for all tasks is found to be at the lower spectrum.
A considerable drop in reconstruction performance can be observed for $ T \rightarrow 1 $ in all tasks, highlighting the inferiority of an unscaled posterior with $ T=1 $.
For completeness, we additionally performed the CT reconstruction experiment for test image 1 with $ T=1 $ (and $ \sigma = 1.7\mathrm{e}{-7} $ as reported in Tab.~\ref{tab:bo_parameters}) and received poor results with $ \mathrm{PSNR} = 15.1\,\mathrm{dB} $.
A lower temperature emphasizes the contribution of the negative log-likelihood (i.e., MSE) and places weight away from the epistemic towards the aleatoric uncertainty; a narrow prior prevents weights from growing too large, effectively avoiding overfitting the corrupted image (note that we use a zero-mean prior).
The reconstruction accuracy landscape is highly non-linear as can be seen in the presented figure, which emphasizes the need for a derivative-free GP model as surrogate.
Additional figures showing intermediate BO steps can be found in the supplemental material to this paper.

\paragraph{CT Reconstruction}

In contrast to denoising or super-resolution, we do not observe overfitting of the low-resolution target image during optimization with any method by means of PSNR (cf.\ Fig.~\ref{fig:overfitting}).
However, the right amount of regularization introduced by the optimized posterior temperature in POTOBIM helps to safely converge with highest reconstruction accuracy.
We observed this consistently for all test images, shown by the quantitative results in Tab.~\ref{tab:quantitative_ct}
POTOBIM shows the least amount of sparse-view related artifacts as introduced by FBP and non-Bayesian DIP (see Fig.~\ref{fig:uncertainty_ct}).
The uncertainty for all methods is high at intensity discontinuities (i.e., image edges), which correlates well with the reconstruction error.
All methods exhibit a higher reconstruction error around the ribs, where only POTOBIM shows high predictive uncertainty.

\paragraph{Super-resolution}

In super-resolution we opt for a smooth reconstruction with high uncertainties in regions with edges, i.e., high Fourier frequencies, while anatomical details are reserved as well.
Further, as we employed nearest neighbor downsampling as forward operator the reconstructions are prone to aliasing, of which POTOBIM's reconstructions exhibit the least expressed in the higher PSNR values (cf. Tab.~\ref{tab:quantitative_sr}).
As indicated by the narrow peaks for non-Bayesian DIP and SGLD, the overfitting behavior starts early in the training phase again (Fig.~\ref{fig:overfitting}).
While overfitting can not be observed for MCD neither, POTOBIM generally yields higher reconstruction accuracies, which can be interpreted as an indicator for better detail preservation as well.
Additionally, both, POTOBIM and MCD, exhibit high uncertainties in edged regions as needed, but MCD unfortunately predicts high uncertainty in the area of the lateral ventricles as well (Fig.~\ref{fig:uncertainty_sr}).
The uncertainty estimate of SGLD is qualitatively and in terms of UCE inferior to POTOBIM's.

\paragraph{Denoising}

Even with BO applied to optimize the related hyperparameters, it is not possible for SGLD and MC dropout to eliminate overfitting the corrupting noisy patterns.
Compared to the non-Bayesian DIP, both converge to a reconstruction that has less noise, but is inferior to the reconstruction of POTOBIM, which safely converges to its highest value (see Fig.~\ref{fig:overfitting}).
The non-Bayesian DIP requires manually applied early stopping, which is indicated by the narrow PSNR peak.
SGLD shows almost identical overfitting behaviour without performing BO, while MC dropout exhibits less severe overfitting, as we showed in our MIDL submission \citep{Toelle2021}. 
With BO applied, the peaks for both SGLD and MC dropout get wider before overfitting starts making early stopping still essential for obtaining the best reconstruction accuracy.
However, our approach using MFVI does not overfit the corrupting patterns and outperforms the other approaches in terms of PSNR on all test images (cf.\ Tab.~\ref{tab:quantitative_results_denoising}).
Further, our approach provides well calibrated predictive uncertainty maps by means of UCE, indicating a greater robustness towards eventual hallucinations compared to DIP with SGLD and MC dropout (Fig.~\ref{fig:uncertainty_denoising}).
We additionally provide results from wavelet denoising with BayesShrink for comparison \citep{Chang2000}.

\paragraph{Inpainting}

For the task of hair inpainting we expect our reconstruction to be smooth with a corresponding high uncertainty in regions that were occluded by hair.
As can be seen in Fig.~\ref{fig:uncertainty_inpainting}--\ref{fig:uncertainty_inpainting2} POTOBIM yields smooth reconstructions without still present hairs as is the case for the other methods.
DIP shows color artifacts and MCD tends to overly smooth out the inpainted areas.
Even worse, MCD and especially SGLD sometimes exhibit artifacts in their reconstructions, which must be avoided at all costs.
Qualitatively, POTOBIM's predicted uncertainty is high in regions corresponding to hair as desired and low in the region of the chloasma, as this is important for the downstream task of classifying the skin lesion (Fig.~\ref{fig:uncertainty_inpainting}).


\section{Discussion}

This work presents a novel uncertainty-aware methodology to solve inverse tasks in medical imaging.
We use the tools of deep learning but avoid supervised learning to alleviate the failure modes of existing data-driven methods, such as hallucinations or their application to out-of-domain data.
By optimizing a randomly-initialized convolutional network as neural representation for each image, the reconstruction does not depend on learned image features or implicit regularization priors.
The optimization of the posterior temperature ensures convergence with highest accuracy.
However, selecting a subpar posterior temperature in practice does not result in a failed reconstruction, as shown by the PSNR landscape from the GP posterior.

\subsection{Limitations \& Future Work}

Optimizing a network with several million parameters for each image anew entails a considerable amount of computational effort.
Depending on the task, the reconstruction of a single image can take up to one hour on recent hardware (i.e., NVIDIA Titan RTX), which prevents POTOBIM to be deployed in clinical routine.
To address this, we envision a combination of supervised learning and image-based optimization by training a network on a large-scale dataset, which acts as a prior distribution (instead of a zero-mean Gaussian) during optimization of a network that has been initialized by a draw from the ``prior network''.
This is possible within our presented Bayesian framework and should enable much faster convergence without showing the aforementioned pitfalls of methods trained with full supervision.
Besides the addressed inverse tasks, POTOBIM could further be applied to unsupervised deformable registration \citep{Laves2019deformable} or any other inverse task as long as the forward operator (e.g., point spread function for fluorescence microscopy) can be implemented in a differentiable manner w.r.t.\ the network weights to obtain a solution for the ill-posed reverse operator.

\section{Conclusion}

We presented Posterior Temperature Optimized Bayesian Inverse Models (POTOBIM), a mean-field variational inference approach to deep image prior with a temperature scaled posterior. 
In general, Bayesian methods are more robust to overfitting due to their inbuilt regularization.
However, as shown empirically for SGLD and MC dropout, overfitting can still be caused even with an optimized weight prior, i.e., a partially tempered posterior.
Since MFVI allows for a broader selection of weight prior distributions, we can obtain a fully tempered posterior making it the method of choice for POTOBIM.
We demonstrated its superiority on four inverse imaging post-processing tasks in the medical domain, namely CT reconstruction, super-resolution, denoising, and inpainting.
Although BO with reconstruction accuracy as quality criterion was performed for all methods to obtain temperature scaled posteriors, POTOBIM does still outperform SGLD and MC dropout, which we attribute to the more detailed options of prior selection.
Our approach yields uncertainty estimates that are qualitatively appealing to a human observer and well-calibrated in terms of UCE as well, producing a tool that is well-equipped against hallucinations.
BO could also be employed with respect to a calibration criterion to further increase the predictive uncertainty maps' expressiveness.

\section*{Acknowledgements}

The authors gratefully acknowledge the data storage service SDS@hd supported by the Ministry of Science, Research and the Arts  Baden-Württemberg (MWK) and the German Research Foundation (DFG) through grant INST 35/1314-1 FUGG and INST 35/1503-1 FUGG.

ML and AS received funding from the Interdisciplinary Competence Center for Interface Research (ICCIR) Hamburg.

\bibliographystyle{model2-names.bst}\biboptions{authoryear}
\bibliography{refs}

\begin{thebibliography}{64}
\expandafter\ifx\csname natexlab\endcsname\relax\def\natexlab#1{#1}\fi
\providecommand{\url}[1]{\texttt{#1}}
\providecommand{\href}[2]{#2}
\providecommand{\path}[1]{#1}
\providecommand{\DOIprefix}{doi:}
\providecommand{\ArXivprefix}{arXiv:}
\providecommand{\URLprefix}{URL: }
\providecommand{\Pubmedprefix}{pmid:}
\providecommand{\doi}[1]{\href{http://dx.doi.org/#1}{\path{#1}}}
\providecommand{\Pubmed}[1]{\href{pmid:#1}{\path{#1}}}
\providecommand{\bibinfo}[2]{#2}
\ifx\xfnm\relax \def\xfnm[#1]{\unskip,\space#1}\fi
\bibitem[{Abbas et~al.(2011)Abbas, Celebi and Garc{\'\i}a}]{Abbas2011}
\bibinfo{author}{Abbas, Q.}, \bibinfo{author}{Celebi, M.E.},
  \bibinfo{author}{Garc{\'\i}a, I.F.}, \bibinfo{year}{2011}.
\newblock \bibinfo{title}{Hair removal methods: A comparative study for
  dermoscopy images}.
\newblock \bibinfo{journal}{Biomedical Signal Processing and Control}
  \bibinfo{volume}{6}, \bibinfo{pages}{395--404}.
\bibitem[{Adler and {\"O}ktem(2019)}]{Adler2019}
\bibinfo{author}{Adler, J.}, \bibinfo{author}{{\"O}ktem, O.},
  \bibinfo{year}{2019}.
\newblock \bibinfo{title}{Deep posterior sampling: Uncertainty quantification
  for large scale inverse problems}, in: \bibinfo{booktitle}{MIDL--Extended
  Abstract Track}.
\bibitem[{Aitchison(2021)}]{Aitchison2021}
\bibinfo{author}{Aitchison, L.}, \bibinfo{year}{2021}.
\newblock \bibinfo{title}{A statistical theory of cold posteriors in deep
  neural networks}, in: \bibinfo{booktitle}{International Conference on
  Learning Representations}.
\bibitem[{Antun et~al.(2020)Antun, Renna, Poon, Adcock and Hansen}]{Antun2020}
\bibinfo{author}{Antun, V.}, \bibinfo{author}{Renna, F.},
  \bibinfo{author}{Poon, C.}, \bibinfo{author}{Adcock, B.},
  \bibinfo{author}{Hansen, A.C.}, \bibinfo{year}{2020}.
\newblock \bibinfo{title}{On instabilities of deep learning in image
  reconstruction and the potential costs of ai}.
\newblock \bibinfo{journal}{Proceedings of the National Academy of Sciences}
  \bibinfo{volume}{117}, \bibinfo{pages}{30088--30095}.
\bibitem[{Armanious et~al.(2020)Armanious, Kumar, Abdulatif, Hepp, Gatidis and
  Yang}]{Armanious2020}
\bibinfo{author}{Armanious, K.}, \bibinfo{author}{Kumar, V.},
  \bibinfo{author}{Abdulatif, S.}, \bibinfo{author}{Hepp, T.},
  \bibinfo{author}{Gatidis, S.}, \bibinfo{author}{Yang, B.},
  \bibinfo{year}{2020}.
\newblock \bibinfo{title}{ipa-medgan: Inpainting of arbitrary regions in
  medical imaging}, in: \bibinfo{booktitle}{IEEE International Conference on
  Image Processing (ICIP)}, pp. \bibinfo{pages}{3005--3009}.
\bibitem[{Arnold et~al.(2010)Arnold, Ghosh, Ameling and Lacey}]{Arnold2010}
\bibinfo{author}{Arnold, M.}, \bibinfo{author}{Ghosh, A.},
  \bibinfo{author}{Ameling, S.}, \bibinfo{author}{Lacey, G.},
  \bibinfo{year}{2010}.
\newblock \bibinfo{title}{Automatic segmentation and inpainting of specular
  highlights for endoscopic imaging}.
\newblock \bibinfo{journal}{EURASIP Journal on Image and Video Processing}
  \bibinfo{volume}{2010}, \bibinfo{pages}{1--12}.
\bibitem[{Ashukha et~al.(2020)Ashukha, Lyzhov, Molchanov and
  Vetrov}]{Ashukha2020}
\bibinfo{author}{Ashukha, A.}, \bibinfo{author}{Lyzhov, A.},
  \bibinfo{author}{Molchanov, D.}, \bibinfo{author}{Vetrov, D.},
  \bibinfo{year}{2020}.
\newblock \bibinfo{title}{Pitfalls of in-domain uncertainty estimation and
  ensembling in deep learning}, in: \bibinfo{booktitle}{International
  Conference on Learning Representations}.
\bibitem[{Baguer et~al.(2020)Baguer, Leuschner and Schmidt}]{Baguer2020}
\bibinfo{author}{Baguer, D.O.}, \bibinfo{author}{Leuschner, J.},
  \bibinfo{author}{Schmidt, M.}, \bibinfo{year}{2020}.
\newblock \bibinfo{title}{Computed tomography reconstruction using deep image
  prior and learned reconstruction methods}.
\newblock \bibinfo{journal}{Inverse Problems} \bibinfo{volume}{36},
  \bibinfo{pages}{094004}.
\bibitem[{Bernardes et~al.(2010)Bernardes, Maduro, Serranho, Ara{\'u}jo,
  Barbeiro and Cunha-Vaz}]{Bernardes2010}
\bibinfo{author}{Bernardes, R.}, \bibinfo{author}{Maduro, C.},
  \bibinfo{author}{Serranho, P.}, \bibinfo{author}{Ara{\'u}jo, A.},
  \bibinfo{author}{Barbeiro, S.}, \bibinfo{author}{Cunha-Vaz, J.},
  \bibinfo{year}{2010}.
\newblock \bibinfo{title}{Improved adaptive complex diffusion despeckling
  filter}.
\newblock \bibinfo{journal}{Optics Express} \bibinfo{volume}{18},
  \bibinfo{pages}{24048--24059}.
\bibitem[{Bhadra et~al.(2020)Bhadra, Kelkar, Brooks and Anastasio}]{Bhadra2020}
\bibinfo{author}{Bhadra, S.}, \bibinfo{author}{Kelkar, V.A.},
  \bibinfo{author}{Brooks, F.J.}, \bibinfo{author}{Anastasio, M.A.},
  \bibinfo{year}{2020}.
\newblock \bibinfo{title}{On hallucinations in tomographic image
  reconstruction}, in: \bibinfo{booktitle}{arXiv Preprint}.
\newblock \bibinfo{note}{ArXiv:2012.00646}.
\bibitem[{Blei et~al.(2017)Blei, Kucukelbir and McAuliffe}]{Blei2017}
\bibinfo{author}{Blei, D.M.}, \bibinfo{author}{Kucukelbir, A.},
  \bibinfo{author}{McAuliffe, J.D.}, \bibinfo{year}{2017}.
\newblock \bibinfo{title}{Variational inference: A review for statisticians}.
\newblock \bibinfo{journal}{Journal of the American Statistical Association}
  \bibinfo{volume}{112}, \bibinfo{pages}{859--877}.
\newblock \DOIprefix\doi{10.1080/01621459.2017.1285773}.
\bibitem[{Blundell et~al.(2015)Blundell, Cornebise, Kavukcuoglu and
  Wierstra}]{Blundell2015}
\bibinfo{author}{Blundell, C.}, \bibinfo{author}{Cornebise, J.},
  \bibinfo{author}{Kavukcuoglu, K.}, \bibinfo{author}{Wierstra, D.},
  \bibinfo{year}{2015}.
\newblock \bibinfo{title}{Weight uncertainty in neural network}, in:
  \bibinfo{booktitle}{ICML}, pp. \bibinfo{pages}{1613--1622}.
\bibitem[{Brosse et~al.(2018)Brosse, Moulines and Durmus}]{Brosse2018}
\bibinfo{author}{Brosse, N.}, \bibinfo{author}{Moulines, E.},
  \bibinfo{author}{Durmus, A.}, \bibinfo{year}{2018}.
\newblock \bibinfo{title}{The promises and pitfalls of stochastic gradient
  langevin dynamics}, in: \bibinfo{booktitle}{Bayesian Deep Learning Workshop
  (NeurIPS)}.
\newblock \bibinfo{note}{ArXiv:1811.10072}.
\bibitem[{Carrillo et~al.(2021)Carrillo, Millardet, Carlier and
  Mateus}]{Carrillo2021}
\bibinfo{author}{Carrillo, H.}, \bibinfo{author}{Millardet, M.},
  \bibinfo{author}{Carlier, T.}, \bibinfo{author}{Mateus, D.},
  \bibinfo{year}{2021}.
\newblock \bibinfo{title}{{Low-count PET image reconstruction with Bayesian
  inference over a Deep Prior}}, in: \bibinfo{booktitle}{SPIE Medical Imaging
  2021}, pp. \bibinfo{pages}{227--235}.
\newblock \DOIprefix\doi{10.1117/12.2580169}.
\bibitem[{Chakrabarty and Maji(2019)}]{Chakrabarty2019}
\bibinfo{author}{Chakrabarty, P.}, \bibinfo{author}{Maji, S.},
  \bibinfo{year}{2019}.
\newblock \bibinfo{title}{The spectral bias of the deep image prior}, in:
  \bibinfo{booktitle}{4th workshop on Bayesian Deep Learning (NeurIPS 2019)}.
\bibitem[{Chang et~al.(2000)Chang, Yu and Vetterli}]{Chang2000}
\bibinfo{author}{Chang, S.G.}, \bibinfo{author}{Yu, B.},
  \bibinfo{author}{Vetterli, M.}, \bibinfo{year}{2000}.
\newblock \bibinfo{title}{{Adaptive wavelet thresholding for image denoising
  and compression}}.
\newblock \bibinfo{journal}{{IEEE Transactions on Image Processing}}
  \bibinfo{volume}{9}, \bibinfo{pages}{1532--1546}.
\newblock \DOIprefix\doi{10.1109/83.862633}.
\bibitem[{Cheng et~al.(2019)Cheng, Gadelha, Maji and Sheldon}]{Cheng2019}
\bibinfo{author}{Cheng, Z.}, \bibinfo{author}{Gadelha, M.},
  \bibinfo{author}{Maji, S.}, \bibinfo{author}{Sheldon, D.},
  \bibinfo{year}{2019}.
\newblock \bibinfo{title}{A bayesian perspective on the deep image prior}, in:
  \bibinfo{booktitle}{IEEE/CVF Conference on Computer Vision and Pattern
  Recognition}, pp. \bibinfo{pages}{5443--5451}.
\bibitem[{Edupuganti et~al.(2021)Edupuganti, Mardani, Vasanawala and
  Pauly}]{Edupuganti2021}
\bibinfo{author}{Edupuganti, V.}, \bibinfo{author}{Mardani, M.},
  \bibinfo{author}{Vasanawala, S.}, \bibinfo{author}{Pauly, J.},
  \bibinfo{year}{2021}.
\newblock \bibinfo{title}{Uncertainty quantification in deep mri
  reconstruction}.
\newblock \bibinfo{journal}{IEEE Transactions on Medical Imaging}
  \bibinfo{volume}{40}, \bibinfo{pages}{239--250}.
\newblock \DOIprefix\doi{10.1109/TMI.2020.3025065}.
\bibitem[{Frazier(2018)}]{Frazier2018}
\bibinfo{author}{Frazier, P.I.}, \bibinfo{year}{2018}.
\newblock \bibinfo{title}{A tutorial on bayesian optimization}, in:
  \bibinfo{booktitle}{arXiv Preprint}.
\newblock \bibinfo{note}{ArXiv:1807.02811}.
\bibitem[{Gal and Ghahramani(2015)}]{Gal2015Bernoulli}
\bibinfo{author}{Gal, Y.}, \bibinfo{author}{Ghahramani, Z.},
  \bibinfo{year}{2015}.
\newblock \bibinfo{title}{Bayesian convolutional neural networks with bernoulli
  approximate variational inference}, in: \bibinfo{booktitle}{arXiv Preprint}.
\newblock \bibinfo{note}{ArXiv:1506.02158}.
\bibitem[{Gal and Ghahramani(2016)}]{Gal2016}
\bibinfo{author}{Gal, Y.}, \bibinfo{author}{Ghahramani, Z.},
  \bibinfo{year}{2016}.
\newblock \bibinfo{title}{Dropout as a bayesian approximation: Representing
  model uncertainty in deep learning}, in: \bibinfo{booktitle}{ICML}, pp.
  \bibinfo{pages}{1050--1059}.
\bibitem[{Gardner et~al.(2018)Gardner, Pleiss, Weinberger, Bindel and
  Wilson}]{Gardner2018}
\bibinfo{author}{Gardner, J.}, \bibinfo{author}{Pleiss, G.},
  \bibinfo{author}{Weinberger, K.Q.}, \bibinfo{author}{Bindel, D.},
  \bibinfo{author}{Wilson, A.G.}, \bibinfo{year}{2018}.
\newblock \bibinfo{title}{{GPyTorch}: Blackbox matrix-matrix gaussian process
  inference with {GPU} acceleration}, in: \bibinfo{booktitle}{Advances in
  Neural Information Processing Systems}.
\bibitem[{Graves(2011)}]{Graves2011}
\bibinfo{author}{Graves, A.}, \bibinfo{year}{2011}.
\newblock \bibinfo{title}{Practical variational inference for neural networks},
  in: \bibinfo{booktitle}{NeurIPS}, pp. \bibinfo{pages}{2348--2356}.
\bibitem[{Green\-span(2009)}]{Greenspan2009}
\bibinfo{author}{Green\-span, H.}, \bibinfo{year}{2009}.
\newblock \bibinfo{title}{Super-resolution in medical imaging}.
\newblock \bibinfo{journal}{The Computer Journal} \bibinfo{volume}{52},
  \bibinfo{pages}{43--63}.
\bibitem[{Hammernik et~al.(2018)Hammernik, Klatzer, Kobler, Recht, Sodickson,
  Pock and Knoll}]{Hammernik2018}
\bibinfo{author}{Hammernik, K.}, \bibinfo{author}{Klatzer, T.},
  \bibinfo{author}{Kobler, E.}, \bibinfo{author}{Recht, M.P.},
  \bibinfo{author}{Sodickson, D.K.}, \bibinfo{author}{Pock, T.},
  \bibinfo{author}{Knoll, F.}, \bibinfo{year}{2018}.
\newblock \bibinfo{title}{Learning a variational network for reconstruction of
  accelerated mri data}.
\newblock \bibinfo{journal}{Magnetic Resonance in Medicine}
  \bibinfo{volume}{79}, \bibinfo{pages}{3055--3071}.
\bibitem[{Heckel and Soltanolkotabi(2020)}]{Heckel2020}
\bibinfo{author}{Heckel, R.}, \bibinfo{author}{Soltanolkotabi, M.},
  \bibinfo{year}{2020}.
\newblock \bibinfo{title}{Denoising and regularization via exploiting the
  structural bias of convolutional generators}, in: \bibinfo{booktitle}{ICLR}.
\bibitem[{Huang et~al.(2021)Huang, Xian, Yang, Qu, Yi, Wu and
  Metaxas}]{Huang2021}
\bibinfo{author}{Huang, Q.}, \bibinfo{author}{Xian, Y.}, \bibinfo{author}{Yang,
  D.}, \bibinfo{author}{Qu, H.}, \bibinfo{author}{Yi, J.}, \bibinfo{author}{Wu,
  P.}, \bibinfo{author}{Metaxas, D.N.}, \bibinfo{year}{2021}.
\newblock \bibinfo{title}{Dynamic mri reconstruction with end-to-end
  motion-guided network}.
\newblock \bibinfo{journal}{Medical Image Analysis} \bibinfo{volume}{68},
  \bibinfo{pages}{101901}.
\bibitem[{Huang et~al.(2019)Huang, Yang, Qu, Yi, Wu and Metaxas}]{Huang2019}
\bibinfo{author}{Huang, Q.}, \bibinfo{author}{Yang, D.}, \bibinfo{author}{Qu,
  H.}, \bibinfo{author}{Yi, J.}, \bibinfo{author}{Wu, P.},
  \bibinfo{author}{Metaxas, D.}, \bibinfo{year}{2019}.
\newblock \bibinfo{title}{Dynamic mri reconstruction with motion-guided
  network}, in: \bibinfo{booktitle}{Proceedings of The 2nd International
  Conference on Medical Imaging with Deep Learning}, pp.
  \bibinfo{pages}{275--284}.
\bibitem[{Jain and Seung(2009)}]{Jain2009}
\bibinfo{author}{Jain, V.}, \bibinfo{author}{Seung, S.}, \bibinfo{year}{2009}.
\newblock \bibinfo{title}{Natural image denoising with convolutional networks},
  in: \bibinfo{booktitle}{Advances in Neural Information Processing Systems},
  pp. \bibinfo{pages}{769--776}.
\bibitem[{Jones et~al.(1998)Jones, Schonlau and Welch}]{Jones1998}
\bibinfo{author}{Jones, D.R.}, \bibinfo{author}{Schonlau, M.},
  \bibinfo{author}{Welch, W.J.}, \bibinfo{year}{1998}.
\newblock \bibinfo{title}{Efficient global optimization of expensive black-box
  functions}.
\newblock \bibinfo{journal}{Journal of Global Optimization}
  \bibinfo{volume}{13}, \bibinfo{pages}{455--492}.
\bibitem[{Kendall and Gal(2017)}]{Kendall2017}
\bibinfo{author}{Kendall, A.}, \bibinfo{author}{Gal, Y.}, \bibinfo{year}{2017}.
\newblock \bibinfo{title}{What uncertainties do we need in bayesian deep
  learning for computer vision?}, in: \bibinfo{booktitle}{NeurIPS}, pp.
  \bibinfo{pages}{5574--5584}.
\bibitem[{Kermany et~al.(2018)Kermany, Goldbaum, Cai, Valentim, Liang, Baxter,
  McKeown, Yang, Wu, Yan, Dong, Prasadha, Pei, Ting, Zhu, Li, Hewett, Dong,
  Ziyar, Shi, Zhang, Zheng, Hou, Shi, Fu, Duan, Huu, Wen, Zhang, Zhang, Li,
  Wang, Singer, Sun, Xu, Tafreshi, Lewis, Xia and Zhang}]{Kermany2018}
\bibinfo{author}{Kermany, D.S.}, \bibinfo{author}{Goldbaum, M.},
  \bibinfo{author}{Cai, W.}, \bibinfo{author}{Valentim, C.C.},
  \bibinfo{author}{Liang, H.}, \bibinfo{author}{Baxter, S.L.},
  \bibinfo{author}{McKeown, A.}, \bibinfo{author}{Yang, G.},
  \bibinfo{author}{Wu, X.}, \bibinfo{author}{Yan, F.}, \bibinfo{author}{Dong,
  J.}, \bibinfo{author}{Prasadha, M.K.}, \bibinfo{author}{Pei, J.},
  \bibinfo{author}{Ting, M.Y.}, \bibinfo{author}{Zhu, J.}, \bibinfo{author}{Li,
  C.}, \bibinfo{author}{Hewett, S.}, \bibinfo{author}{Dong, J.},
  \bibinfo{author}{Ziyar, I.}, \bibinfo{author}{Shi, A.},
  \bibinfo{author}{Zhang, R.}, \bibinfo{author}{Zheng, L.},
  \bibinfo{author}{Hou, R.}, \bibinfo{author}{Shi, W.}, \bibinfo{author}{Fu,
  X.}, \bibinfo{author}{Duan, Y.}, \bibinfo{author}{Huu, V.A.},
  \bibinfo{author}{Wen, C.}, \bibinfo{author}{Zhang, E.D.},
  \bibinfo{author}{Zhang, C.L.}, \bibinfo{author}{Li, O.},
  \bibinfo{author}{Wang, X.}, \bibinfo{author}{Singer, M.A.},
  \bibinfo{author}{Sun, X.}, \bibinfo{author}{Xu, J.},
  \bibinfo{author}{Tafreshi, A.}, \bibinfo{author}{Lewis, M.A.},
  \bibinfo{author}{Xia, H.}, \bibinfo{author}{Zhang, K.}, \bibinfo{year}{2018}.
\newblock \bibinfo{title}{Identifying medical diagnoses and treatable diseases
  by image-based deep learning}.
\newblock \bibinfo{journal}{Cell} \bibinfo{volume}{172},
  \bibinfo{pages}{1122--1131}.
\newblock \DOIprefix\doi{10.1016/j.cell.2018.02.010}.
\bibitem[{Kudo et~al.(2013)Kudo, Suzuki and Rashed}]{Kudo2013}
\bibinfo{author}{Kudo, H.}, \bibinfo{author}{Suzuki, T.},
  \bibinfo{author}{Rashed, E.A.}, \bibinfo{year}{2013}.
\newblock \bibinfo{title}{Image reconstruction for sparse-view ct and interior
  ct—introduction to compressed sensing and differentiated backprojection}.
\newblock \bibinfo{journal}{Quantitative imaging in medicine and surgery}
  \bibinfo{volume}{3}, \bibinfo{pages}{147}.
\bibitem[{Laves et~al.(2020a)Laves, Ihler, Fast, Kahrs and
  Ortmaier}]{Laves2020}
\bibinfo{author}{Laves, M.H.}, \bibinfo{author}{Ihler, S.},
  \bibinfo{author}{Fast, J.F.}, \bibinfo{author}{Kahrs, L.A.},
  \bibinfo{author}{Ortmaier, T.}, \bibinfo{year}{2020}a.
\newblock \bibinfo{title}{Well-calibrated regression uncertainty in medical
  imaging with deep learning}, in: \bibinfo{booktitle}{Medical Imaging with
  Deep Learning}.
\bibitem[{Laves et~al.(2021)Laves, Ihler, Fast, Kahrs and Ortmaier}]{Laves2021}
\bibinfo{author}{Laves, M.H.}, \bibinfo{author}{Ihler, S.},
  \bibinfo{author}{Fast, J.F.}, \bibinfo{author}{Kahrs, L.A.},
  \bibinfo{author}{Ortmaier, T.}, \bibinfo{year}{2021}.
\newblock \bibinfo{title}{Recalibration of aleatoric and epistemic regression
  uncertainty in medical imaging}.
\newblock \bibinfo{journal}{Journal of Machine Learning for Biomedical Imaging}
  , \bibinfo{pages}{1--26}.
\bibitem[{Laves et~al.(2019)Laves, Ihler and Ortmaier}]{Laves2019deformable}
\bibinfo{author}{Laves, M.H.}, \bibinfo{author}{Ihler, S.},
  \bibinfo{author}{Ortmaier, T.}, \bibinfo{year}{2019}.
\newblock \bibinfo{title}{Deformable medical image registration using a
  randomly-initialized {CNN} as regularization prior}, in:
  \bibinfo{booktitle}{Medical Imaging with Deep Learning--Extended Abstract
  Track}.
\newblock
  \bibinfo{note}{\href{https://arxiv.org/abs/1908.00788}{arXiv:1908.00788}}.
\bibitem[{Laves et~al.(2020b)Laves, T{\"o}lle and Ortmaier}]{Laves2020MCDIP}
\bibinfo{author}{Laves, M.H.}, \bibinfo{author}{T{\"o}lle, M.},
  \bibinfo{author}{Ortmaier, T.}, \bibinfo{year}{2020}b.
\newblock \bibinfo{title}{Uncertainty estimation in medical image denoising
  with bayesian deep image prior}, in: \bibinfo{booktitle}{Uncertainty for Safe
  Utilization of Machine Learning in Medical Imaging, and Graphs in Biomedical
  Image Analysis}, pp. \bibinfo{pages}{81--96}.
\bibitem[{Lee et~al.(2018a)Lee, Lee, Kim, Cho and Cho}]{Lee2018deep}
\bibinfo{author}{Lee, H.}, \bibinfo{author}{Lee, J.}, \bibinfo{author}{Kim,
  H.}, \bibinfo{author}{Cho, B.}, \bibinfo{author}{Cho, S.},
  \bibinfo{year}{2018}a.
\newblock \bibinfo{title}{Deep-neural-network-based sinogram synthesis for
  sparse-view ct image reconstruction}.
\newblock \bibinfo{journal}{IEEE Transactions on Radiation and Plasma Medical
  Sciences} \bibinfo{volume}{3}, \bibinfo{pages}{109--119}.
\bibitem[{Lee et~al.(2018b)Lee, Lee and Kang}]{Lee2018}
\bibinfo{author}{Lee, S.}, \bibinfo{author}{Lee, M.S.}, \bibinfo{author}{Kang,
  M.G.}, \bibinfo{year}{2018}b.
\newblock \bibinfo{title}{Poisson--gaussian noise analysis and estimation for
  low-dose x-ray images in the nsct domain}.
\newblock \bibinfo{journal}{Sensors} \bibinfo{volume}{18},
  \bibinfo{pages}{1019}.
\bibitem[{Lempitsky et~al.(2018)Lempitsky, Vedaldi and Ulyanov}]{Lempitsky2018}
\bibinfo{author}{Lempitsky, V.}, \bibinfo{author}{Vedaldi, A.},
  \bibinfo{author}{Ulyanov, D.}, \bibinfo{year}{2018}.
\newblock \bibinfo{title}{{Deep Image Prior}}, in: \bibinfo{booktitle}{IEEE/CVF
  Conference on Computer Vision and Pattern Recognition}, pp.
  \bibinfo{pages}{9446--9454}.
\newblock \DOIprefix\doi{10.1109/CVPR.2018.00984}.
\bibitem[{Lindauer and Hutter(2020)}]{Lindauer2020}
\bibinfo{author}{Lindauer, M.}, \bibinfo{author}{Hutter, F.},
  \bibinfo{year}{2020}.
\newblock \bibinfo{title}{Best practices for scientific research on neural
  architecture search}.
\newblock \bibinfo{journal}{Journal of Machine Learning Research}
  \bibinfo{volume}{21}, \bibinfo{pages}{1--18}.
\bibitem[{Loshchilov and Hutter(2019)}]{Loshchilov2019}
\bibinfo{author}{Loshchilov, I.}, \bibinfo{author}{Hutter, F.},
  \bibinfo{year}{2019}.
\newblock \bibinfo{title}{Decoupled weight decay regularization}, in:
  \bibinfo{booktitle}{ICLR}.
\bibitem[{L{\"u}sebrink et~al.(2017)L{\"u}sebrink, Mattern, Yakupov and
  Speck}]{T1MRI2018}
\bibinfo{author}{L{\"u}sebrink, Falkand~Sciarra, A.}, \bibinfo{author}{Mattern,
  H.}, \bibinfo{author}{Yakupov, R.}, \bibinfo{author}{Speck, O.},
  \bibinfo{year}{2017}.
\newblock \bibinfo{title}{Data from: T1-weighted in vivo human whole brain mri
  dataset with an ultrahigh isotropic resolution of 250 \textmu m}.
\newblock \DOIprefix\doi{10.5061/dryad.38s74}.
\bibitem[{Ma et~al.(2020)Ma, Wei, Feng, He, Guo and Wang}]{Ma2020}
\bibinfo{author}{Ma, Y.}, \bibinfo{author}{Wei, B.}, \bibinfo{author}{Feng,
  P.}, \bibinfo{author}{He, P.}, \bibinfo{author}{Guo, X.},
  \bibinfo{author}{Wang, G.}, \bibinfo{year}{2020}.
\newblock \bibinfo{title}{Low-dose ct image denoising using a generative
  adversarial network with a hybrid loss function for noise learning}.
\newblock \bibinfo{journal}{IEEE Access} \bibinfo{volume}{8},
  \bibinfo{pages}{67519--67529}.
\bibitem[{Michailovich and Tannenbaum(2006)}]{Michailovich2006}
\bibinfo{author}{Michailovich, O.V.}, \bibinfo{author}{Tannenbaum, A.},
  \bibinfo{year}{2006}.
\newblock \bibinfo{title}{Despeckling of medical ultrasound images}.
\newblock \bibinfo{journal}{IEEE Transactions on Ultrasonics, Ferroelectrics,
  and Frequency Control} \bibinfo{volume}{53}, \bibinfo{pages}{64--78}.
\newblock \DOIprefix\doi{10.1109/TUFFC.2006.1588392}.
\bibitem[{Narnhofer et~al.(2021)Narnhofer, Effland, Kobler, Hammernik, Knoll
  and Pock}]{Narnhofer2021}
\bibinfo{author}{Narnhofer, D.}, \bibinfo{author}{Effland, A.},
  \bibinfo{author}{Kobler, E.}, \bibinfo{author}{Hammernik, K.},
  \bibinfo{author}{Knoll, F.}, \bibinfo{author}{Pock, T.},
  \bibinfo{year}{2021}.
\newblock \bibinfo{title}{Bayesian uncertainty estimation of learned
  variational mri reconstruction}.
\newblock \bibinfo{journal}{IEEE Transactions on Medical Imaging}
  \bibinfo{volume}{XX}, \bibinfo{pages}{1--13}.
\newblock \DOIprefix\doi{10.1109/TMI.2021.3112040}.
\bibitem[{Peng et~al.(2020)Peng, Li, Li, Wang, Zhao, Qiu and Chen}]{Peng2020}
\bibinfo{author}{Peng, C.}, \bibinfo{author}{Li, B.}, \bibinfo{author}{Li, M.},
  \bibinfo{author}{Wang, H.}, \bibinfo{author}{Zhao, Z.}, \bibinfo{author}{Qiu,
  B.}, \bibinfo{author}{Chen, D.Z.}, \bibinfo{year}{2020}.
\newblock \bibinfo{title}{An irregular metal trace inpainting network for x-ray
  ct metal artifact reduction}.
\newblock \bibinfo{journal}{Medical Physics} \bibinfo{volume}{47},
  \bibinfo{pages}{4087--4100}.
\bibitem[{Rahaman et~al.(2019)Rahaman, Baratin, Arpit, Draxler, Lin, Hamprecht,
  Bengio and Courville}]{Rahaman2019}
\bibinfo{author}{Rahaman, N.}, \bibinfo{author}{Baratin, A.},
  \bibinfo{author}{Arpit, D.}, \bibinfo{author}{Draxler, F.},
  \bibinfo{author}{Lin, M.}, \bibinfo{author}{Hamprecht, F.},
  \bibinfo{author}{Bengio, Y.}, \bibinfo{author}{Courville, A.},
  \bibinfo{year}{2019}.
\newblock \bibinfo{title}{On the spectral bias of neural networks}, in:
  \bibinfo{booktitle}{ICML}, pp. \bibinfo{pages}{5301--5310}.
\bibitem[{Rudin et~al.(1992)Rudin, Osher and Fatemi}]{rudin1992}
\bibinfo{author}{Rudin, L.I.}, \bibinfo{author}{Osher, S.},
  \bibinfo{author}{Fatemi, E.}, \bibinfo{year}{1992}.
\newblock \bibinfo{title}{Nonlinear total variation based noise removal
  algorithms}.
\newblock \bibinfo{journal}{Physica D: Nonlinear Phenomena}
  \bibinfo{volume}{60}, \bibinfo{pages}{259--268}.
\bibitem[{Snoek et~al.(2015)Snoek, Rippel, Swersky, Kiros, Satish, Sundaram,
  Patwary, Prabhat and Adams}]{Snoek2015}
\bibinfo{author}{Snoek, J.}, \bibinfo{author}{Rippel, O.},
  \bibinfo{author}{Swersky, K.}, \bibinfo{author}{Kiros, R.},
  \bibinfo{author}{Satish, N.}, \bibinfo{author}{Sundaram, N.},
  \bibinfo{author}{Patwary, M.}, \bibinfo{author}{Prabhat, M.},
  \bibinfo{author}{Adams, R.}, \bibinfo{year}{2015}.
\newblock \bibinfo{title}{Scalable bayesian optimization using deep neural
  networks}, in: \bibinfo{booktitle}{International Conference on Machine
  Learning}, pp. \bibinfo{pages}{2171--2180}.
\bibitem[{Sotiras et~al.(2013)Sotiras, Davatzikos and Paragios}]{Sotiras2013}
\bibinfo{author}{Sotiras, A.}, \bibinfo{author}{Davatzikos, C.},
  \bibinfo{author}{Paragios, N.}, \bibinfo{year}{2013}.
\newblock \bibinfo{title}{Deformable medical image registration: A survey}.
\newblock \bibinfo{journal}{IEEE Trans Med Imag} \bibinfo{volume}{32},
  \bibinfo{pages}{1153--1190}.
\newblock \DOIprefix\doi{10.1109/TMI.2013.2265603}.
\bibitem[{Tanno et~al.(2017)Tanno, Worrall, Ghosh, Kaden, Sotiropoulos,
  Criminisi and Alexander}]{Tanno2017}
\bibinfo{author}{Tanno, R.}, \bibinfo{author}{Worrall, D.E.},
  \bibinfo{author}{Ghosh, A.}, \bibinfo{author}{Kaden, E.},
  \bibinfo{author}{Sotiropoulos, S.N.}, \bibinfo{author}{Criminisi, A.},
  \bibinfo{author}{Alexander, D.C.}, \bibinfo{year}{2017}.
\newblock \bibinfo{title}{Bayesian image quality transfer with cnns: exploring
  uncertainty in dmri super-resolution}, in: \bibinfo{booktitle}{MICCAI}, pp.
  \bibinfo{pages}{611--619}.
\bibitem[{Tezcan et~al.(2018)Tezcan, Baumgartner, Luechinger, Pruessmann and
  Konukoglu}]{Tezcan2018}
\bibinfo{author}{Tezcan, K.C.}, \bibinfo{author}{Baumgartner, C.F.},
  \bibinfo{author}{Luechinger, R.}, \bibinfo{author}{Pruessmann, K.P.},
  \bibinfo{author}{Konukoglu, E.}, \bibinfo{year}{2018}.
\newblock \bibinfo{title}{Mr image reconstruction using deep density priors}.
\newblock \bibinfo{journal}{IEEE Transactions on Medical Imaging}
  \bibinfo{volume}{38}, \bibinfo{pages}{1633--1642}.
\bibitem[{T{\"o}lle et~al.(2021)T{\"o}lle, Laves and Schlaefer}]{Toelle2021}
\bibinfo{author}{T{\"o}lle, M.}, \bibinfo{author}{Laves, M.H.},
  \bibinfo{author}{Schlaefer, A.}, \bibinfo{year}{2021}.
\newblock \bibinfo{title}{A mean-field variational inference approach to deep
  image prior for inverse problems in medical imaging}, in:
  \bibinfo{booktitle}{Medical Imaging with Deep Learning}.
\bibitem[{Tschandl et~al.(2018)Tschandl, Rosendahl and Kittler}]{Tschandl2018}
\bibinfo{author}{Tschandl, P.}, \bibinfo{author}{Rosendahl, C.},
  \bibinfo{author}{Kittler, H.}, \bibinfo{year}{2018}.
\newblock \bibinfo{title}{The ham10000 dataset, a large collection of
  multi-source dermatoscopic images of common pigmented skin lesions}.
\newblock \bibinfo{journal}{Scientific Data} \bibinfo{volume}{5},
  \bibinfo{pages}{1--9}.
\bibitem[{Wang et~al.(2018)Wang, Yu, Wang, Zu, Lalush, Lin, Wu, Zhou, Shen and
  Zhou}]{Wang2018}
\bibinfo{author}{Wang, Y.}, \bibinfo{author}{Yu, B.}, \bibinfo{author}{Wang,
  L.}, \bibinfo{author}{Zu, C.}, \bibinfo{author}{Lalush, D.S.},
  \bibinfo{author}{Lin, W.}, \bibinfo{author}{Wu, X.}, \bibinfo{author}{Zhou,
  J.}, \bibinfo{author}{Shen, D.}, \bibinfo{author}{Zhou, L.},
  \bibinfo{year}{2018}.
\newblock \bibinfo{title}{3d conditional generative adversarial networks for
  high-quality pet image estimation at low dose}.
\newblock \bibinfo{journal}{NeuroImage} \bibinfo{volume}{174},
  \bibinfo{pages}{550--562}.
\bibitem[{Welling and Teh(2011)}]{Welling2011}
\bibinfo{author}{Welling, M.}, \bibinfo{author}{Teh, Y.W.},
  \bibinfo{year}{2011}.
\newblock \bibinfo{title}{Bayesian learning via stochastic gradient langevin
  dynamics}, in: \bibinfo{booktitle}{ICML}, pp. \bibinfo{pages}{681--688}.
\bibitem[{Wenzel et~al.(2020)Wenzel, Roth, Veeling, Swiatkowski, Tran, Mandt,
  Snoek, Salimans, Jenatton and Nowozin}]{Wenzel2020}
\bibinfo{author}{Wenzel, F.}, \bibinfo{author}{Roth, K.},
  \bibinfo{author}{Veeling, B.}, \bibinfo{author}{Swiatkowski, J.},
  \bibinfo{author}{Tran, L.}, \bibinfo{author}{Mandt, S.},
  \bibinfo{author}{Snoek, J.}, \bibinfo{author}{Salimans, T.},
  \bibinfo{author}{Jenatton, R.}, \bibinfo{author}{Nowozin, S.},
  \bibinfo{year}{2020}.
\newblock \bibinfo{title}{How good is the {B}ayes posterior in deep neural
  networks really?}, in: \bibinfo{booktitle}{International Conference on
  Machine Learning}, pp. \bibinfo{pages}{10248--10259}.
\bibitem[{Wilson and Izmailov(2020)}]{Wilson2020}
\bibinfo{author}{Wilson, A.G.}, \bibinfo{author}{Izmailov, P.},
  \bibinfo{year}{2020}.
\newblock \bibinfo{title}{Bayesian deep learning and a probabilistic
  perspective of generalization}.
\newblock \bibinfo{journal}{arXiv preprint arXiv:2002.08791} .
\bibitem[{Wol\-terink et~al.(2017)Wol\-terink, Leiner, Viergever and
  I{\v{s}}gum}]{Wolterink2017}
\bibinfo{author}{Wol\-terink, J.M.}, \bibinfo{author}{Leiner, T.},
  \bibinfo{author}{Viergever, M.A.}, \bibinfo{author}{I{\v{s}}gum, I.},
  \bibinfo{year}{2017}.
\newblock \bibinfo{title}{Generative adversarial networks for noise reduction
  in low-dose ct}.
\newblock \bibinfo{journal}{IEEE Transactions on Medical Imaging}
  \bibinfo{volume}{36}, \bibinfo{pages}{2536--2545}.
\bibitem[{Yang et~al.(2018)Yang, Yan, Zhang, Yu, Shi, Mou, Kalra, Zhang, Sun
  and Wang}]{Yang2018}
\bibinfo{author}{Yang, Q.}, \bibinfo{author}{Yan, P.}, \bibinfo{author}{Zhang,
  Y.}, \bibinfo{author}{Yu, H.}, \bibinfo{author}{Shi, Y.},
  \bibinfo{author}{Mou, X.}, \bibinfo{author}{Kalra, M.K.},
  \bibinfo{author}{Zhang, Y.}, \bibinfo{author}{Sun, L.},
  \bibinfo{author}{Wang, G.}, \bibinfo{year}{2018}.
\newblock \bibinfo{title}{Low-dose ct image denoising using a generative
  adversarial network with wasserstein distance and perceptual loss}.
\newblock \bibinfo{journal}{IEEE Trans Med Imag} \bibinfo{volume}{37},
  \bibinfo{pages}{1348--1357}.
\bibitem[{Yi and Babyn(2018)}]{Yi2018}
\bibinfo{author}{Yi, X.}, \bibinfo{author}{Babyn, P.}, \bibinfo{year}{2018}.
\newblock \bibinfo{title}{Sharpness-aware low-dose ct denoising using
  conditional generative adversarial network}.
\newblock \bibinfo{journal}{Journal of Digital Imaging} \bibinfo{volume}{31},
  \bibinfo{pages}{655--669}.
\bibitem[{{\v{Z}}abi{\'c} et~al.(2013){\v{Z}}abi{\'c}, Wang, Morton and
  Brown}]{Zabic2013}
\bibinfo{author}{{\v{Z}}abi{\'c}, S.}, \bibinfo{author}{Wang, Q.},
  \bibinfo{author}{Morton, T.}, \bibinfo{author}{Brown, K.M.},
  \bibinfo{year}{2013}.
\newblock \bibinfo{title}{A low dose simulation tool for ct systems with energy
  integrating detectors}.
\newblock \bibinfo{journal}{Medical Physics} \bibinfo{volume}{40},
  \bibinfo{pages}{031102}.
\newblock \DOIprefix\doi{10.1118/1.4789628}.
\bibitem[{Zhu et~al.(2018)Zhu, Liu, Cauley, Rosen and Rosen}]{Zhu2018}
\bibinfo{author}{Zhu, B.}, \bibinfo{author}{Liu, J.Z.},
  \bibinfo{author}{Cauley, S.F.}, \bibinfo{author}{Rosen, B.R.},
  \bibinfo{author}{Rosen, M.S.}, \bibinfo{year}{2018}.
\newblock \bibinfo{title}{Image reconstruction by domain-transform manifold
  learning}.
\newblock \bibinfo{journal}{Nature} \bibinfo{volume}{555},
  \bibinfo{pages}{487--492}.

\end{thebibliography}

\section*{Supplementary Material}

\subsection*{KL Divergence Between Two Gaussians}
\label{app:two_gaussians}

If a Gaussian prior is chosen for convenience, the KL divergence is analytically tractable (cf.\ Eq.\,(\ref{eq:kl_mc})).
Let $ p(x) = \mathcal{N}(\mu_{p}, \sigma_{p}^{2}) $ and $ q(x) = \mathcal{N}(\mu_{q}, \sigma_{q}^{2}) $.
It is known that
\begin{align*}
    &\kl [ q(x) \Given p(x) ] = \int q(x) \log \frac{q(x)}{p(x)} \, \mathrm{d}x \\
    & = \int q(x) \log q(x) \, \mathrm{d}x - \int q(x) \log p(x) \, \mathrm{d}x \\
    & = - \frac{1}{2} \left( 1 + \log 2 \pi \sigma_{q}^{2} \right) + \frac{1}{2} \log 2 \pi \sigma_{p}^{2} + \frac{\sigma_{q}^{2} + ( \mu_{q} - \mu_{p})^{2}}{2\sigma_{p}^{2}} \\
    & = \log \frac{\sigma_p}{\sigma_q} + \frac{\sigma_{q}^{2} + ( \mu_q - \mu_p )^{2}}{2 \sigma_{p}^{2}} - \frac{1}{2} ~ .
\end{align*}

\subsection*{Detailed Experimental Settings}

To correspond to best practices in AutoML \cite{Lindauer2020} and to ensure reproducibility, we provide more details on our training procedure and experimental settings:
\begin{itemize}
    \setlength{\itemsep}{0pt}
    \item Code for training pipeline and evaluation is available at \href{https://github.com/Cardio-AI/mfvi-dip-mia}{github.com/Cardio-AI/mfvi-dip-mia}.
    \item The code includes all hyperparameters and random seeds.
    \item We used Python 3.8.8, PyTorch 1.8.1 and GPyTorch 1.4.1 on Ubuntu Linux 18.04.5.
    \item The identical network architecture as proposed by \citet{Lempitsky2018} in the original DIP paper was employed; a U-Net like autoencoder with skip-connections.
    \item We used the AdamW optimizer with a constant learn rate between $ 1\mathrm{e}{-3} $ and $ 3\mathrm{e}{-3} $, depending on the task.
    \item The following number of iterations were used: $ 1\mathrm{e}{5} $ (CT reconstruction), $ 5\mathrm{e}{4} $ (denoising and inpainting), and $ 3\mathrm{e}{4} $ (super-resolution). The BOs were initialized with the following candidates: MC dropout $ p \in \{ 0.02, 0.2 \} $ and $ \lambda \in \{ 0.1, 1\mathrm{e}{-6} \} $, POTOBIM $ T \in \{ 1\mathrm{e}{-4}, 1\mathrm{e}{-7} \} $ and $ \sigma \in \{ 0.1, 1\mathrm{e}{-6} \} $, SGLD $ \gamma \in \{ 0.9995, 0.999999 \} $ and $ \lambda \in \{ 1\mathrm{e}{-4}, 1\mathrm{e}{-8} \} $.
    \item All compared models were evaluated on the same computer using two Nvidia Titan RTX GPUs and an AMD Ryzen Threadripper 1950X CPU.
    \item A single BO step took approx.\ 75\,min and the BO was manually terminated after 11 steps, resulting in an overall runtime of approx.\ 13.75\,hours per BO. Note that per BO step, up to 4 temperature candidates are evaluated in parallel.
\end{itemize}


\begin{figure*}
    \centering
    \includegraphics[width=1\textwidth]{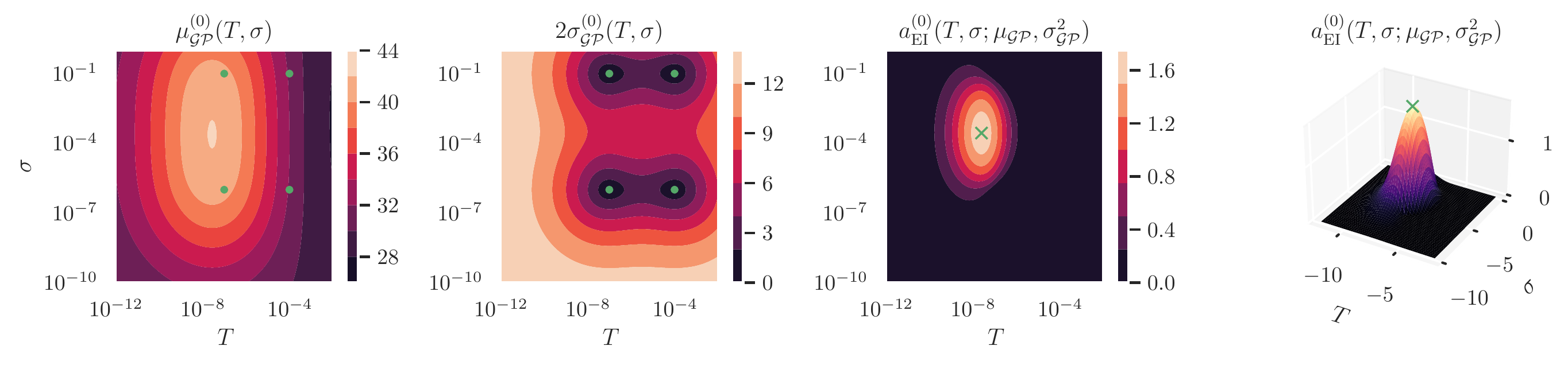}
    \includegraphics[width=1\textwidth]{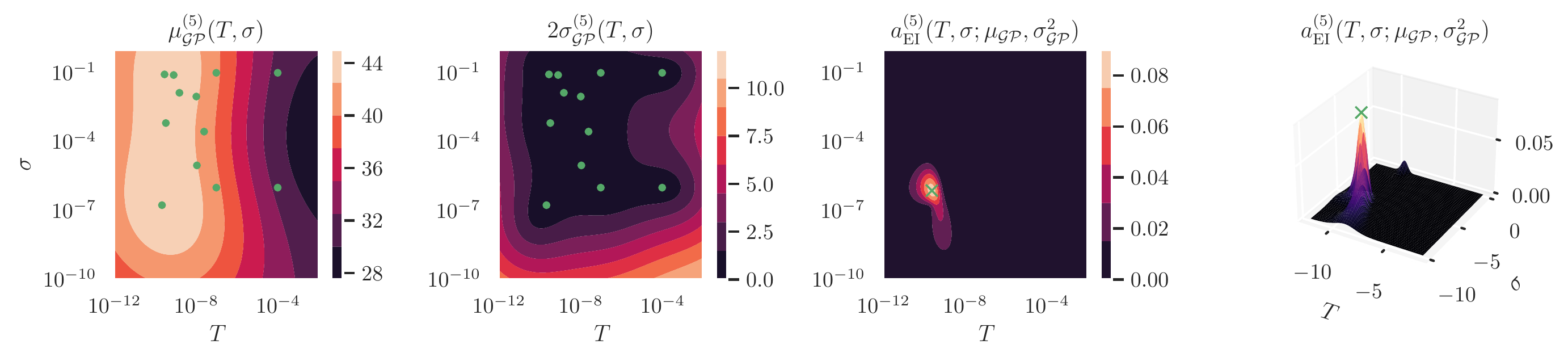}
    \includegraphics[width=1\textwidth]{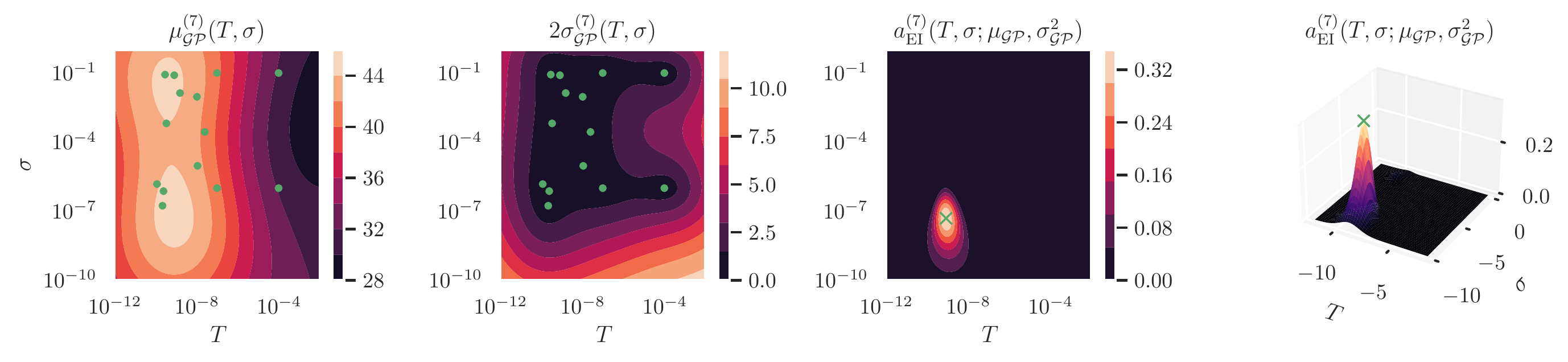}
    \includegraphics[width=1\textwidth]{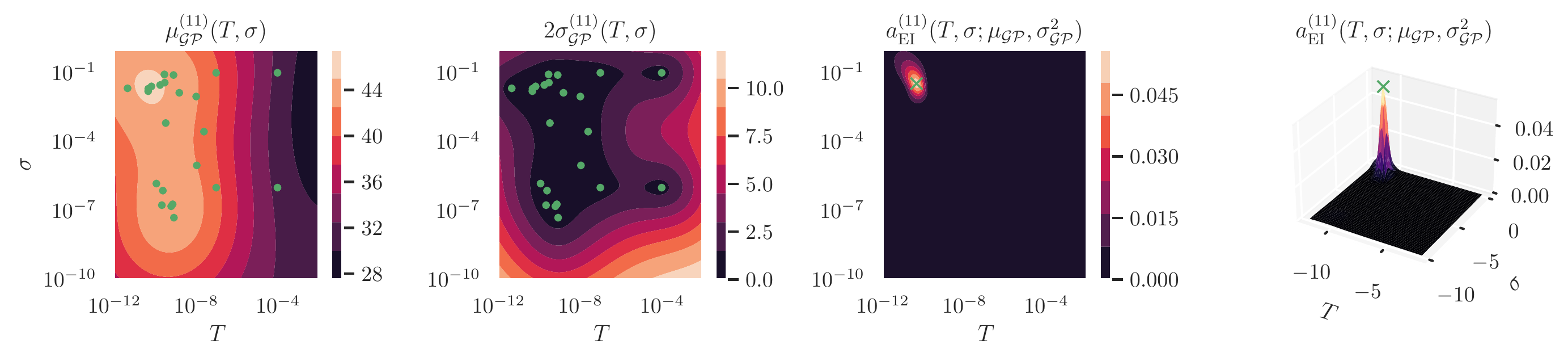}
    \caption{CT reconstruction with POTOBIM: GP mean, confidence (2 standard deviations) and expected improvement acquisition function after BO iteration $ i \in \{0, 5, 7, 11\} $. Green dots denote observed points and green crosses show candidates for the next BO iteration. Note that per BO step, up to 4 candidates are evaluated in parallel.}
    \label{fig:app_bo_mfvi_ct}
\end{figure*}

\begin{figure*}
    \centering
    \includegraphics[width=1\textwidth]{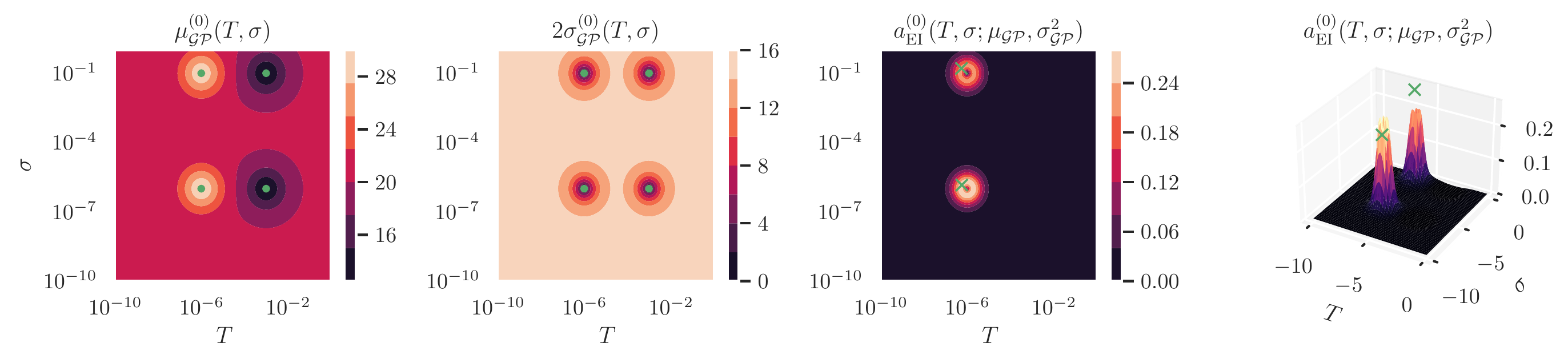}
    \includegraphics[width=1\textwidth]{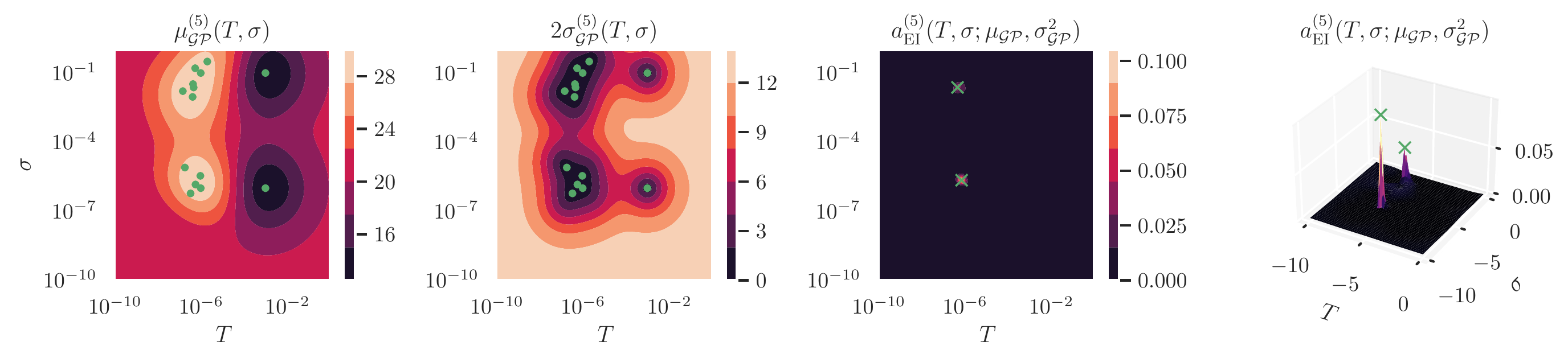}
    \includegraphics[width=1\textwidth]{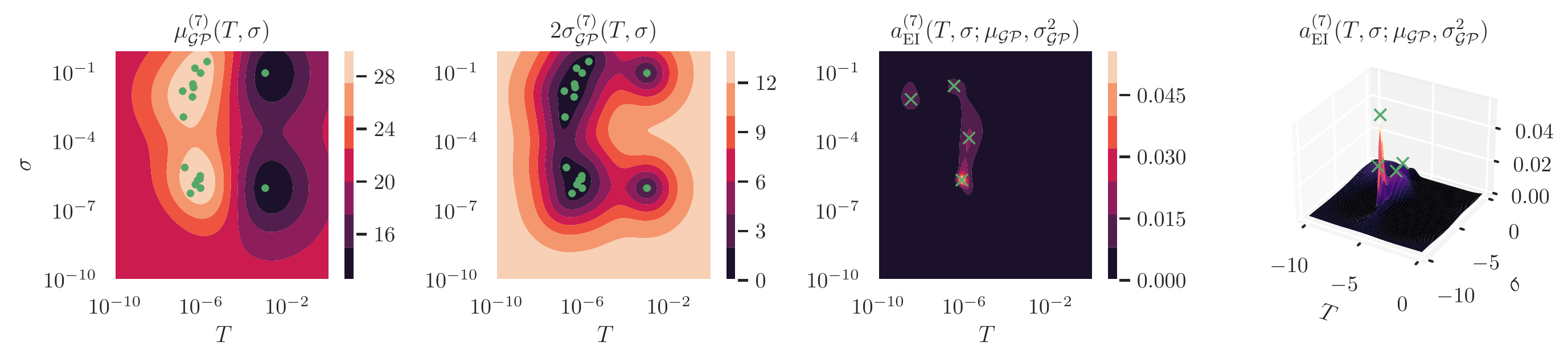}
    \includegraphics[width=1\textwidth]{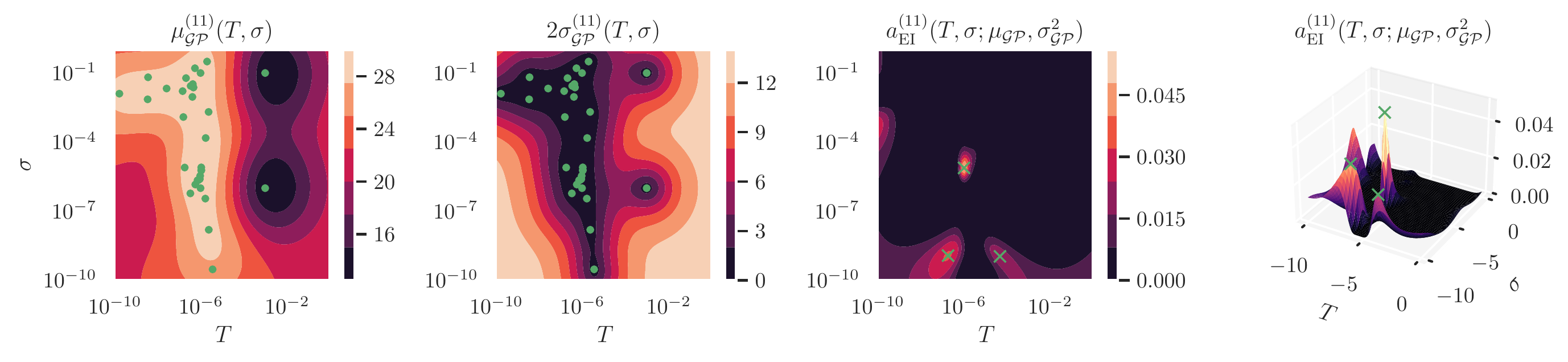}
    \caption{MRI super-resolution with POTOBIM: GP mean, confidence (2 standard deviations) and expected improvement acquisition function after BO iteration $ i \in \{0, 5, 7, 11\} $. Green dots denote observed points and green crosses show candidates for the next BO iteration. Note that per BO step, up to 4 candidates are evaluated in parallel.}
    \label{fig:app_bo_mfvi_sr}
\end{figure*}

\begin{figure*}
    \centering
    \includegraphics[width=1\textwidth]{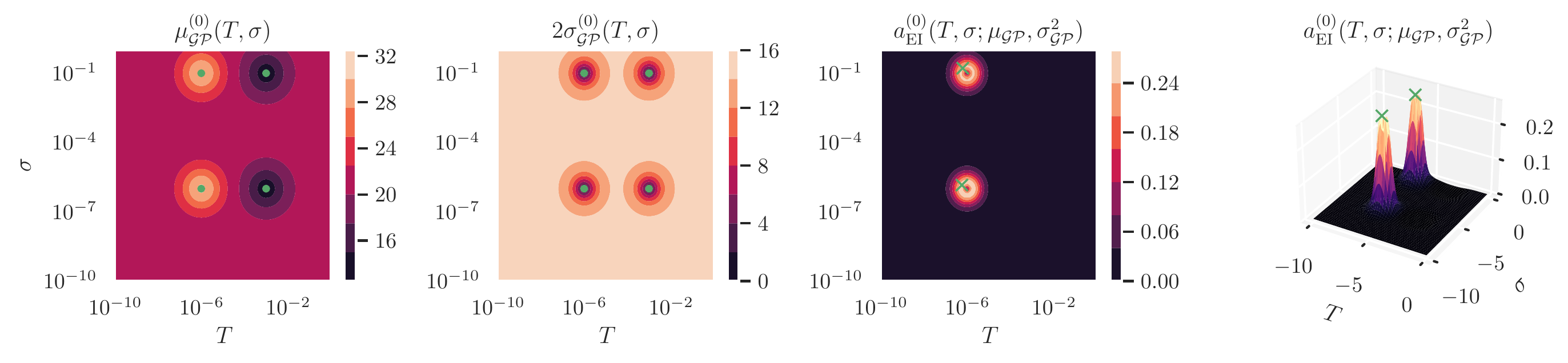}
    \includegraphics[width=1\textwidth]{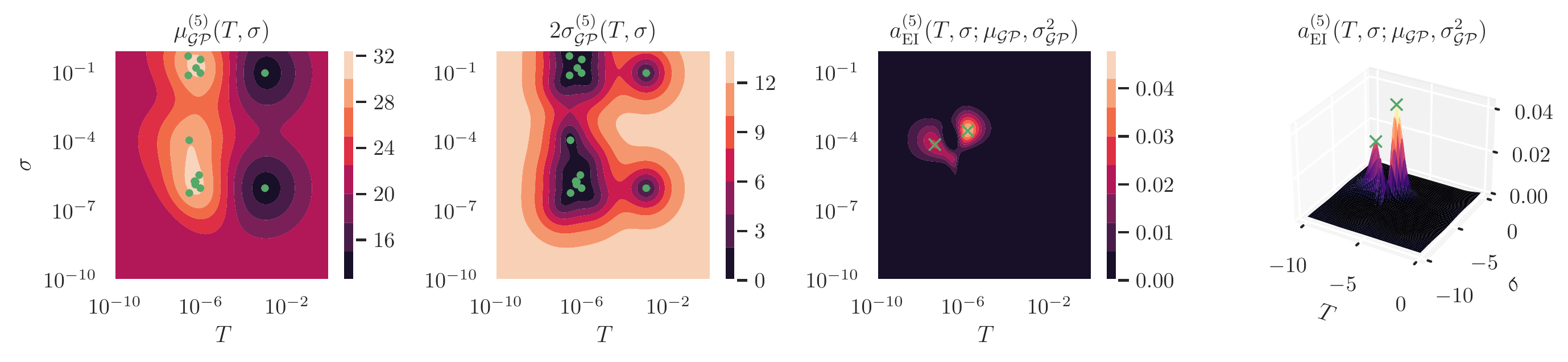}
    \includegraphics[width=1\textwidth]{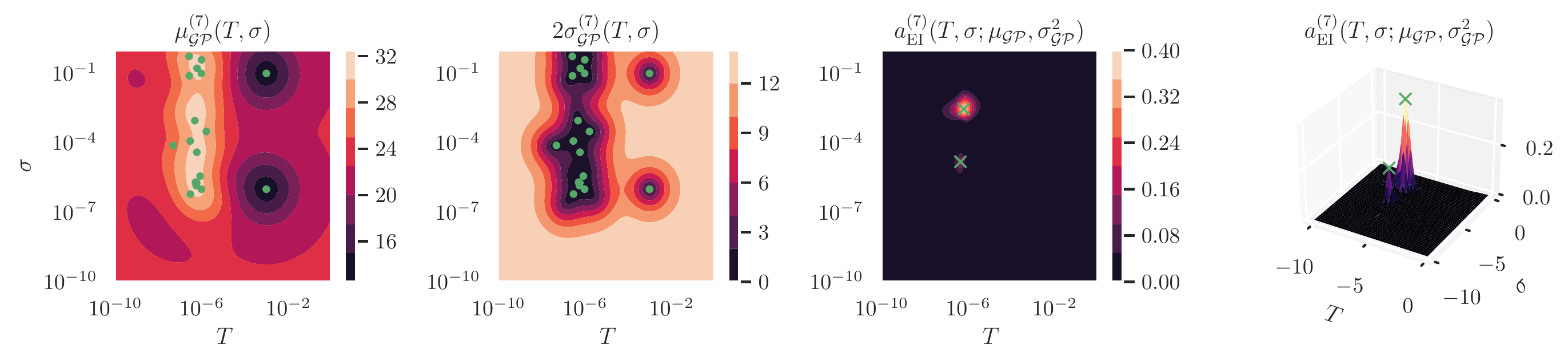}
    \includegraphics[width=1\textwidth]{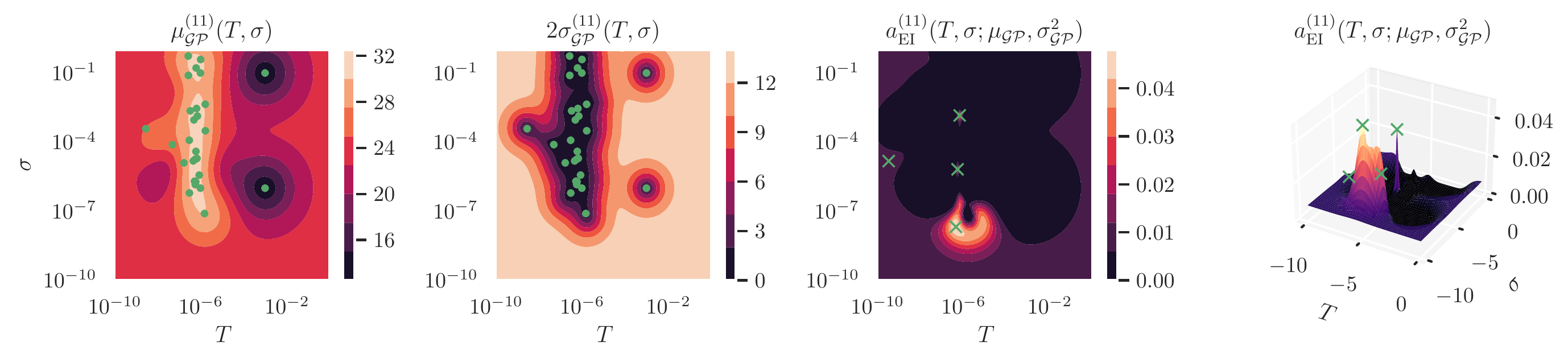}
    \caption{Denoising with POTOBIM: GP mean, confidence (2 standard deviations) and expected improvement acquisition function after BO iteration $ i \in \{0, 5, 7, 11\} $. Green dots denote observed points and green crosses show candidates for the next BO iteration. Note that per BO step, up to 4 candidates are evaluated in parallel.}
    \label{fig:app_bo_mfvi_den}
\end{figure*}

\begin{figure*}
    \centering
    \includegraphics[width=1\textwidth]{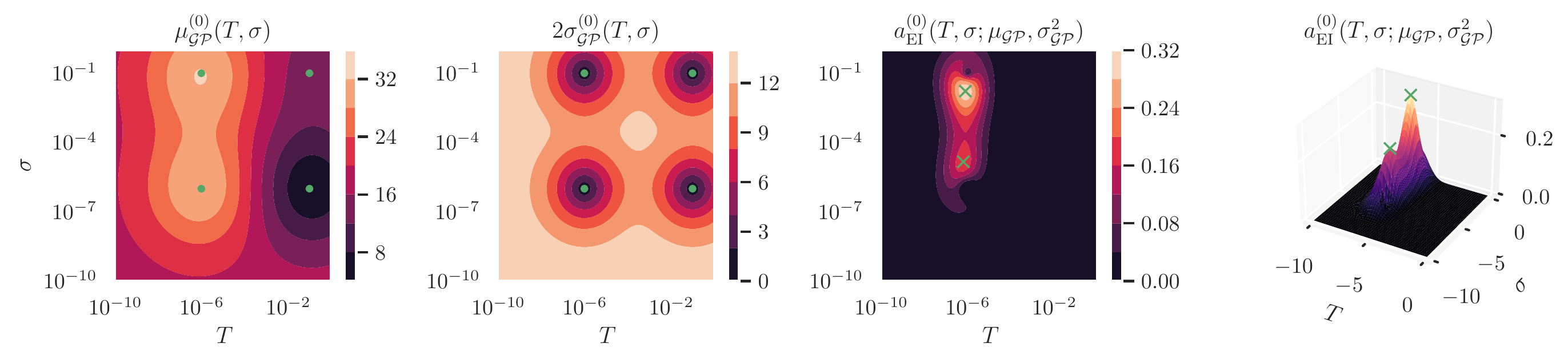}
    \includegraphics[width=1\textwidth]{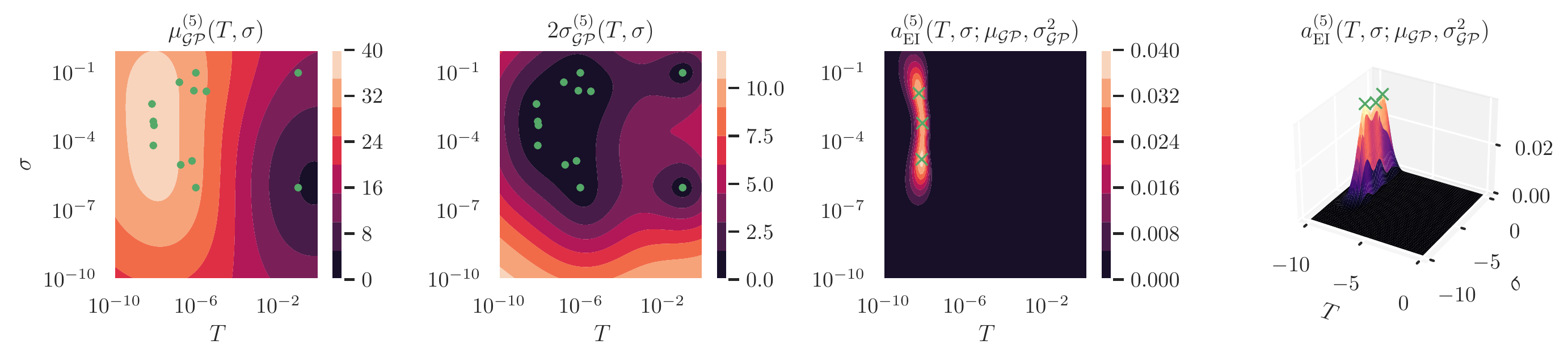}
    \includegraphics[width=1\textwidth]{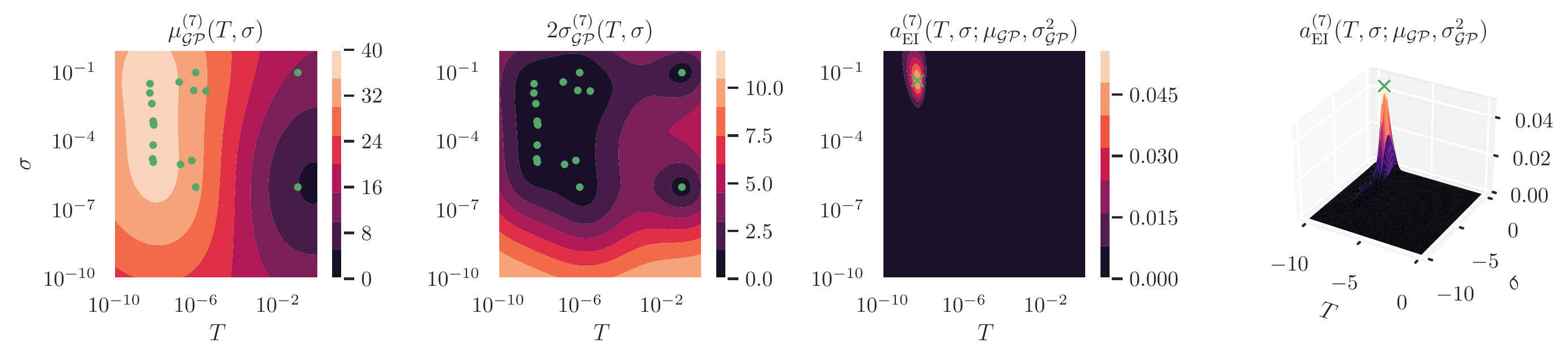}
    \includegraphics[width=1\textwidth]{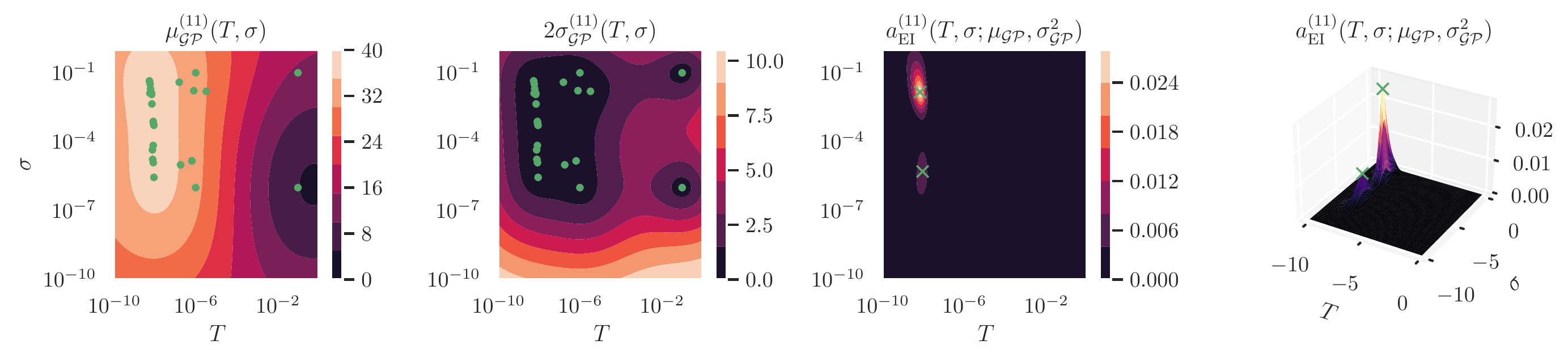}
    \caption{Hair inpainting with POTOBIM: GP mean, confidence (2 standard deviations) and expected improvement acquisition function after BO iteration $ i \in \{0, 5, 7, 11\} $. Green dots denote observed points and green crosses show candidates for the next BO iteration. Note that per BO step, up to 4 candidates are evaluated in parallel.}
    \label{fig:app_bo_mfvi_inp}
\end{figure*}

\begin{figure*}
    \centering
    \includegraphics[width=1\textwidth]{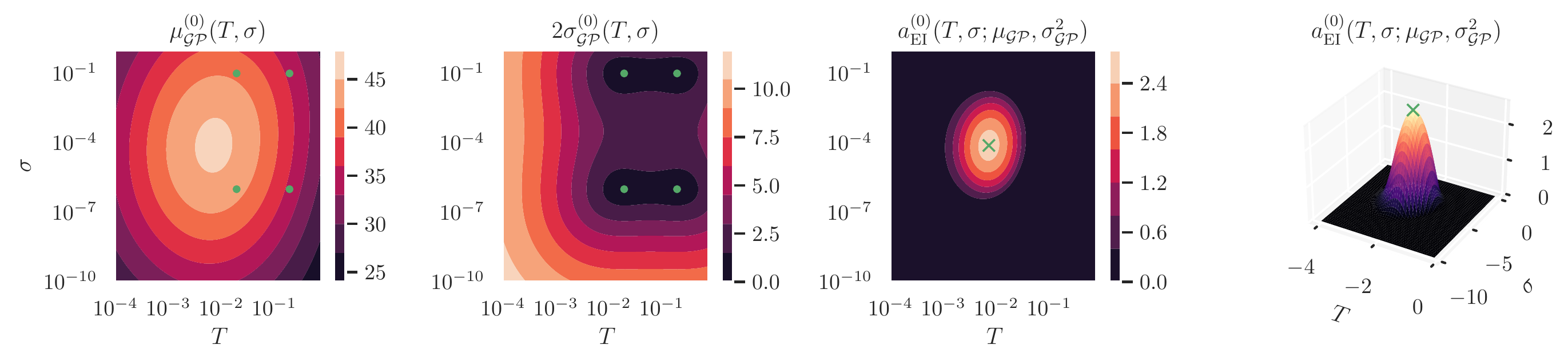}
    \includegraphics[width=1\textwidth]{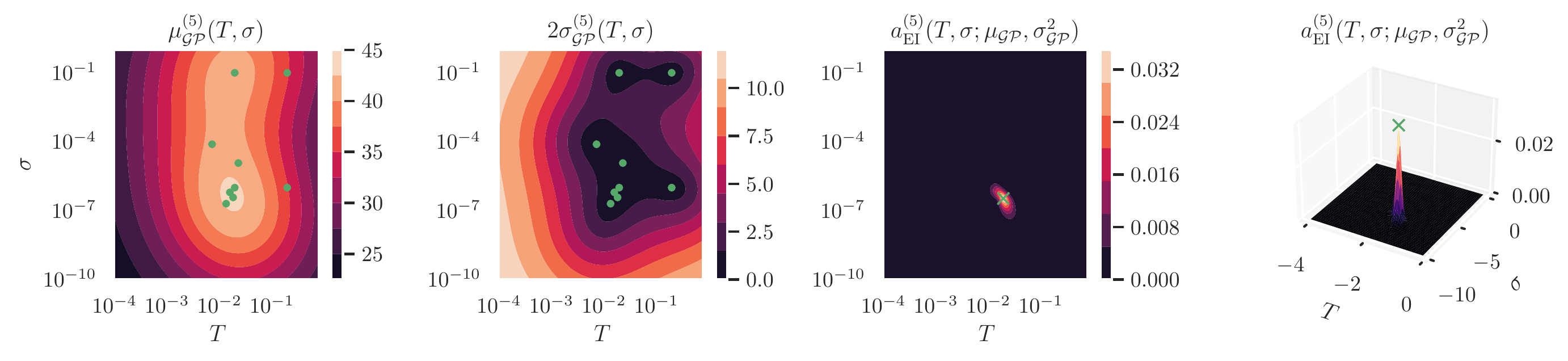}
    \includegraphics[width=1\textwidth]{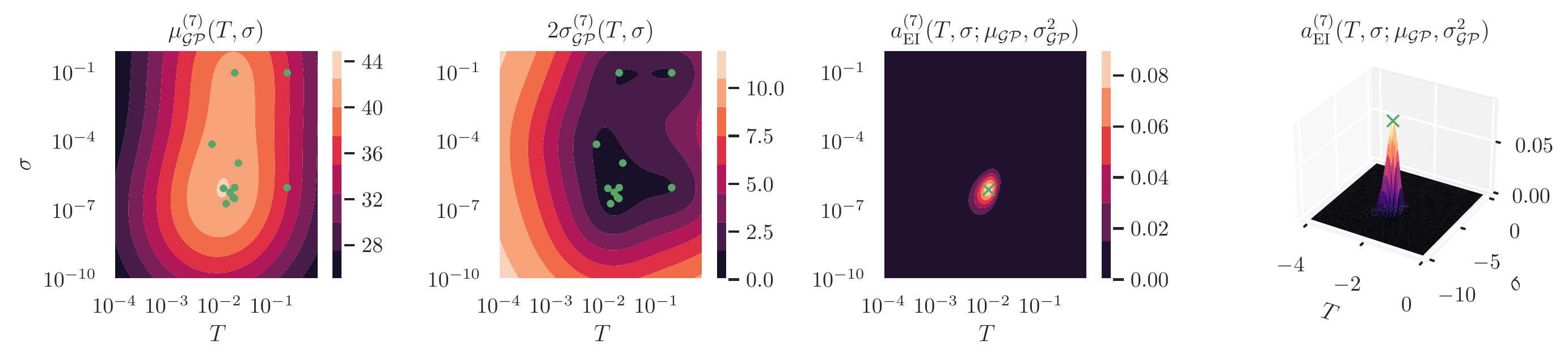}
    \includegraphics[width=1\textwidth]{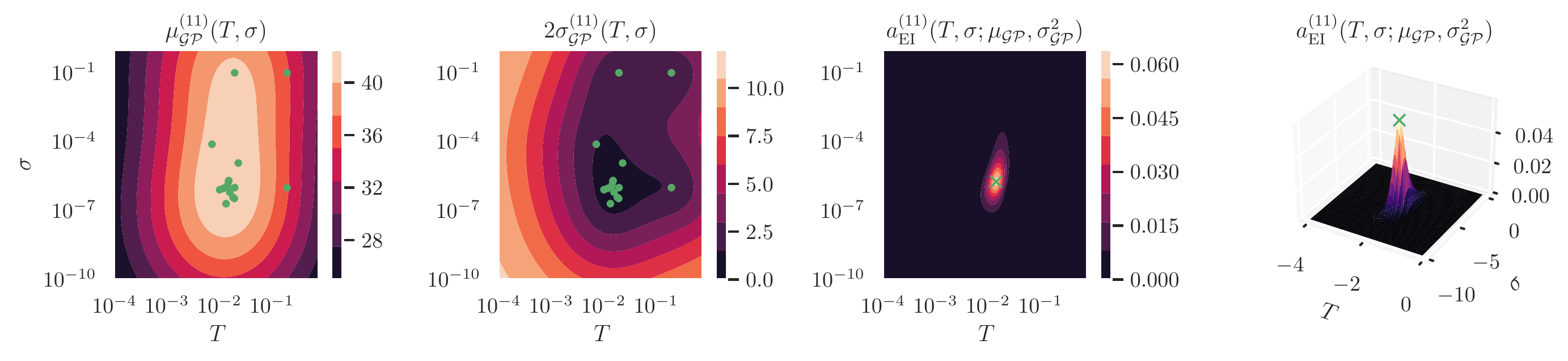}
    \caption{CT reconstruction with MCD: GP mean, confidence (2 standard deviations) and expected improvement acquisition function after BO iteration $ i \in \{0, 5, 7, 11\} $. Green dots denote observed points and green crosses show candidates for the next BO iteration. Note that per BO step, up to 4 candidates are evaluated in parallel.}
    \label{fig:app_bo_mcd_ct}
\end{figure*}

\begin{figure*}
    \centering
    \includegraphics[width=1\textwidth]{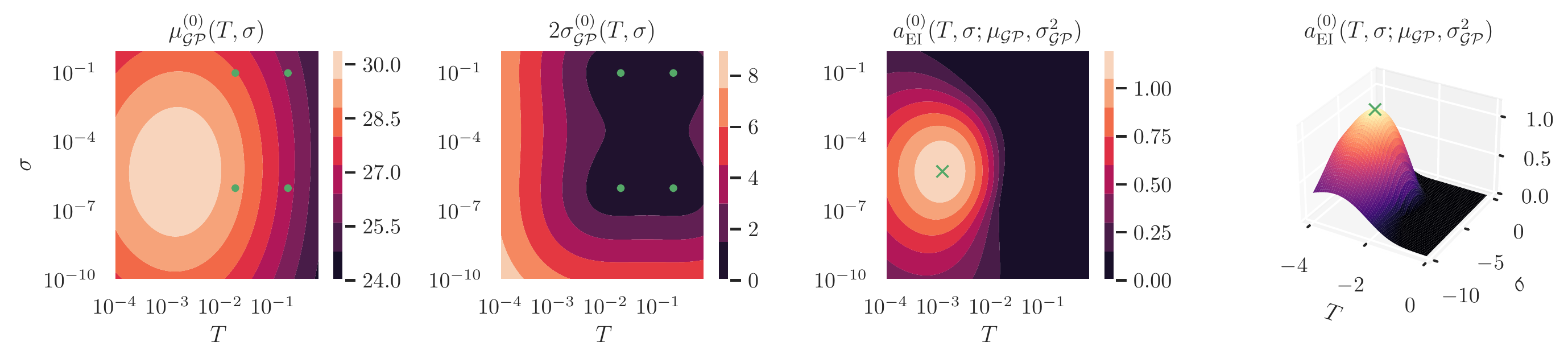}
    \includegraphics[width=1\textwidth]{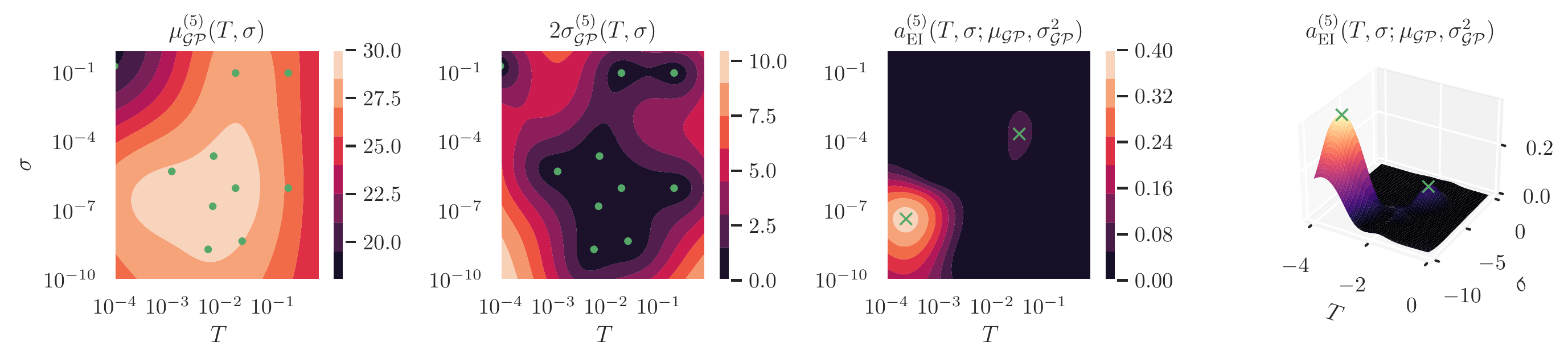}
    \includegraphics[width=1\textwidth]{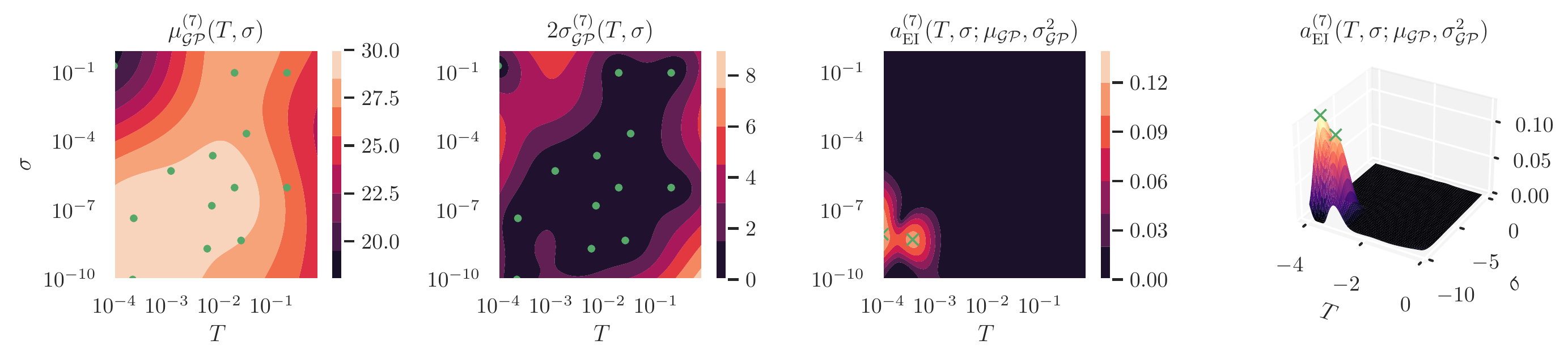}
    \includegraphics[width=1\textwidth]{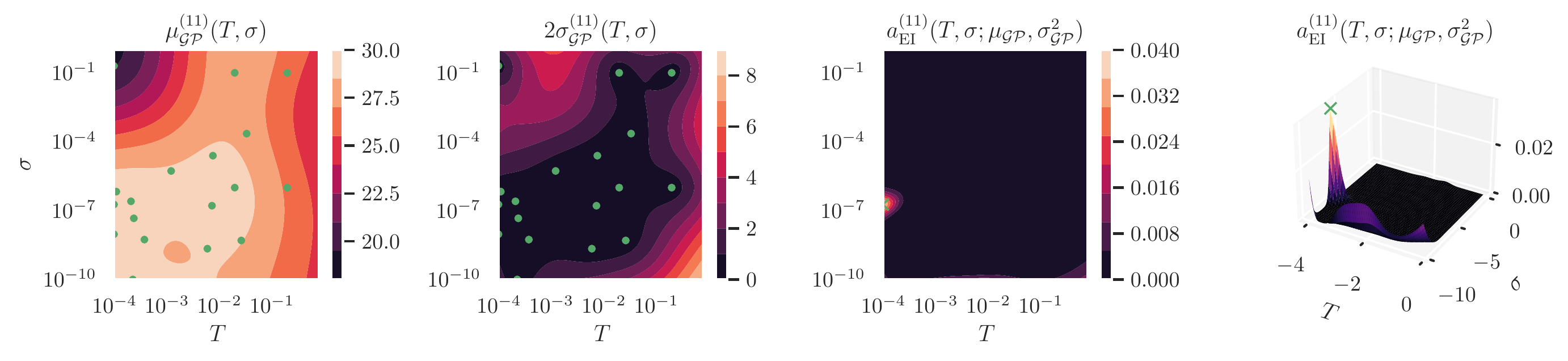}
    \caption{MRI super-resolution with MCD: GP mean, confidence (2 standard deviations) and expected improvement acquisition function after BO iteration $ i \in \{0, 5, 7, 11\} $. Green dots denote observed points and green crosses show candidates for the next BO iteration. Note that per BO step, up to 4 candidates are evaluated in parallel.}
    \label{fig:app_bo_mcd_sr}
\end{figure*}

\begin{figure*}
    \centering
    \includegraphics[width=1\textwidth]{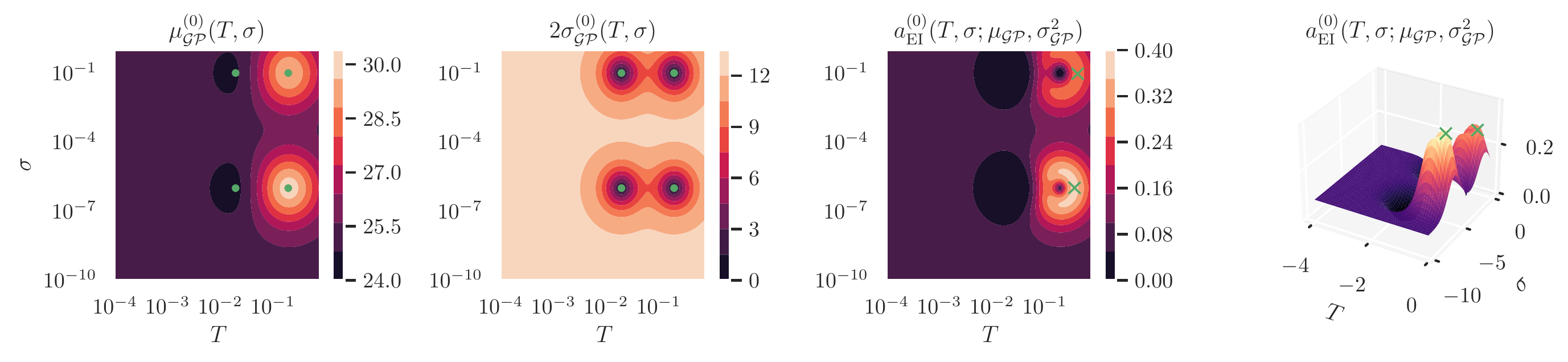}
    \includegraphics[width=1\textwidth]{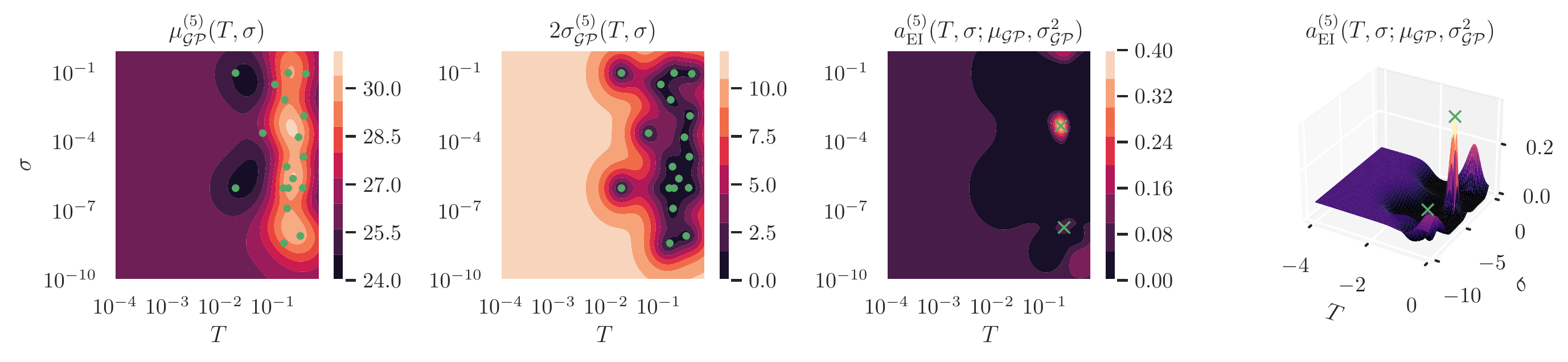}
    \includegraphics[width=1\textwidth]{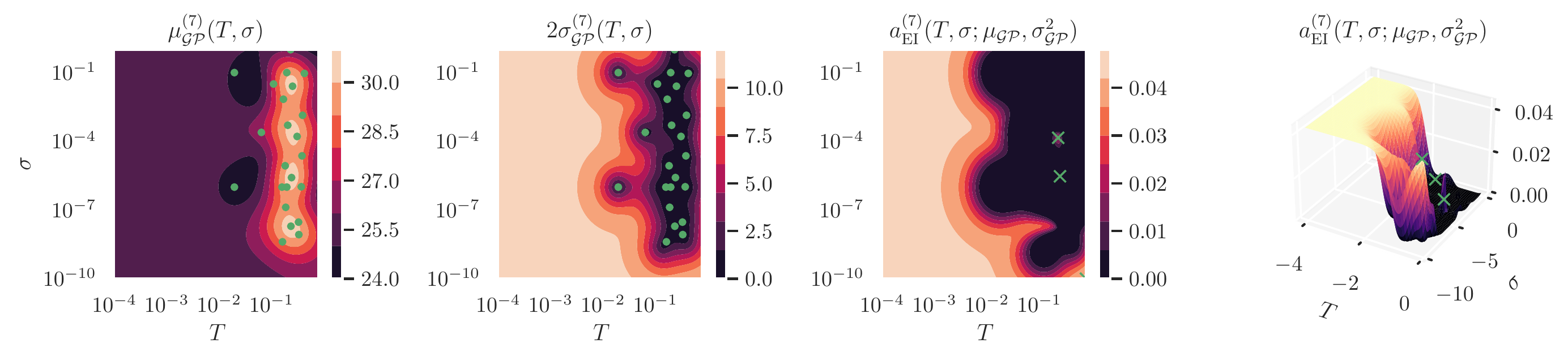}
    \includegraphics[width=1\textwidth]{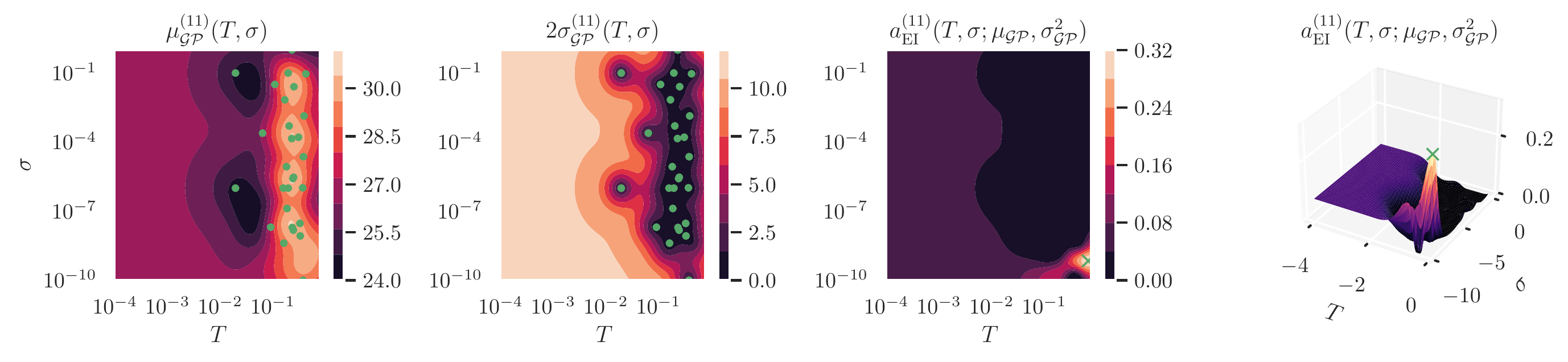}
    \caption{Denoising with MCD: GP mean, confidence (2 standard deviations) and expected improvement acquisition function after BO iteration $ i \in \{0, 5, 7, 11\} $. Green dots denote observed points and green crosses show candidates for the next BO iteration. Note that per BO step, up to 4 candidates are evaluated in parallel.}
    \label{fig:app_bo_mcd_den}
\end{figure*}

\begin{figure*}
    \centering
    \includegraphics[width=1\textwidth]{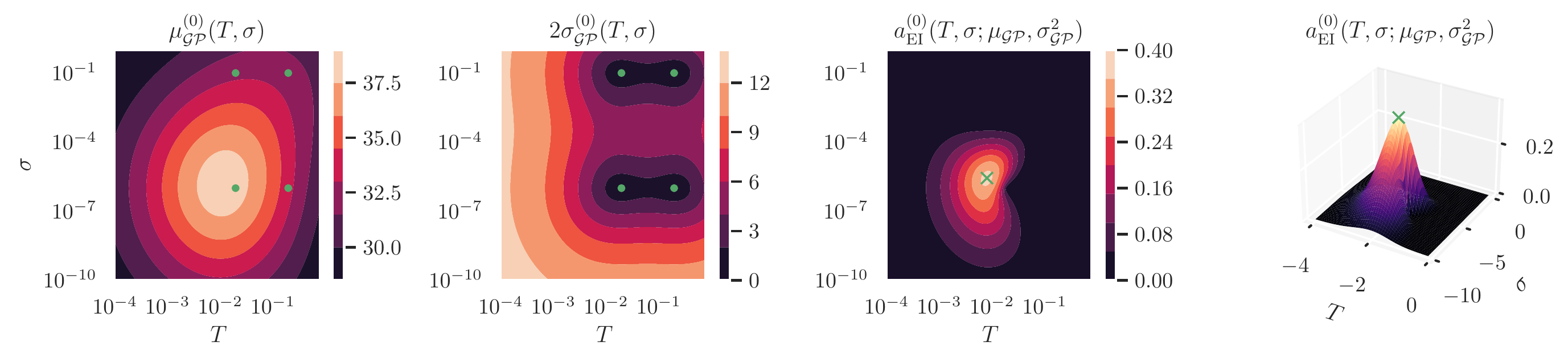}
    \includegraphics[width=1\textwidth]{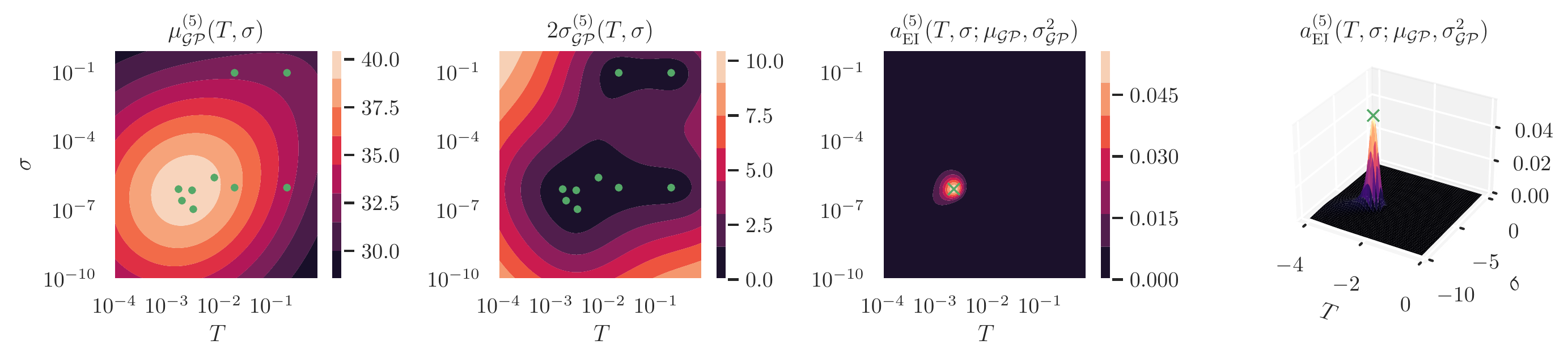}
    \includegraphics[width=1\textwidth]{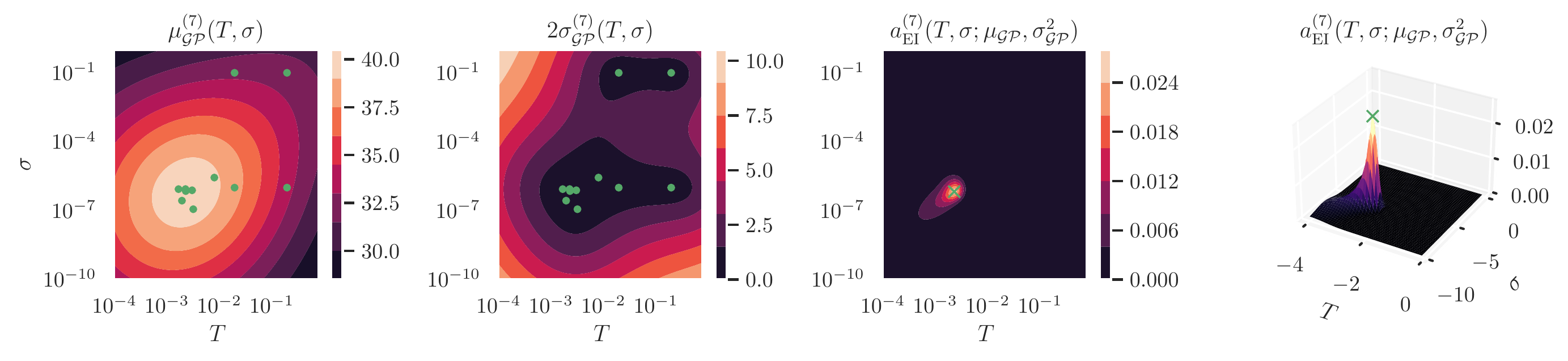}
    \includegraphics[width=1\textwidth]{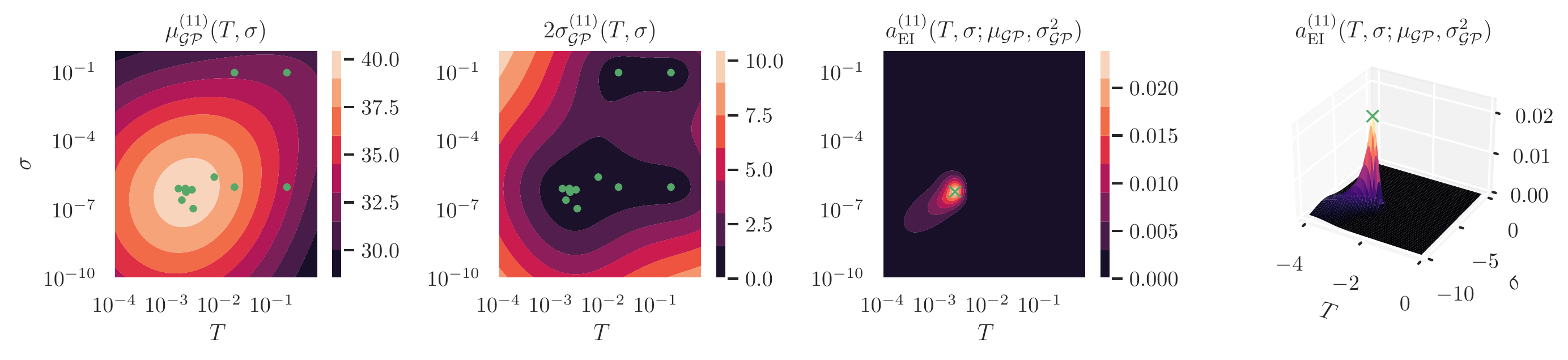}
    \caption{Hair inpainting with MCD: GP mean, confidence (2 standard deviations) and expected improvement acquisition function after BO iteration $ i \in \{0, 5, 7, 11\} $. Green dots denote observed points and green crosses show candidates for the next BO iteration. Note that per BO step, up to 4 candidates are evaluated in parallel.}
    \label{fig:app_bo_mcd_inp}
\end{figure*}

\begin{figure*}
    \centering
    \includegraphics[width=1\textwidth]{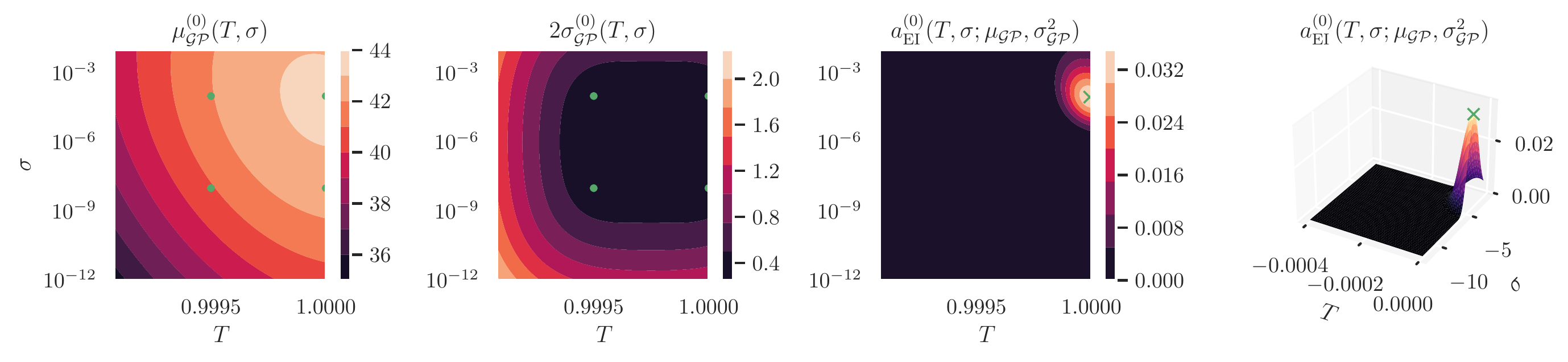}
    \includegraphics[width=1\textwidth]{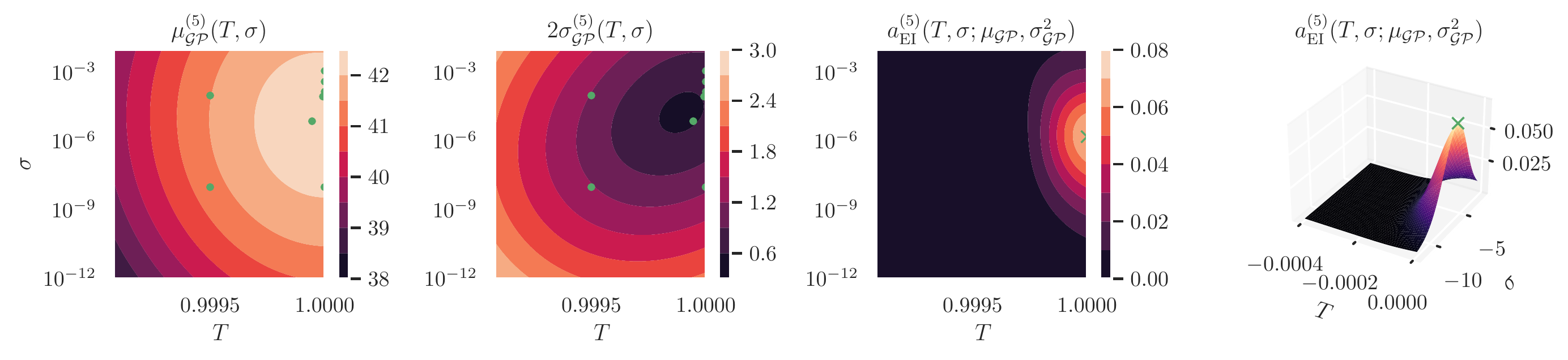}
    \includegraphics[width=1\textwidth]{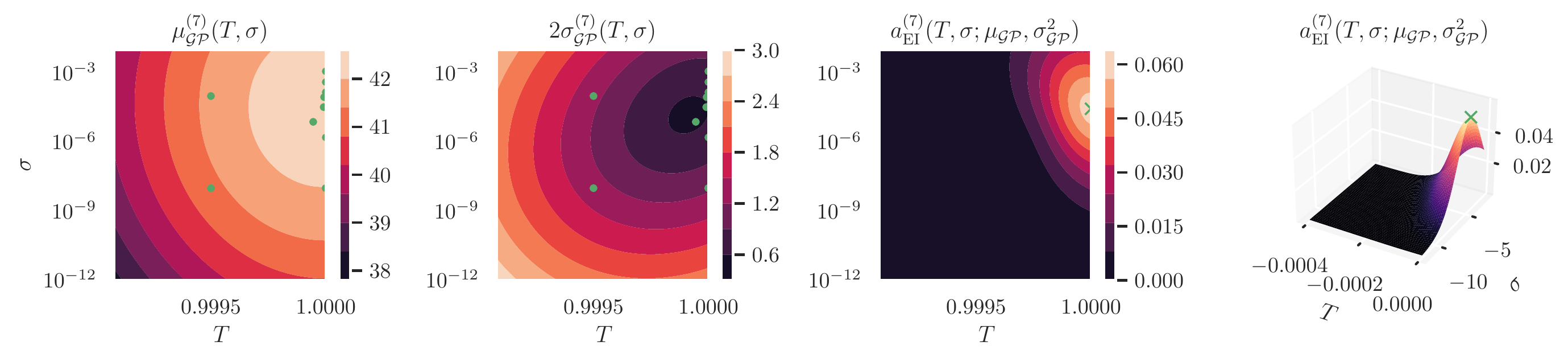}
    \includegraphics[width=1\textwidth]{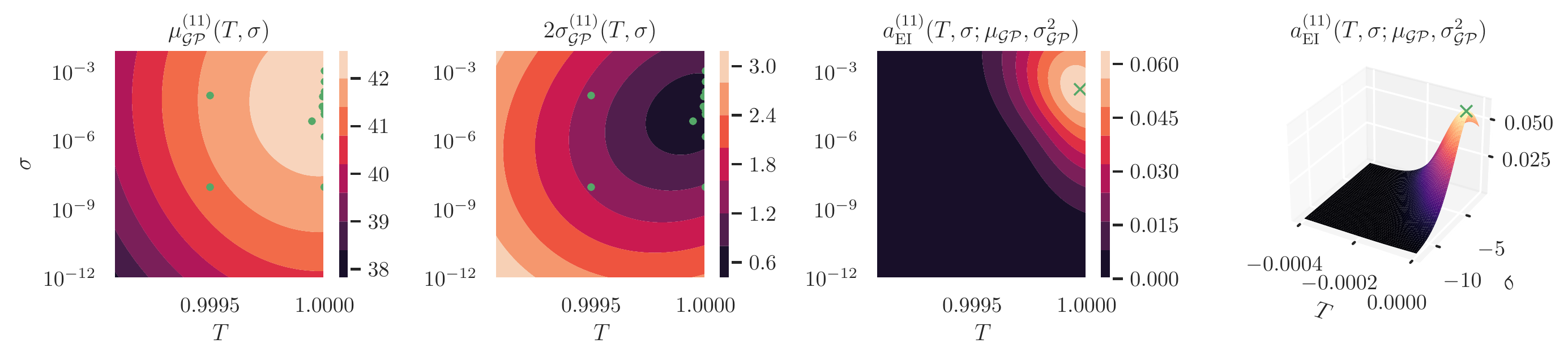}
    \caption{CT reconstruction with SGLD: GP mean, confidence (2 standard deviations) and expected improvement acquisition function after BO iteration $ i \in \{0, 5, 7, 11\} $. Green dots denote observed points and green crosses show candidates for the next BO iteration. Note that per BO step, up to 4 candidates are evaluated in parallel.}
    \label{fig:app_bo_sgld_ct}
\end{figure*}

\begin{figure*}
    \centering
    \includegraphics[width=1\textwidth]{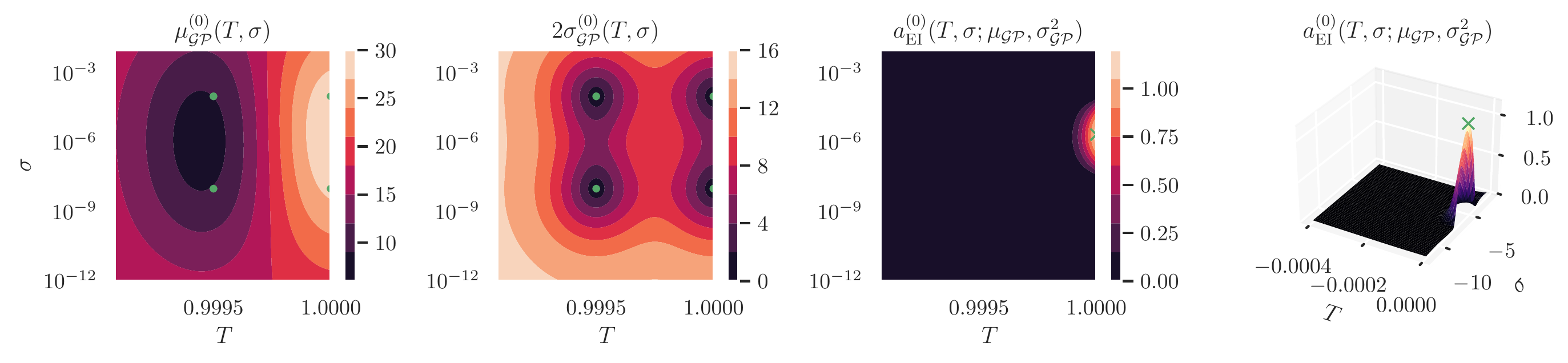}
    \includegraphics[width=1\textwidth]{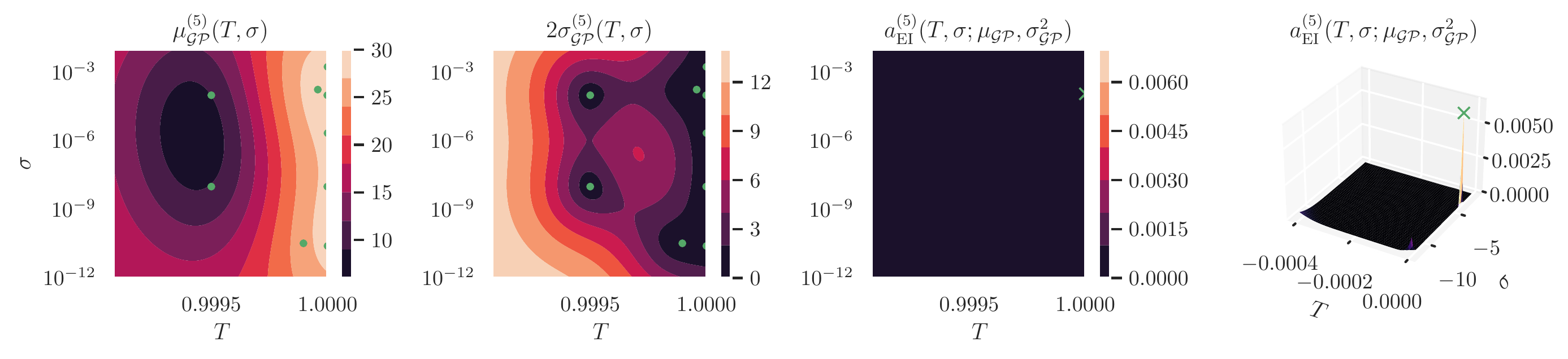}
    \includegraphics[width=1\textwidth]{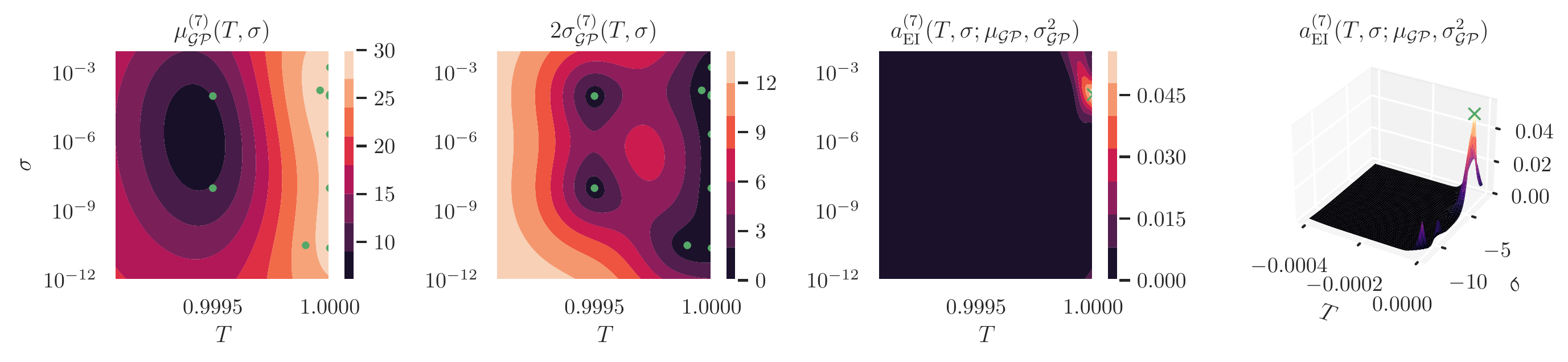}
    \includegraphics[width=1\textwidth]{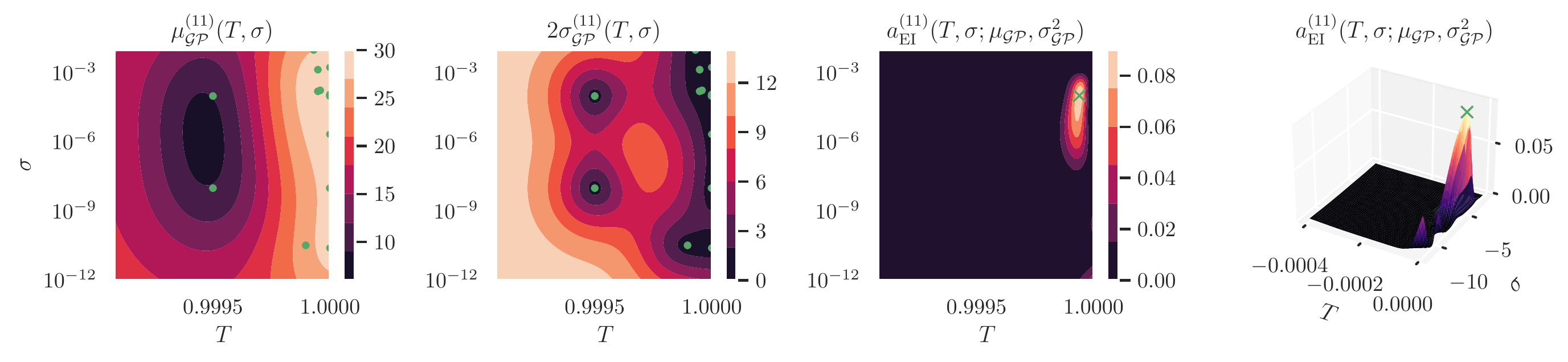}
    \caption{MRI super-resolution with SGLD: GP mean, confidence (2 standard deviations) and expected improvement acquisition function after BO iteration $ i \in \{0, 5, 7, 11\} $. Green dots denote observed points and green crosses show candidates for the next BO iteration. Note that per BO step, up to 4 candidates are evaluated in parallel.}
    \label{fig:app_bo_sgld_sr}
\end{figure*}

\begin{figure*}
    \centering
    \includegraphics[width=1\textwidth]{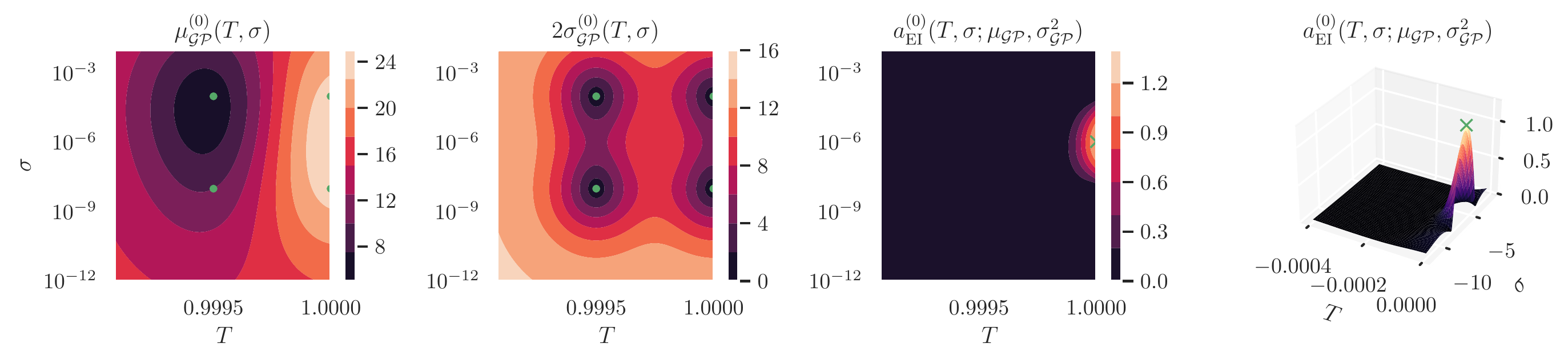}
    \includegraphics[width=1\textwidth]{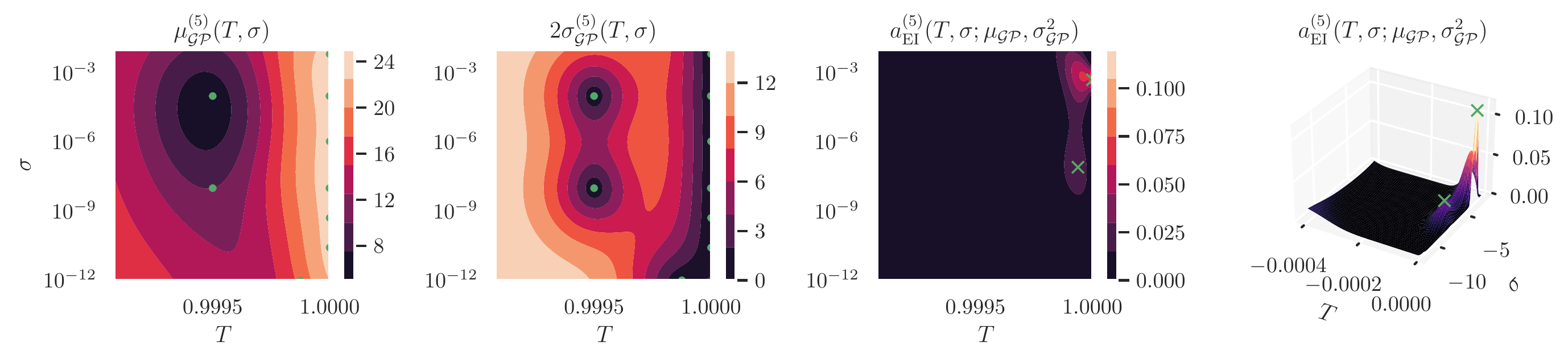}
    \includegraphics[width=1\textwidth]{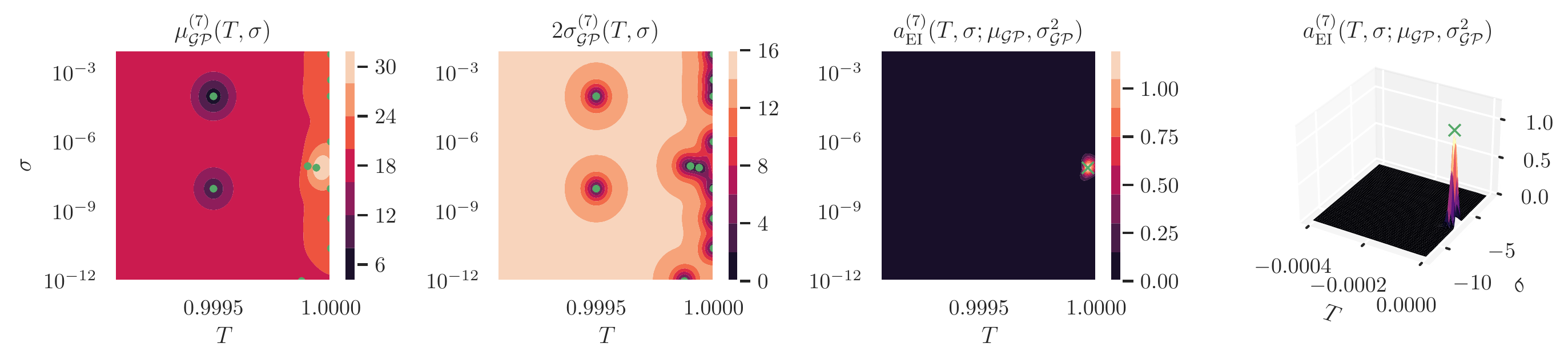}
    \includegraphics[width=1\textwidth]{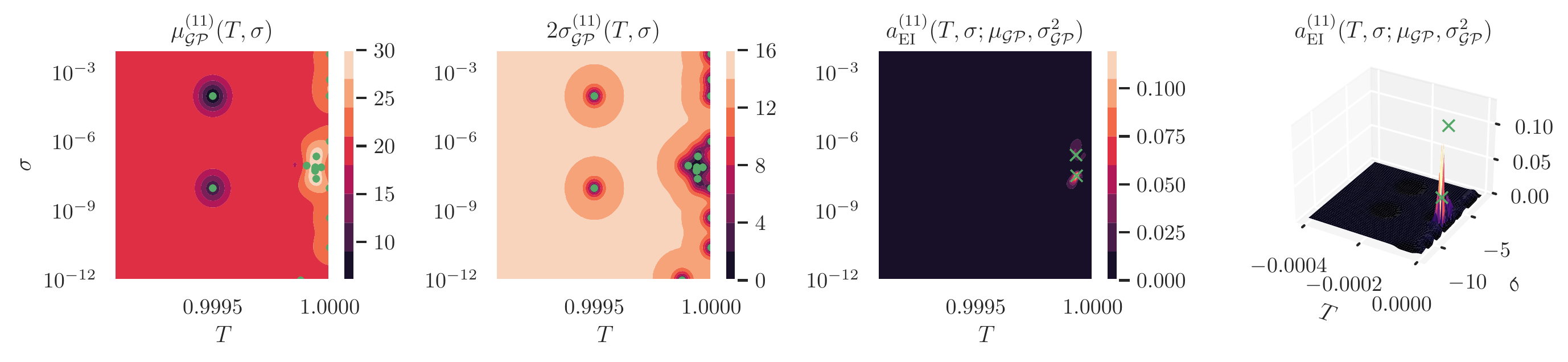}
    \caption{Denoising with SGLD: GP mean, confidence (2 standard deviations) and expected improvement acquisition function after BO iteration $ i \in \{0, 5, 7, 11\} $. Green dots denote observed points and green crosses show candidates for the next BO iteration. Note that per BO step, up to 4 candidates are evaluated in parallel.}
    \label{fig:app_bo_sgld_den}
\end{figure*}

\begin{figure*}
    \centering
    \includegraphics[width=1\textwidth]{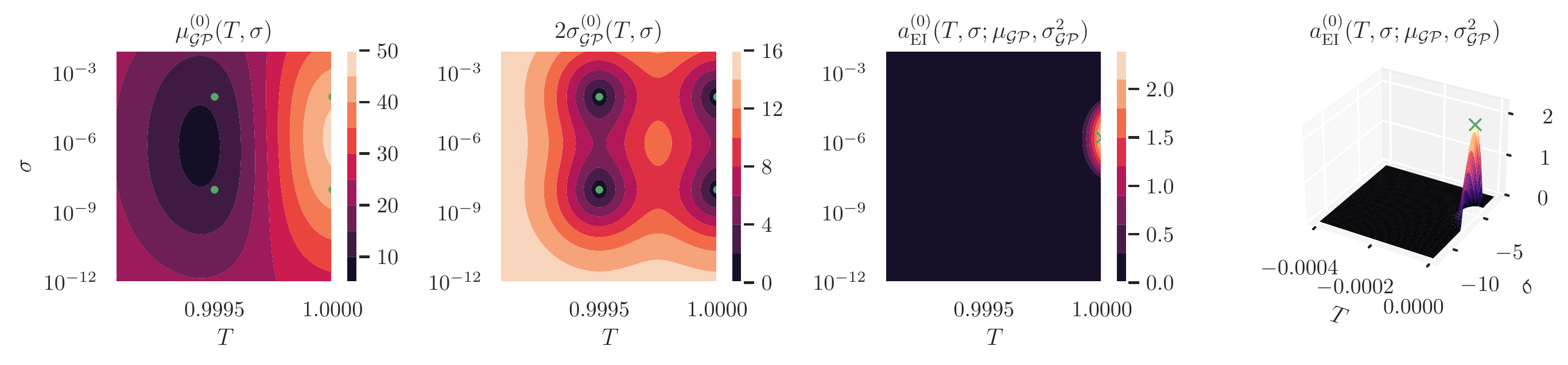}
    \includegraphics[width=1\textwidth]{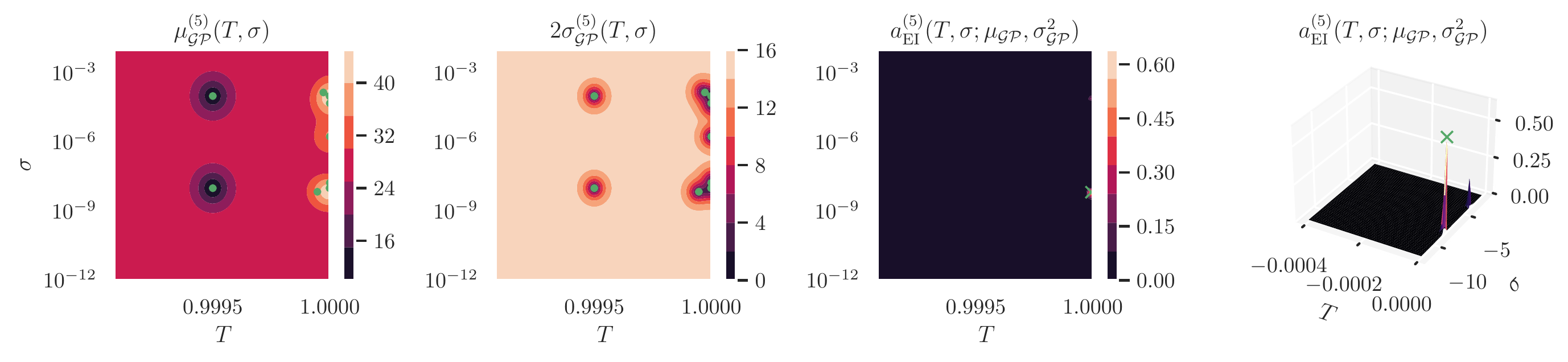}
    \includegraphics[width=1\textwidth]{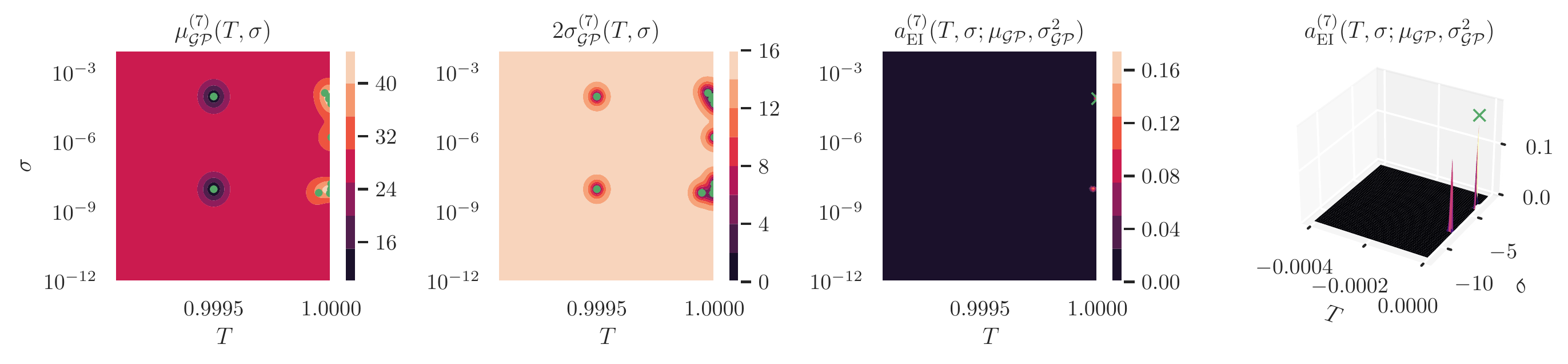}
    \includegraphics[width=1\textwidth]{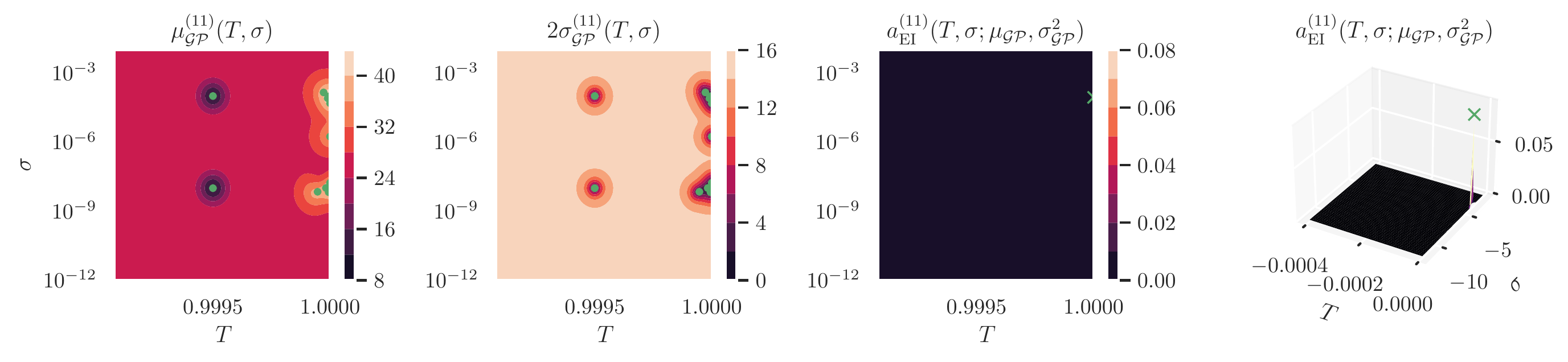}
    \caption{Hair inpainting with SGLD: GP mean, confidence (2 standard deviations) and expected improvement acquisition function after BO iteration $ i \in \{0, 5, 7, 11\} $. Green dots denote observed points and green crosses show candidates for the next BO iteration. Note that per BO step, up to 4 candidates are evaluated in parallel.}
    \label{fig:app_bo_sgld_inp}
\end{figure*}

\end{document}